\PassOptionsToPackage{names,dvipsnames}{xcolor} %

\documentclass[acmsmall]{acmart}
\usepackage{booktabs}   %
\usepackage{subcaption} %
\usepackage[english]{babel}
\usepackage[inference]{semantic} 
\usepackage{amsmath}\allowdisplaybreaks
\usepackage{mathpartir}
\usepackage{centernot}
\usepackage{amssymb}
\usepackage{stmaryrd} 
\usepackage{xspace}
\usepackage{latexsym}
\usepackage{ifthen}
\usepackage{mathtools}
\usepackage[names,dvipsnames]{xcolor}
\usepackage{color}
\usepackage{listings}
\usepackage{verbatim}
\usepackage[colorinlistoftodos,textsize=footnotesize]{todonotes}
\usepackage{tikz}
\usetikzlibrary{positioning,shadows,arrows,calc,backgrounds,fit,shapes,snakes,shapes.multipart,decorations.pathreplacing,shapes.misc,patterns}
\tikzset{myiff/.style={double,implies-implies,,double equal sign distance}}
\tikzset{myimpl/.style={,double,-implies,double equal sign distance}}
\usepackage{tikzscale}
\usepackage[T1]{fontenc}
\usepackage[scaled=.83]{beramono}
\usepackage{epigraph}
\usepackage{booktabs}
\usepackage{dashrule}
\usepackage{float}
\usepackage{etoolbox}
\usepackage{wrapfig}
\usepackage{nameref}
\usepackage{scalerel}
\usepackage{hyperref}
\usepackage[capitalize]{cleveref}
\usepackage{enumitem}
\usepackage{bm}
\usepackage[framemethod=TikZ]{mdframed}
\usepackage{msc}

\hypersetup{ pdfpagemode=UseOutlines, colorlinks=true, linkcolor=black, citecolor=blue } %

\AtBeginEnvironment{tikzpicture}{\catcode`$=3 } %

\DeclareMathAlphabet{\mathsfit}{T1}{\sfdefault}{\mddefault}{\sldefault}
\SetMathAlphabet{\mathsfit}{bold}{T1}{\sfdefault}{\bfdefault}{\sldefault}

\newcommand{\mi}[1]{\ensuremath{\mathit{#1}}}
\newcommand{\mr}[1]{\ensuremath{\mathrm{#1}}}

\newcommand{\mtt}[1]{\ensuremath{\mathtt{#1}}}
\newcommand{\mf}[1]{\ensuremath{\mathbf{#1}}}

\newcommand{\mc}[1]{\ensuremath{\mathcal{#1}}}
\newcommand{\ms}[1]{\ensuremath{\mathsf{#1}}}

\newcommand{\mb}[1]{\ensuremath{\mathbb{#1}}}

\newcommand{\isdef}[0]{\ensuremath{\mathrel{\overset{\makebox[0pt]{\mbox{\normalfont\tiny\sffamily def}}}{=}}}}

\newcommand{\relmiddle}[1]{\mathrel{}\middle#1\mathrel{}}

\newcommand\bnfdef{\ensuremath{\mathrel{::=}}}

\newcommand{\OB}[1]{\ensuremath{\overline{#1}}}

\newcommand{\Ra}{\ensuremath{\Rightarrow}}

\newcommand{\La}{\ensuremath{\Leftarrow}}

\newcommand{\myset}[2]{\ensuremath{\left\{#1 ~\relmiddle|~ #2\right\}}}

\newcommand{\partialmapsto}[0]{\rightharpoonup}

\newcommand{\termsl}[0]{\ensuremath{{\src{\Downarrow}}}\xspace}
\newcommand{\termt}[0]{\ensuremath{{\trg{\Downarrow}}}\xspace}
\newcommand{\termc}[0]{\ensuremath{{\com{\Downarrow}}}\xspace}

\newcommand{\isterm}[1]{\ensuremath{\vdash #1 : \mi{term}}}

\AtEndEnvironment{example}{\null\hfill$\boxdot$}
\AtEndEnvironment{axiom}{\null\hfill$\boxdot$}

\Crefname{lstlisting}{Listing}{Listings}
\Crefname{problem}{Problem}{Problems}

\Crefname{equation}{Rule}{Rules}

\newcommand{\compskel}[3]{\ensuremath{\bl{\left\llbracket \mr{#1} \right\rrbracket^{#2}_{#3}}}}
\newcommand{\comp}[1]{\compskel{#1}{}{}}
\newcommand{\compgen}[1]{\compskel{#1}{\S}{\T}}
\newcommand{\compgento}[1]{\compskel{#1}{\T}{\O}}

\newcommand{\compgents}[1]{\compskel{#1}{\T}{\S}}

\newcommand{\compll}[1]{\compskel{#1}{\Lo}{\Lt}}

\newcommand{\compversion}[1]{\phantom{}_{#1}}
\newcommand{\compo}[0]{\compversion{1}}
\newcommand{\compt}[0]{\compversion{2}}

\newcommand{\compucl}[1]{\compskel{#1}{{\UCLang}}{{\UCLang}}}
\newcommand{\compuclo}[1]{\compskel{#1}{{\UCLango}}{{\UCLango}}}

\newcommand{\compsucl}[1]{\compskel{#1}{\S}{{\UCLango}}}
\newcommand{\compucls}[1]{\compskel{#1}{{\UCLango}}{\S}}
\newcommand{\comptucl}[1]{\compskel{#1}{\T}{{\UCLango}}}
\newcommand{\compuclt}[1]{\compskel{#1}{{\UCLango}}{\T}}

\newcommand{\compcomm}[1]{\compskel{#1}{}{\comsc{comm}}}
\newcommand{\compcommad}[1]{\compskel{#1}{\comsc{adap}}{\comsc{comm}}}
\newcommand{\prgscomm}[0]{\prgs{comm}}
\newcommand{\prgtcomm}[0]{\prgt{comm}}
\newcommand{\idfucomm}[0]{\idfu{comm}}
\newcommand{\protcomm}[0]{\prot{comm}}

\newcommand{\funname}[1]{\mtt{#1}}
\newcommand{\fun}[2]{\ensuremath{\bl{\funname{#1}\left( #2 \right)}}\xspace}
\newcommand{\dom}[1]{\fun{dom}{#1}}

\newcommand{\ilc}[0]{\text{ILC}\xspace}
\newcommand{\ILC}[0]{\ilc}
\newcommand{\rrilc}[0]{\text{RILC}\xspace}

\newcommand{\rilc}[0]{\rrilc}

\newcommand{\itm}[0]{\text{ITM}\xspace}
\newcommand{\ITM}[0]{\itm}
\newcommand{\itms}[0]{\text{ITMs}\xspace}
\newcommand{\ITMs}[0]{\itms}

\newcommand{\UCLang}[0]{\text{UCLang}\xspace}
\newcommand{\UCLango}[0]{\UCLang} %

\newcommand{\deepsec}[0]{\text{DEEPSEC}\xspace}
\newcommand{\DEEPSEC}[0]{\deepsec}

\definecolor{mydarkgrey}{rgb}{0.3,0.3,0.4}

\newcommand{\langtypeset}[1]{\mc{#1}}

\let\Ss\S
\renewcommand{\S}[0]{\src{\langtypeset{S}}\xspace}
\newcommand{\T}[0]{\trg{\langtypeset{T}}\xspace}

\renewcommand{\O}[0]{\oth{{O}}\xspace}

\newcommand{\Lo}[0]{\comsc{\langtypeset{\langlett_{\mr{1}}}}\xspace}
\newcommand{\Lt}[0]{\comsc{\langtypeset{\langlett_{\mr{2}}}}\xspace}

\newcommand{\langlett}[0]{L} %

\newcommand{\contextletter}[0]{A}
\newcommand{\ctx}[1]{#1}%
\newcommand{\ctxs}[1]{\src{\ctx{\contextletter}#1}\xspace} 
\newcommand{\ctxt}[1]{\trg{\ctx{\contextletter}#1}\xspace}%
\newcommand{\ctxc}[1]{\com{\ctx{\contextletter}#1}\xspace}%
\newcommand{\ctxo}[1]{\oth{\ctx{\contextletter}#1}\xspace}%

\newcommand{\ctxtdummy}[0]{\ctxt{_\dummyannot}}

\newcommand{\prgtdummy}[0]{\prgt{\dummyannot}}

\newcommand{\ctxhs}[2]{\ctxs{#1}{\holes{#2}}\xspace}

\newcommand{\ctxhc}[2]{\ctxc{#1}{\hole{#2}}\xspace}

\newcommand{\evalctx}[0]{\ensuremath{{E}}}

\newcommand{\programletter}[0]{P}
\newcommand{\prggen}[2]{\programletter_{#1}^{#2}\xspace}

\newcommand{\prgs}[1]{\src{\prggen{#1}{}}\xspace}
\newcommand{\prgt}[1]{\trg{\prggen{#1}{}}\xspace}
\newcommand{\prgc}[1]{\com{\prggen{#1}{}}\xspace}
\newcommand{\prgo}[1]{\oth{\prggen{#1}{}}\xspace}

\newcommand{\wholeprogramletter}[0]{W}
\newcommand{\wprggen}[2]{\wholeprogramletter_{\ensuremath{#1}}^{\ensuremath{#2}}}

\newcommand{\wprgs}[1]{\src{\wprggen{#1}{}}\xspace}
\newcommand{\wprgt}[1]{\trg{\wprggen{#1}{}}\xspace}
\newcommand{\wprgc}[1]{\com{\wprggen{#1}{}}\xspace}

\newcommand{\holev}[1]{\ensuremath{\left[#1\right]}}
\newcommand{\linksymbol}[0]{\bowtie}

\newcommand{\linktt}[0]{\, \trg{\ensuremath{{{\linksymbol}}}}\, }
\newcommand{\links}[0]{\mathrel{\src{\linksymbol}}}
\newcommand{\linko}[0]{\mathrel{\oth{\linksymbol}}}
\newcommand{\linkc}[0]{\mathrel{\com{\linksymbol}}}
\newcommand{\hole}[1]{\ensuremath{ \linksymbol #1 }}
\newcommand{\holes}[1]{\ensuremath{ \links #1 }} %

\newcommand{\trues}[0]{\src{{true}}\xspace}
\newcommand{\falses}[0]{\src{{false}}\xspace}

\newcommand{\come}[0]{\com{\emptyset}\xspace}

\newcommand{\neutcol}[0]{black}
\newcommand{\srccol}[0]{RoyalBlue}
\newcommand{\trgcol}[0]{RedOrange}
\newcommand{\othercol}[0]{YellowOrange}
\newcommand{\idecol}[0]{Emerald}
\newcommand{\commoncol}[0]{black}    %

\newcommand{\col}[2]{\ensuremath{{\color{#1}{#2}}}}

\newcommand{\src}[1]{\mi{\col{\srccol}{#1}}}
\newcommand{\trg}[1]{\mf{\col{\trgcol }{#1}}}
\newcommand{\trgnb}[1]{\trg{#1}}

\newcommand{\oth}[1]{\mi{\col{CarnationPink}{#1}}}

\newcommand{\ide}[1]{\mtt{\col{\idecol }{#1}}}
\newcommand{\con}[1]{\ms{\col{\othercol }{#1}}}

\newcommand{\bl}[1]{\col{\neutcol}{#1}}
\newcommand{\com}[1]{\mi{\col{\commoncol }{#1}}}
\newcommand{\comuc}[1]{\ms{\col{\commoncol }{#1}}}
\newcommand{\comsc}[1]{\mi{\col{\commoncol }{#1}}}

\newcounter{typerule}
\crefname{typerule}{rule}{rules}

\newcommand{\typeruleInt}[5]{%
	\def\thetyperule{#1}%
	\refstepcounter{typerule}%
	\label{tr:#4}%
  \ensuremath{\begin{array}{c}#5 \inference{#2}{#3}\end{array}} 
}
\newcommand{\typerule}[4]{%
  \typeruleInt{#1}{#2}{#3}{#4}{\textsf{\scriptsize ({#1})} \\      }
}

\pgfdeclarelayer{background}
\pgfdeclarelayer{threadground}
\pgfdeclarelayer{veryback}
\pgfdeclarelayer{veryback2}
\pgfdeclarelayer{veryback3}
\pgfdeclarelayer{back2}
\pgfdeclarelayer{foreground}

\newcommand{\BREAK}[0]{
\botrule
\begin{center}$\spadesuit$\end{center}
\botrule}

\def\botrule{\vspace{0mm}\hrule\vspace{2mm}}

\DeclareMathOperator\compat{\ensuremath{\sim}} %
\DeclareMathOperator\precompat{\ensuremath{\lesssim}}

\newcommand{\extcompat}[4]{\ensuremath{#1, #2 \vdash #3 \compat #4}}
\newcommand{\extprecompat}[3]{\ensuremath{#1 \vdash #2 \precompat #3}}

\newcommand{\xto}[1]{\ensuremath{~\mathrel{\xrightarrow{~#1~}}~}}
\newcommand{\Xto}[1]{\ensuremath{~\mathrel{\xRightarrow{~#1~}}~}}

\definecolor{mygreen}{rgb}{0,0.6,0}
\definecolor{mygray}{rgb}{0.5,0.5,0.5}
\definecolor{mymauve}{rgb}{0.58,0,0.82}

\lstdefinelanguage{SRC} %
{morekeywords={abstract, all, and, as, assert, but, disj, else, exactly, extends, fact, for, fun, iden, if, iff, implies, in, Int, int, let, lone, module, no, none, not, one, open, or, part, pred, run, seq, set, sig, some, sum, then, univ, package, class, public, private, null, return, new, interface, extern, object, implements, System, static, super, try , catch, throw, throws, Unit, var, val, of, principal, trust, rd, wr, to, from, Read, Write, imports, exports, definitions, processes, loop, fwd, takernd, prg, invert, keygen, xors, choice, error},
sensitive=true,
keywordstyle=\itshape\color{\srccol}, %
commentstyle=\color{purple!40!black},
morecomment=[l][\small\itshape\color{purple!40!black}]{//},
identifierstyle=\color{\srccol},
stringstyle=\color{orange},
basicstyle=\small,
basicstyle={\small\ttfamily},
numbers=left,
numberstyle=\tiny\color{mygray},
tabsize=2,
numbersep=3pt,
breaklines=true,
lineskip=-2pt,
stepnumber=1,
captionpos=b,
breaklines=true,
breakatwhitespace=false,
showspaces=false,
showtabs=false,
float=!h,
columns=fullflexible,escapeinside={(*@}{@*)},
moredelim=**[is][\color{red!60}]{@}{@},
literate={->}{{$\to$}}1 {^}{{$\mspace{-3mu}\widehat{\quad}\mspace{-3mu}$}}1
{<}{$\langle$}2 {>}{$\rangle$}2 {>=}{$\geq$}2 {=<}{$\leq$}2
{<:}{{$<\mspace{-3mu}:$}}2 {:>}{{$:\mspace{-3mu}>$}}2
{=>}{{$\Rightarrow$ }}2 {+}{$+$ }2 {++}{{$+\mspace{-8mu}+$ }}2
{<=>}{{$\Leftrightarrow$ }}2 {+}{$+$ }2 {++}{{$+\mspace{-8mu}+$ }}2
{\~}{{$\mspace{-3mu}\widetilde{\quad}\mspace{-3mu}$}}1
{!=}{$\neq$ }2 {*}{${}^{\ast}$}1 %
{\#}{$\#$}1
}

\lstdefinelanguage{TRG}
{morekeywords={abstract, all, and, as, assert, but, check, disj, else, exactly, extends, fact, for, fun, iden, if, iff, implies, in, Int, int, let, lone, module, no, none, not, one, open, or, part, pred, run, seq, set, sig, some, sum, then, univ, package, class, public, private, null, return, new, interface, extern, object, implements, System, static, super, try , catch, throw, throws, Unit, var, val, principal, trust, label, load, add, addi, into, test, rd, wr, to, from, Read, Write, imports, exports, definitions, processes, loop, fwd, takernd, prg, invert, keygen, xors, choice, error},
sensitive=true,
keywordstyle=\bfseries\color{\trgcol}, %
commentstyle=\color{purple!40!black},
morecomment=[l][\small\itshape\color{purple!40!black}]{//},
identifierstyle=\color{\trgcol},
stringstyle=\color{orange},
basicstyle=\small,
basicstyle={\small\ttfamily},
numbers=left,
numberstyle=\tiny\color{mygray},
tabsize=2,
numbersep=3pt,
breaklines=true,
lineskip=-2pt,
stepnumber=1,
captionpos=b,
breaklines=true,
breakatwhitespace=false,
showspaces=false,
showtabs=false,
float=!h,
columns=fullflexible,escapeinside={(*@}{@*)},
moredelim=**[is][\color{red!60}]{@}{@},
literate={->}{{$\to$}}1 {^}{{$\mspace{-3mu}\widehat{\quad}\mspace{-3mu}$}}1
{<}{$<$ }2 {>}{$>$ }2 {>=}{$\geq$ }2 {=<}{$\leq$ }2
{<:}{{$<\mspace{-3mu}:$}}2 {:>}{{$:\mspace{-3mu}>$}}2
{=>}{{$\Rightarrow$ }}2 {+}{$+$ }2 {++}{{$+\mspace{-8mu}+$ }}2
{<=>}{{$\Leftrightarrow$ }}2 {+}{$+$ }2 {++}{{$+\mspace{-8mu}+$ }}2
{\~}{{$\mspace{-3mu}\widetilde{\quad}\mspace{-3mu}$}}1
{!=}{$\neq$ }2 {*}{${}^{\ast}$}1 %
{\#}{$\#$}1
}
\DeclareMathOperator\ceq{\ensuremath{\mathrel{\simeq_{ctx}}}}

\DeclareMathOperator\ceqs{\src{\simeq_{ctx}}}
\DeclareMathOperator\ceqt{\trgnb{\bm{\simeq}_{\trg{ctx}}}}

\DeclareMathOperator\aeq{\bl{=_{\alpha}}}

\newcommand{\contradiction}[0]{\hfill\ensuremath{\lightning}}

\def\teqaux#1{\vcenter{\hbox{\ooalign{\hfil
       \raise6pt \hbox{\scriptsize{T}}\hfil\cr\hfil
       $=$}}}}

\def\ceqwaux#1{\vcenter{\hbox{\ooalign{\hfil
       \raise6pt \hbox{\scriptsize{w-b}}\hfil\cr\hfil
       $\ceq$}}}}

\def\praux#1{\vcenter{\hbox{\ooalign{\hfil
       \raise4pt \hbox{$\subset$}\hfil\cr\hfil
       $\sim$}}}}

\newcommand{\labelfont}[1]{\ensuremath{\mathlabel{#1}}}

\newcommand{\sndgen}[2]{\ensuremath{\labelfont{send}~ #1~ \labelfont{on}~#2}}
\newcommand{\snd}[2]{\ensuremath{\sndgen{#1}{#2}?}}
\newcommand{\sndb}[2]{\ensuremath{\sndgen{#1}{#2}!}}

\newcommand{\behav}[1]{{Behav}\left(#1\right)}
\newcommand{\behavs}[1]{\src{\behav{#1}}}
\newcommand{\behavt}[1]{\trg{\behav{#1}}}
\newcommand{\behavc}[1]{\com{\behav{#1}}}

\DeclareMathOperator\relateAbs{\bl{\bumpeq}}%
\DeclareMathOperator\beheq{\relateAbs}

\DeclareMathOperator\isprefix{\bl{\leq}}

\newcommand{\sem}[0]{\mathrel{\rightsquigarrow}}
\newcommand{\sems}[0]{\mathrel{\src{\sem}}}
\newcommand{\semt}[0]{\, \trgnb{\boldsymbol{\sem}}\, }
\newcommand{\semc}[0]{\mathrel{\com{\sem}}}

\newcommand{\tracesymbol}[0]{t}
\newcommand{\trace}[0]{\com{\alphaseq}}
\newcommand{\prefixsymbol}[0]{m}
\newcommand{\prefix}[0]{\com{\prefixseq}}

\newcommand{\actsymbol}[0]{\tau}

\newcommand{\intrace}[0]{\com{\OB{\actsymbol}}}
\newcommand{\inprefix}[0]{\com{\OB{\mu}}}

\newcommand{\alphaseq}[0]{\OB{\tracesymbol}}

\newcommand{\prefixseq}[0]{\OB{\prefixsymbol}}

\newcommand{\at}[0]{\trg{\actsymbol}\xspace}
\newcommand{\ac}[0]{\com{\actsymbol}\xspace}

\newcommand{\noact}[0]{\epsilon}
\newcommand{\errt}[0]{\maltese}

\newcommand{\lam}[2]{\ensuremath{\lambda #1\ldotp #2}}
\newcommand{\pair}[1]{\ensuremath{\left\langle#1\right\rangle}}

\newcommand{\true}{\ensuremath{{true}}}
\newcommand{\false}{\ensuremath{{false}}}
\newcommand{\unit}{\ensuremath{{unit}}}

\newcommand{\inl}[1]{\ensuremath{{inl}~#1}}
\newcommand{\inr}[1]{\ensuremath{{inr}~#1}}

\newcommand{\bop}[0]{\ensuremath{\otimes}}

\newcommand{\ifte}[3]{{if}~#1~{then}~#2~{else}~#3}

\def\formatCompilers#1{\mi{#1}\xspace}
\newcounter{criteria}
\crefname{criteria}{}{}
\newcommand{\criteria}[2]{%
	\def\thecriteria{\detokenize{#1}}%
  	\refstepcounter{criteria}%
  	\label{cr:#2}%
  	#1%
}
\def\porc{C}	%
\def\porcc{P}

\def\pf#1{\ensuremath{\let\porc\porcc #1}}

\newcommand{\rhpcomp}[0]{\formatCompilers{RHP}}
\newcommand{\rhplabel}[0]{rhp}
\newcommand{\rhpref}[0]{\Cref{cr:\rhplabel}\xspace}

\newcommand{\rhpdef}[0]{\criteria{\rhpcomp}{\rhplabel}}

\newcommand{\rtpcomp}[0]{\formatCompilers{RTP}}
\newcommand{\rtplabel}[0]{rtp}
\newcommand{\rtpref}[0]{\Cref{cr:\rtplabel}\xspace}

\newcommand{\rtpdef}[0]{\criteria{\rtpcomp}{\rtplabel}}

\newcommand{\faccomp}[0]{\formatCompilers{FAC}}
\newcommand{\faclabel}[0]{fac}
\newcommand{\facref}[0]{\Cref{cr:\faclabel}\xspace}

\newcommand{\facdef}[0]{\criteria{\faccomp}{\faclabel}}

\newcommand{\rghspcomp}[0]{\formatCompilers{RGHSP}}
\newcommand{\rghsplabel}[0]{rghsp}
\newcommand{\rghspref}[0]{\Cref{cr:\rghsplabel}\xspace}

\newcommand{\rghspdef}[0]{\criteria{\rghspcomp}{\rghsplabel}}

\newcommand{\myred}[0]{\ensuremath{\ \hookrightarrow\ }}
\newcommand{\myredx}[1]{\ensuremath{\ \lhook\joinrel\xrightarrow{#1}}}

\newcommand{\nred}[0]{\ensuremath{\not\hookrightarrow}}
\newcommand{\redapp}[1]{\ensuremath{\!\!\!\!^{#1}\ }}

\newcommand{\redp}[0]{\ensuremath{\ \hookrightarrow_0\ }}

\newcommand{\inps}[1]{#1|_I}
\newcommand{\outs}[1]{#1|_O}

\makeatletter
\xdef\@thefnmark{\@empty}

\newcommand{\Thmref}[1]{\Cref{#1}~(\nameref{#1})}
\makeatother

\newcommand{\subst}[2]{\ensuremath{\bl{\left[#1\relmiddle/#2\right]}}} %

\renewcommand{\emptyset}[0]{\varnothing}

\newcounter{hps}
\crefname{hps}{}{}

\newcommand{\proven}[1]{\ensuremath{\checkmark}}

\makeatletter
\DeclareRobustCommand{\defeq}{\mathrel{\rlap{%
  \raisebox{0.3ex}{$\m@th\cdot$}}%
  \raisebox{-0.3ex}{$\m@th\cdot$}}%
  =}
\DeclareRobustCommand{\eqdef}{=\mathrel{\rlap{%
  \raisebox{0.3ex}{$\m@th\cdot$}}%
  \raisebox{-0.3ex}{$\m@th\cdot$}}%
  }
\makeatother

\makeatletter
\newcommand\footnoteref[1]{\protected@xdef\@thefnmark{\ref{#1}}\@footnotemark}
\makeatother

\newtheorem{axiom}{Axiom}

\newcommand{\ledot}{\mathrel{\ooalign{\hss\raise.200ex\hbox{$\cdot$}\hss\cr$\le$}}}
\newcommand{\gedot}{\mathrel{\ooalign{\hss\raise.200ex\hbox{$\cdot$}\hss\cr$\ge$}}}

\newcommand{\Bits}[0]{\mi{Bits}}

\newcommand{\mathcmd}[1]{{\normalfont\ensuremath{#1}}\xspace}

\newcommand{\mathnamesc}[1]{\mathcmd{\textsc{#1}}}
\newcommand{\mathlabel}[1]{\mathcmd{\textsf{#1}}}

\newcommand{\textop}[1]{\relax\ifmmode\mathop{\text{#1}}\else\text{#1}\fi}

\newcommand{\UC}[0]{\comuc{UC}\xspace}

\newcommand{\RC}[0]{\comsc{RC}\xspace}
\newcommand{\uc}[0]{\UC}

\newcommand{\dummyannot}[0]{d}

\newcommand{\protocolletter}[0]{\ensuremath{\pi}} %
\newcommand{\idealfuncletter}[0]{\ensuremath{F}}
\newcommand{\simulatorletter}[0]{\ensuremath{S}}
\newcommand{\attackerletter}[0]{\ensuremath{A}}
\newcommand{\environmentletter}[0]{\ensuremath{Z}}

\newcommand{\prot}[1]{\protg{#1}{}\xspace}
\newcommand{\idfu}[1]{\idfug{#1}{}\xspace}
\newcommand{\patt}[1]{\con{\attackerletter_{#1}}\xspace}
\newcommand{\simu}[1]{\ide{\simulatorletter_{#1}}\xspace}
\newcommand{\env}[1]{\comuc{\environmentletter_{#1}}\xspace}

\newcommand{\pattdummy}[0]{\patt{\dummyannot}}

\newcommand{\protg}[2]{\con{\protocolletter_{#1}^{#2}}\xspace}
\newcommand{\idfug}[2]{\ide{\idealfuncletter_{#1}^{#2}}\xspace}

\newcommand{\Exec}{\mathnamesc{Exec}}
\newcommand{\ExecT}{\mathnamesc{ExecT}}

\newcommand{\Execfun}[1]{\mathnamesc{Exec}\left(#1\right)}
\newcommand{\ExecTfun}[1]{\mathnamesc{ExecT}\left(#1\right)}

\DeclareMathOperator\vdashuc{\ensuremath{\bl{\vdash_{\uc}}}}
\DeclareMathOperator\vdashucp{\ensuremath{\bl{\vdash_{\uc}}}}

\DeclareMathOperator\vdashrs{\ensuremath{\bl{\vdash_{\comuc{RS}}}}}

\newcommand{\wrty}[1]{\ensuremath{\com{Wr\ #1}}}
\newcommand{\rdty}[1]{\ensuremath{\com{Rd\ #1}}}

\newcommand{\myto}[1]{\ensuremath{\to_{#1}}}
\newcommand{\lolli}[1]{\multimap}

\newcommand{\wrtk}[0]{\varpi}

\newcommand{\lamlb}[3]{\ensuremath{\lambda_{#1} #2\ldotp #3}}
\newcommand{\applb}[3]{\ensuremath{\left(#2\ #3\right)_{#1}}}
\newcommand{\pairlb}[2]{\ensuremath{\left\langle#2\right\rangle_{#1}}}
\newcommand{\inllb}[2]{\ensuremath{{inl}_{#1}~#2}}
\newcommand{\inrlb}[2]{\ensuremath{{inr}_{#1}~#2}}
\newcommand{\splitlb}[2]{\ensuremath{{split}_{#1}\left(#2\right)}}
\newcommand{\fixlb}[2]{\ensuremath{{fix}_{#1}\left(#2\right)}}
\newcommand{\caseoflb}[4]{\ensuremath{{case}_{#1}~#2~{of}~\inl{x_1}\mapsto #3\mid\inr{x_2}\mapsto #4}}
\newcommand{\letinlb}[4]{{let}_{#1}~#2=#3~{in}~#4}
\newcommand{\newch}[2]{\nu\left(#1\right).#2}
\newcommand{\wrex}[2]{\ensuremath{{wr\left(#1,#2\right)}}}
\newcommand{\rdex}[2]{\ensuremath{{rd\left(#1,#2\right)}}}
\newcommand{\chanex}[4]{\ensuremath{{ch\left(#1,#2,#3,#4\right)}}}
\newcommand{\forkex}[2]{\ensuremath{#1\left(\parallel #2\right)}}

\newcommand{\rdchan}[1]{\ensuremath{{\mi{Read}\left(#1\right)}}}
\newcommand{\wrchan}[1]{\ensuremath{{\mi{Write}\left(#1\right)}}}

\newcommand{\bang}[1]{!#1}

\DeclareMathSymbol{\mathinvertedexclamationmark}{\mathord}{operators}{'074}
\DeclareMathSymbol{\mathexclamationmark}{\mathord}{operators}{'041}
\makeatletter
\newcommand{\raisedmathinvertedexclamationmark}{%
  \mathord{\mathpalette\raised@mathinvertedexclamationmark\relax}%
}
\newcommand{\raised@mathinvertedexclamationmark}[2]{%
  \raisebox{\depth}{$\m@th#1\mathinvertedexclamationmark$}%
}
\makeatother

\newcommand{\gnab}[1]{\raisedmathinvertedexclamationmark#1}

\newcommand{\keygen}[1]{{keygen\left(#1\right)}}

\newcommand{\invert}[2]{{invert\left(#1,#2\right)}}

\newcommand{\prgen}[2]{{prg\left(#1,#2\right)}}

\newcommand{\takerand}[0]{{takernd}}

\newcommand{\nrnd}[0]{n^{\mi{rnd}}}
\newcommand{\secpam}[0]{\lambda}
\newcommand{\totrand}[0]{\fun{NR}{\secpam}}

\newcommand{\canZ}{\mathit{z}}
\newcommand{\canZfun}[1]{\canZ\left(#1\right)}

\newcommand{\formatmymess}[3]{\mess{{\small#1}}{#2}{#3}}

\newcommand{\mymesstau}[3]{\formatmymess{ {#2} }{#1}{#3}\nextlevel}
\newcommand{\mymessalphanl}[3]{\formatmymess{#2}{#1}{#3}\nextlevel}

\newcommand{\newinst}[3]{\declinst{#1}{#2}{#3}}

\newcommand{\initstatesym}[0]{\Omega_0}
\newcommand{\SInit}[1]{\ensuremath{{\initstatesym}\left({#1}\right)}\xspace}

\newcommand{\linkatk}[0]{\parcomst}
\newcommand{\linkatkos}[0]{\parcomos}
\newcommand{\linkatkot}[0]{\parcomot}

\newcommand{\linkatktt}[0]{\parcomtt}
\newcommand{\linkatkts}[0]{\parcomts}

\newcommand{\linkprog}[0]{\ffist}
\newcommand{\linkprogos}[0]{\ffios}

\newcommand{\linkprogtt}[0]{\ffitt}
\newcommand{\linkprogts}[0]{\ffits}
\newcommand{\linkprogot}[0]{\ffiot}

\newcommand{\linkwhole}[0]{\wholecomst}
\newcommand{\linkwholeos}[0]{\wholecomos}
\newcommand{\linkwholeot}[0]{\wholecomot}
\newcommand{\linkwholett}[0]{\wholecomtt}
\newcommand{\linkwholets}[0]{\wholecomts}

\newcommand{\ffisymbol}[0]{\bl{\varoast}}%
\newcommand{\ffigen}[2]{\ensuremath{\mathrel{\ffisymbol^{#1}_{#2}}}}
\newcommand{\ffist}[0]{\ffigen{\S}{\T}}
\newcommand{\ffios}[0]{\ffigen{\O}{\S}}
\newcommand{\ffiot}[0]{\ffigen{\O}{\T}}

\newcommand{\ffitt}[0]{\ffigen{\T}{\T}}
\newcommand{\ffits}[0]{\ffigen{\T}{\S}}

\newcommand{\parcomsymbol}[0]{\bl{\varobar}}%
\newcommand{\parcomgen}[2]{\ensuremath{\mathrel{\parcomsymbol^{#1}_{#2}}}}
\newcommand{\parcomst}[0]{\parcomgen{\S}{\T}}
\newcommand{\parcomos}[0]{\parcomgen{\O}{\S}}
\newcommand{\parcomot}[0]{\parcomgen{\O}{\T}}

\newcommand{\parcomtt}[0]{\parcomgen{\T}{\T}}
\newcommand{\parcomts}[0]{\parcomgen{\T}{\S}}

\newcommand{\wholecomsymbol}[0]{\bl{\varocircle}}%
\newcommand{\wholecomgen}[2]{\ensuremath{\mathrel{\wholecomsymbol^{#1}_{#2}}}}
\newcommand{\wholecomst}[0]{\wholecomgen{\S}{\T}}
\newcommand{\wholecomos}[0]{\wholecomgen{\O}{\S}}
\newcommand{\wholecomot}[0]{\wholecomgen{\O}{\T}}

\newcommand{\wholecomtt}[0]{\wholecomgen{\T}{\T}}
\newcommand{\wholecomts}[0]{\wholecomgen{\T}{\S}}

\newcommand{\chans}[0]{D}
\newcommand{\procs}[0]{\Pi}
\newcommand{\intfs}[0]{I}
\newcommand{\expos}[0]{X}

\newcommand{\chan}[1]{\ensuremath{c#1}}

\newcommand{\moduleletter}[0]{M}
\newcommand{\modgen}[1]{\moduleletter{\ensuremath{#1}}}
\newcommand{\modc}[1]{\com{\modgen{#1}}}

\newcommand{\exprletter}[0]{e}
\newcommand{\exprgen}[1]{\ensuremath{\exprletter{#1}}}
\newcommand{\exc}[1]{\com{\exprgen{#1}}}

\newcommand{\configletter}[0]{\Omega}
\newcommand{\configgen}[1]{\ensuremath{\configletter{#1}}}
\newcommand{\confc}[1]{\com{\configgen{#1}}}

\newcommand{\dummy}[2]{\ensuremath{#2\vdash#1 : \mi{dummy}}}

\newcommand{\card}[1]{\ensuremath{|\!|{#1}|\!|}}

\newcommand{\stackht}[2]{\ensuremath{#1\cdot#2}}

\newcommand{\myarray}[1]{\ensuremath{\left[#1\right]}}
\newcommand{\mylength}[1]{\card{#1}}

\newcommand{\pif}[0]{\ensuremath{\text{ if }}}
\newcommand{\piff}[0]{\ensuremath{\text{ iff }}}
\newcommand{\pthen}[0]{\ensuremath{\text{ then }}}
\newcommand{\pand}[0]{\ensuremath{\text{ and }}}
\newcommand{\peither}[0]{\ensuremath{\text{ either }}}
\newcommand{\por}[0]{\ensuremath{\text{ or }}}

\newcommand\xrsquigarrow[1]{%
\mathrel{%
\begin{tikzpicture}[baseline= {( $ (current bounding box.south) + (0,-0.5ex) $ )}]
  \node[inner sep=.5ex] (a) {$\scriptstyle #1$};
  \path[draw,implies-,double distance between line centers=1.5pt,decorate,
    decoration={zigzag,amplitude=0.7pt,segment length=1.2mm,pre=lineto,
    pre   length=4pt}] 
    (a.south east) -- (a.south west);
\end{tikzpicture}}%
}
\DeclareMathOperator\semfat{\ensuremath{\xrsquigarrow{\phantom{a.}}}}

\newcommand{\envredkk}[3]{
	\vdash_{env} #1 \rightsquigarrow #2 \triangleright #3
}

\newcommand{\prob}[0]{\ensuremath{\rho}}

\DeclareMathOperator\indistp{\bl{\approx}}

\DeclareMathOperator\nindistp{\bl{\not\approx}}

\newcommand{\hyperprop}[1]{\comsc{#1}\xspace}
\newcommand{\schp}{\hyperprop{SCHP}}
\newcommand{\hp}{\hyperprop{HP}}

\newcommand{\hs}{\hyperprop{HSP}}
\newcommand{\rhs}{\hyperprop{XHSP}}

\newcommand{\theoryBaseURL}{https://uc-is-sc.github.io/UCisRC.html}
\newcommand{\theoryURL}[1]{\theoryBaseURL\##1}
\newcommand{\theoryTRef}[1]{\href{\theoryURL{#1}}{Theorem \texttt{#1}}}
\newcommand{\theoryProof}[1]{(Proven in Isabelle/HOL: {#1})}

\newcommand{\compos}[2]{ \ensuremath{#1^{#2} }}

\newcommand{\xorbinding}[3]{ \ensuremath{\mi{xor}\left(#1,#2,#3 \right)} }

\newcommand{\cryptobinder}[3]{\pair{#1,#2,#3}}

\newcommand{\cic}[2]{\ensuremath{\vdash #1 \mid\mid #2 : \mi{CIC}}}

\definecolor{sandybrown}{rgb}{0.96, 0.64, 0.38}
\colorlet{sqBlue}{RoyalBlue}
\colorlet{sqGreen}{Emerald}
\colorlet{sqRed}{RubineRed}
\colorlet{sqViolet}{RedViolet}
\colorlet{sqBrown}{sandybrown}
\colorlet{sqGray}{Gray}
\lstdefinelanguage{sapic}{
  morekeywords=[1]{
	out, in, if, then, else, event, insert, delete, lookup, as, in, lock, unlock, let,
        new, get, inj-event
      },
  morekeywords		= [2]{bitstring,pkey,skey,time,channel},
  morekeywords		= [3]{snd,fst,dec,enc,pk,diff},
  morekeywords		= [4]{free, type, fun, reduc, equation, end,
    query, equivalence, lemma, theory, begin,    builtins, functions, equations, +, lemma, process, table,
    exists-trace,},
  morekeywords		= [5]{All, Ex, forall},
  morekeywords		= [6]{attacker, mess,K,KU,KD},
  morekeywords		= [7]{Accept,Honest},
  sensitive=true,
  morecomment		= [n][\itshape]{(*}{*)},
  morecomment		= [n][\bfseries]{(**}{*)},
  literate=
        {||}{{$\mid$}}1
        {==>}{{$\implies$}}2
        {\&}{{\textsf{\&}}}1
	{:=}{{$\defeq$}}2
        {'c}{{\textsl{c}}}3
        {'k}{{\textit{k}}}3
        {'lk}{\textit{lk}}3
        {'key}{{\textit{key}}}3
        {'sk}{\textit{sk}}3
        {'sk1}{\textit{sk1}}3
        {'sk2}{\textit{sk2}}3
}

\lstdefinestyle{sapic}{
  language={sapic},
  basicstyle		= \footnotesize\ttfamily,
  keywordstyle		= [2]{\mdseries\color{sqGreen}},
  keywordstyle		= [4]{\bfseries\color{sqViolet}},
  keywordstyle		= [3]{\slshape},
  keywordstyle		= [1]{\bfseries\color{\commoncol}},
  keywordstyle		= [5]{\bfseries\color{sqBrown}},
  keywordstyle		= [6]{\bfseries},
  keywordstyle		= [7]{\mdseries},
  mathescape		= false,
  columns		= fullflexible,
  keepspaces		= true,
  identifierstyle=\color{\commoncol},
}

\theoremstyle{definition}

\newtheorem{definition}{Definition}
\newtheorem{theorem}{Theorem}
\newtheorem{lemma}{Lemma}

\newtheorem{corollary}{Corollary}
\Crefname{corollary}{Corollary}{Corollaries}
\Crefname{informal}{Definition}{Definition}
\Crefname{assumption}{Assumption}{Assumptions}
\crefname{assumption}{Assumption}{Assumptions}
\Crefname{property}{Property}{Properties}
\crefname{property}{Property}{Properties}

\Crefname{paragraph}{Section}{Sections}
\renewcommand{\rghspcomp}[0]{\formatCompilers{RSCHP}}
\colorlet{NAVYBLUE}{NavyBlue}%
\setcopyright{rightsretained}
\acmJournal{TOPLAS}
\acmYear{2022} \acmVolume{1} \acmNumber{1} \acmArticle{1} \acmMonth{1} \acmPrice{}\acmDOI{10.1145/3436809}

\begin{document}

\title{
	Universal Composability is Robust Compilation
} 

\author{Marco Patrignani}
\orcid{0000-0003-3411-9678}             %
\affiliation{
  \department{disi}             %
  \institution{University of Trento}
  \city{Trento}
  \country{Italy}
}
\authornote{Part of this work was conducted while at MPI-SWS, at Cispa, and at Stanford University.}
\email{mp@cs.stanford.edu}         %

\author{Robert K\"unnemann}
\affiliation{%
  \institution{CISPA Helmholz Center for Information Security}
  \city{Saarbr\"ucken}
  \country{Germany}
}
\email{robert.kuennemann@cispa.de}

\author{Riad S. Wahby}
\affiliation{%
  \institution{Carnegie Mellon University}
  \city{Pittsburgh, PA}
  \country{USA}
}
\authornote{Part of this work was conducted while at Stanford University.}
\email{riad@cmu.edu}



\begin{abstract}
This paper discusses the relationship between two frameworks: universal composability (\UC) and robust compilation (\RC).
In cryptography, \UC is a framework for the specification and analysis of cryptographic protocols with a strong compositionality guarantee: \UC protocols remain secure even when composed with other protocols. 
In programming language security, \RC is a novel framework for determining secure compilation by proving whether compiled programs are as secure as their source-level counterparts no matter what target-level code they interact with.
Presently, these disciplines are studied in isolation, though we argue that there is a deep connection between them and exploring this connection will benefit both research fields.

This paper formally proves the connection between \UC and \RC and then it explores the benefits of this connection.
For this, this paper first identifies which conditions must programming languages fulfil in order to possibly attain \UC-like composition.
Then, it proves \UC of both an existing and a new commitment protocol as a corollary of the related compilers attaining \RC.
Finally, it mechanises these proofs in Deepsec, obtaining symbolic guarantees that the protocol is indeed \UC.

Our connection lays the groundwork towards a better and deeper understanding of both \UC and \RC, and the benefits we showcase from this connection provide first evidence of scalable mechanised proofs for \UC.

\begin{center}\small\it
	In order to differentiate various sub-parts of the \UC and \RC frameworks, we use \emph{syntax highlighting} to a degree that colourblind and black\&white readers can benefit from~\cite{patrignani2020use}.

	\UC talks about ideal functionalities (typeset in a \ide{verbatim}, \ide{emerald} font) and protocol implementations (typeset in a \con{sans}-\con{serif}, \con{orange} font).
	\RC deals with source languages (typeset in an \src{italics}, \src{blue} font) and target ones (typeset in a \trg{bold}, \trg{red} font).
	Elements that are common to each framework are typeset in a \comuc{\commoncol}, \comuc{sans}-\comuc{serif} font for \UC and in a \comsc{\commoncol}, \comsc{italicised} font for \RC to avoid repetition.

	For a better experience, please view this paper in colour.
\end{center}
 \end{abstract}

\maketitle

\section{Introduction}\label{sec:intro}
In cryptography, universal composability (\uc) is a framework for the specification and analysis of cryptographic protocols with a key guarantee about compositionality~\cite{%
uc,
uc2,
Hofheinz2011GNUC:-A-New-Uni,
R.Kusters2006Simulation-Base,
DBLP:journals/iandc/BackesPW07,
DBLP:conf/tosca/Maurer11}.
If a cryptographic protocol is proven \uc, that protocol behaves like some high-level, secure-by-construction \emph{ideal functionality} no matter what the protocol interacts with.
As such, if that protocol is used within a larger protocol, in order to reason about the latter, we can replace the former protocol with its ideal functionality and just reason about the rest.
In other words, \uc protocols are secure \emph{even when composed} with other protocols.

In programming language security, robust compilation (\RC)~\cite{rhc,journey-rel} is a framework for studying secure compilation as the \emph{robust} preservation of classes of hyperproperties~\cite{ClarksonS10}.
That is, for each class \mc{H} of hyperproperties (e.g., safety, hypersafety, subset-closed hyperproperties etc.), \RC provides a criterion stating that if attained by a compiler, then the compiler preserves class \mc{H} from the source programs it inputs to the target ones it produces.
Moreover, this preservation is done robustly, i.e., \emph{no matter} what code is linked against the source and target programs.
These two frameworks (\Cref{sec:bg}) belong to different research fields and seem to deal with different notions, however we demonstrate that they are deeply connected.
Both frameworks are concerned with abstract notions which are generally \emph{deemed} secure, and more concrete notions whose security must be \emph{proven against arbitrary attackers}.
In \UC, the abstract notions are called ideal functionalities (\idfu{}) while concrete ones are called protocols (\prot{}).
In \RC, the abstract notions are called source programs (\prgs{}), for the source language of the compiler \comp{\cdot}, while concrete ones are target programs (\prgt{}), for the target language of the compiler.
Moreover, proving that a protocol \uc-realises an idea functionality ($\prot{}\vdashuc\idfu{}$), or that a compiler attains a \RC criterion for class \mc{H} ($\vdash\comp{\cdot}:\mc{H}$), is done by showing that any attack at the concrete level (protocol or target program) can be simulated at the abstract level (functionality or source program).

This paper is the first to witness (and formally prove in Isabelle) the connection between these two frameworks (\Cref{sec:connecting}).
Specifically, we connect \UC to the \RC criterion that corresponds to \mi{R}obust \mi{H}yperproperties \mi{P}reservation (dubbed \rhpref).
We detail all assumptions typically made by the two frameworks and clarify which additional (simple) assumptions need to hold for this connection to hold.
Effectively, we demonstrate that proving $\prot{}\vdashuc\idfu{}$ is equivalent to: code \idfu{} as a program \prgs{}, code \prot{} as a program \prgt{}, and prove that $\vdash\comp{\cdot}:\rhpref$ for the compiler that translates \prgs{} into \prgt{}.
After presenting this connection, this paper discusses the benefits that both frameworks gain from it, focusing on the possibility of obtaining rigorous, scalable, mechanised proofs of \UC from \RC proofs.
For this, we note that \UC results are stated in terms of the semantics of Interactive Turing Machines (ITMs), while \RC ones are stated in terms of arbitrary source and target languages.
Thus, we first identify which conditions arbitrary programming languages must fulfil in order to be used to attain \UC proofs from \RC ones (\Cref{sec:composition}).
These conditions formalise the behaviour of the linking operators of such languages, and they impose some constraint on the languages' semantics.
In addition, these conditions let us derive common \UC corollaries that simplify the mechanisation of \RC proofs.

We then present one language that we prove to fulfil these conditions: the \emph{reactive, interactive $\lambda$-calculus} (\rilc, \Cref{sec:rilc}).
\rilc extends the interactive $\lambda$-calculus of \citet{ilc} with a module system and a trace semantics, which are needed in order to carry out \RC proofs.
The module system is needed to clearly demarcate the boundary between protocol and attacker (resp.\ functionality and simulator).
The trace semantics describes the behaviour of a protocol linked with an attacker from the perspective of an external observer (i.e., the environment perspective, in \UC terms), as commonly done in reactive systems formalisations~\cite{acetoReactiveSystemsModelling2007}.
Together, these additions yield a language where we can express all the concepts required by \UC, but still perform proofs in a \RC style, leveraging the formal language semantics.

Next, we demonstrate how the semantics-based proof techniques used for \RC proofs can be leveraged to obtain rigorous, scalable \UC proofs (\Cref{sec:sec-comp-proofs}).
For this we take a protocol for one-time commitments \protcomm~\cite{uc-comm,ilc} and prove that it \UC-realises some ideal functionality \idfucomm.
We prove this both with static and with dynamic corruption models; to the best of our knowledge the former is a known result~\cite{uc-comm,ilc} but the latter is novel.
By relying on our connection, we code the ideal functionality and the protocols as programs (\prgscomm and \prgtcomm respectively) in \rilc. 
We prove that the compiler \compcomm{\cdot}, which translates \prgscomm to \prgtcomm attains \rhpref, so from our connection, \protcomm \UC-realises \idfucomm.
We carry out this proof for both the static and the dynamic corruption models, and both proofs rely on simplifications of a trace-based backtranslation, a semantics-based proof techniques often used in \RC proofs~\cite{catalinRSC,rsc,rhc,mfac,scoo-j,journey-rel,exorcising,akram}.

Finally, we mechanise both these proofs in \deepsec~\cite{chevalDEEPSECDecidingEquivalence2018}, a fully automated protocol verification tool for
privacy properties. 
We translate \prgscomm and \prgtcomm in the languages of \deepsec (which is very close to \rilc) and
use the tool to prove that the programs are trace equivalent, so in turn, the compiler from the former to the latter is \rhpref.
The connection between \UC and programming languages semantics had already been conjectured, but the programming-languages counterpart had been (wrongly) identified~\cite{conjecture} to be fully-abstract compilation~\cite{abadi,scsurvey} (interestingly, the first formal criterion for secure compilation).
We discuss this conjecture and other elements of our connection in \Cref{sec:disc}, before presenting related work (\Cref{sec:rw}) and concluding (\Cref{sec:conc}).

Concretely, this paper makes the following contributions:
\begin{itemize}
	\item it proves in Isabelle that (under some simple conditions), \UC coincides with \rhpref, the hyperproperty preservation instance of \RC;
	
	\item in doing so, it disproves an existing conjecture~\cite{conjecture} claiming that \UC is a different secure compilation criterion called fully-abstract compilation;

	\item it provides the conditions for arbitrary languages to be used for our connection;

	\item it formalises \rilc, the reactive, interactive $\lambda$-calculus and proves it fulfils the aforementioned conditions (together with other useful properties);

	\item it proves a one-time commitment protocol to \UC-realise some ideal functionality under both static and dynamic corruption models by proving the compiler from functionality to protocol is \rhpref;
	
	\item it mechanises these proofs in \deepsec, obtaining symbolic \UC guarantees.
\end{itemize}

For the sake of simplicity, we delegate much of the auxiliary formalism to the technical report~\cite{ucissc-prisc}, where the interested reader can find full language formalisation, auxiliary lemmas and proofs.
Mechanised proofs for the main results (in Isabelle) and for the commitment protocol (in \deepsec) can be found at our project page:
\begin{center}
	\url{https://uc-is-sc.github.io/}
\end{center}
\section{Background: \UC and \RC}\label{sec:bg}
This section presents the \UC (\Cref{sec:uc}) and \RC (\Cref{sec:rc})
frameworks and the details that we rely on to define our connection
between them.

\smallskip

	\paragraph{A Note on Abstraction.}
	There exists several \UC frameworks and a lot of programming languages that can fit the \RC framework.
	In the following, the details of both frameworks (e.g., the choice of interactive Turing machines of or the language semantics) are deliberately abstract and handled in an axiomatic manner.
	Such an abstract treatment lets us derive the connection in the most general sense, but it may leave the knowledgeable reader concerned about some technical details.
	These technical details are handled and discussed later on, when we show how concrete \UC frameworks and what real languages fulfil these axioms and provide a concrete instance of our connection.
\subsection{Universal Composability}\label{sec:uc}

In \UC~\cite{canettiUniversallyComposableSecurity2001} the involved parties form a collection of \emph{Interactive Turing Machines} (\itm), that is, Turing Machines that operate concurrently and that have shared tapes between each other.
Crucially: in a collection of \itms, only a single machine operates at a time, so computation happens under sequential consistency.

\paragraph{\ITMs}
The framework of \UC talks about entities in two distincts worlds: the \emph{real} and the \emph{ideal} ones.
The entities of each world comprise exactly the same roles: one encodes the protocol of interest, one models the adversary and one that models an environment providing inputs and observing behaviour.
In the real world, these entities are: the protocol $\prot{}$, the adversary $\patt{}$ and the environment $\env{}$.
In the ideal world, these entities are: the ideal functionality $\idfu{}$, the simulator $\simu{}$ and the (same) environment \env{}.
In either world, the three entities form a \emph{network} that executes according to the semantics of \ITMs.

\paragraph{\ITMs Semantics}
\ITMs have a deterministic semantics whose source of randomness comes from a specific tape called the random tape that any machine in the network can read from.
The final outcome of the interaction of a network of \ITMs (as perceived by the environment) is (wlog) a single bit \comuc{0/1}~\cite[\Ss1.1]{uc}.
Thus, the semantics of a network of \ITMs (e.g., \prot{}, \patt{}, \env{}) can be expressed as a random variable \comuc{\Execfun{\prot{},\patt{},\env{}}} whose image is a single bit.
 
Oftentimes, e.g., in \UC proof arguments, it is useful to describe the sequence of messages exchanged by the entities in the network.
We call this sequence of message a \emph{trace}, to uniform the lexicon with the one of \RC (described in \Cref{sec:rc} below).
Thus, the trace of a network of \ITMs (e.g., \prot{}, \patt{}, \env{}) can be expressed as a random variable \comuc{\ExecTfun{\prot{},\patt{},\env{}}} whose image is a sequence of messages.
\begin{definition}[\ITMs Semantics in \UC]\label{def:itm-exec}
    \begin{align*}
        \Execfun{\cdot, \cdot, \cdot}
            &
            : \comuc{0/1}
        &
        \ExecTfun{\cdot, \cdot, \cdot}
            &
            :
            \text{ trace of messages }
    \end{align*}
    The precise definition of $\ExecT$ and $\Exec$ differs from framework to framework, so in \Cref{sec:axioms-connection} we capture their essence with a set of axioms that abstract from the framework-specific (gory) details. 
    We verified that these axioms hold for the \UC-like
    frameworks~\cite{uc,iuc,ilc},
    as we report in \Cref{sec:diff-uc}.
\end{definition}

The semantics of \ITMs plays a role in the security argument of \UC, which is about the real world emulating the ideal one. 
This means indistinguishability of executions i.e., of the random variable \comuc{\Exec} when run on a real and on an ideal network.

\paragraph{\UC-Emulation and Indistinguishability}
\UC distinguishes different forms of emulation, the two most important ones being \emph{perfect} and \emph{computational} emulation~\cite{hofheinz06simulatable,hofheinz2013polynomial}.
In this work we focus on perfect emulation, and we leave discussing the computational for future work.
In perfect emulation, the real and ideal worlds are considered indistinguishable ($\indistp$) if their semantics are equally distributed. 
In general, for two random variables $X,Y$ in the same observation space $E$ (bits in our case), this means that $X \indistp Y \iff \forall e\in E\ldotp \Pr[X = e] = \Pr[Y = e]$, where $\Pr[c]$ indicates the probability of condition $c$.

\paragraph{\UC Security}
We now have all the elements to define the \UC security argument (\Cref{def:uc}).

In \UC, a protocol $\prot{}$ is secure (denoted with $\prot{}\vdashucp \idfu{}$) if it refines an ideal functionality $\idfu{}$ that is secure by construction (aka, the protocol \prot{} \UC-emulates the functionality \idfu{}).
\UC ensures that $\prot{}$ is as secure as $\idfu{}$ by requiring that any attack on $\prot{}$ can be rewritten into an attack on $\idfu{}$. 
This is expressed via simulation: $\prot{}$ emulates $\idfu{}$ if for any adversary $\patt{}$, there is a simulated adversary $\simu{}$ such that, for any environment $\env{}$, the networks $(\env{},\prot{},\patt{})$ and $(\env{},\idfu{},\simu{})$ are indistinguishable.
\begin{definition}[\UC Perfect Emulation]\label{def:uc}
    \begin{align*}
        \prot{}\vdashucp \idfu{} \isdef
        \forall \patt{}\ldotp \exists \simu{}\ldotp \forall \env{}\ldotp
        \Execfun{\env{},\patt{},\prot{}}
        \indistp
        \Execfun{\env{},\simu{},\idfu{}}
    \end{align*}
\end{definition}

\subsection{Robust Compilation}\label{sec:rc}
Robust compilation criteria~\cite{rhc,journey-rel} are defined for an generic notion of a source language \S and a target language \T so long as they provide few elements that we now describe.

\paragraph{Formal Languages, Traces and Semantics}
In \RC, languages must have a notion of partial programs \prgc{} (or, components) and of attackers to the programs (or, program contexts) \comsc{\ctx{\contextletter}}.
The two can be linked together \comsc{\ctx{\contextletter}\hole{\prgc{}}}%
\footnote{
	Linking is often denoted as \comsc{\ctx{\contextletter}\holev{\prgc{}}}, we use a different notation here to drive a cleaner visual connection with \UC.
}
and the result is a another program \wprgc{} that we call \emph{complete}.
Intuitively, a complete program is one that has no external dependencies.
Notice however that a complete program can still be linked with other programs that use the code of the former program.%
\footnote{
	We refrain from using the word `whole' since in programming language semantics it has a slightly different connotation than the one of this paper.
	Essentially, canonical whole programs cannot be extended anymore with other code, while our complete programs can (as we show in \Cref{sec:composition}).
}
For example, a program may be the implementation of an encryption library and it can be linked with an attacker that interacts with said library, yielding a complete program.
This complete program can then be linked within a larger program, say modelling a web browser, that uses the encryption library.
While this setup with complete programs is not canonical in the \RC setting (which is built on programming language semantics theory), we justify it later, when reasoning about composition of programs (and attackers, and complete programs) in \Cref{sec:compos-op}.

Languages must provide a representation of their behaviour through traces \comsc{\intrace} which are an infinite sequence of security-relevant events.
The trace model considered by \citet{rhc} is that of infinite traces performed against an interface \emph{common} to both source and target languages (e.g., syscalls).%
\footnote{
	Having a common trace model across language is a common assumption in compiler works~\cite{Leroy09b,rhc}, though existing work has shown how to lift this assumption~\cite{journey-rel,rsc}.
	We keep this assumption in our development for the sake of simplicity.
}
The trace model also considers prefixes (\inprefix), i.e., finite traces.

To talk meaningfully about cryptographic operations, we expect the language semantics to take a randomness parameter in input.
This, in turn, changes the trace model and traces become a pair of a sequence of actions (\comsc{\intrace}, as before) together with the probability for that sequence to happen (\comsc{\prob}).
Formally, we indicate a trace and a prefix as follows:
\begin{align*}
	\text{ Interaction Trace } \comsc{\intrace} \bnfdef
		&\
		\text{ infinite sequence of messages }
	&
 	\text{ Trace } \comsc{\trace} \bnfdef
 		&\
 		\comsc{(\intrace,\prob)}
 	\\
	\text{ Interaction Prefix } \comsc{\inprefix} \bnfdef
		&\
		\text{ finite sequence of messages }
	&
 	\text{ Prefix } \comsc{\prefix} \bnfdef
 		&\
 		\comsc{(\inprefix,\prob)}
\end{align*}
Technically, the trace model may be something different than what we present here, so long as it is something that agrees with the \UC trace model.

We leave the semantics of our languages abstract, and indicate the reduction relation of the semantics with $\sem$.
However, there must be a way to indicate the behaviour of programs in terms of the set of traces generated by a program according to the semantics of the language:
\[
	\comsc{\behavc{\wprgc{}}} \isdef \myset{\intrace}{\comsc{\wprgc{}\sem \intrace}}
\]
and indicate that two programs \wprgc{1} and \wprgc{2} have the same behaviours as follows:
\[
	\comsc{\wprgc{1}\beheq\wprgc{2}} \isdef \comsc{\behavc{\wprgc{1}} = \behavc{\wprgc{2}}} 
\]
Since in this paper we assume a common trace model, shared between all languages, note that $\beheq$ can also be used to compare whether programs in different languages have the same behaviour or not.
The $\beheq$ relation is reflexive, symmetric and transitive, i.e., it is an equivalence.
As an example, below are the traces that describe the behaviour of a program that returns a single random bit once queried:
\begin{align*}
	&
	\{ (\labelfont{Query?}\cdot\labelfont{Reply\ 0!}, 1/2)
	,
	(\labelfont{Query?}\cdot\labelfont{Reply\ 1!}, 1/2)
	\}
\end{align*}

The semantics of the languages that we consider (which includes the semantics of most programming languages, including \itms) is supposed to have a simple property that, for lack of a better word, we call `being ok'.
Intuitively, for any program \prgc{}, a semantics is \comsc{ok} if, given that it can produce all prefixes of a trace, then it produces the trace too.
In the following, we use $\Rightarrow$ as logical implication. 
\begin{definition}[Ok Semantics]\label{def:sem-ok}
	\begin{align*}
	\vdash\comsc{\sem}:\comsc{ok} \isdef \forall \wprgc{}, \trace\ldotp 
		\pif (\pif \forall \prefix\ldotp \prefix\isprefix\trace \pthen \wprgc{}\sem \prefix) \pthen \wprgc{}\sem\trace
	\end{align*}
\end{definition}

\paragraph{Hyperproperties}
Hyperproperties~\cite{ClarksonS10} are a formal representation of predicates on programs, i.e., they are predicates on sets of traces.
They capture many security-relevant properties including conventional safety and liveness (i.e., predicates on traces), as well as properties like non-interference (i.e., predicates on sets of traces).
The class of hyperproperties we focus on (for now) is \emph{arbitrary hyperproperties} (\hp).

\paragraph{Compilers}
The last element we need to introduce for \RC are compilers.
Given a source language \S and a target language \T, the compiler that translates \S components \prgs{} into \T ones is denoted with \compgen{\cdot}.
When the languages of the compiler are not relevant, we simply write \comp{\cdot}.

We take compilers to be partial functions, and indicate a compiler to be undefined for certain inputs \src{i} as $\comp{\src{i}}=\bot$.
The criteria we define (and use) in this paper only hold for those inputs for which a compiler is defined.

\paragraph{\rhpref}
Below we present \rhpdef (\Cref{def:rhp}), a criterion that is fulfilled by compilers that preserve subset-closed hyperproperties robustly.
Intuitively, a compiler attains \rhpref if the behaviour (\behavt{\cdot}) of the compiled program (\comp{\prgs{}}) linked (\trg{\hole{}}) with an arbitrary target attacker (\ctxt{}) is the same as the behaviour (\behavs{\cdot}) of the source program (\prgs{}) linked (\src{\hole{}}) with a source attacker (\ctxs{}).
If we unfold the definition of behaviour, we obtain that a compiler attains \rhpref if and only if any prefix (\prefix) produced ($\semt$) by the compiled program linked with an arbitrary target attacker can be replicated ($\sems$) by the source program linked with a source attacker.
\begin{definition}[Robust HP Preservation]\label{def:rhp}
	\begin{align*}
		\comp{\cdot}\vdash \rhpref \isdef 
			&\ 
			\forall\prgs{},\ctxt{} \ldotp\exists\ctxs{}\ldotp
				\ctxt{}\hole{\comp{\prgs{}}} \beheq \ctxs{}\hole{\prgs{}}
		\\
		\text{ or equivalently: }
			&\
			\forall\prgs{},\ctxt{} \ldotp\exists\ctxs{}\ldotp
				\behavt{\ctxt{}\hole{\comp{\prgs{}}}} 
				=
				\behavs{\ctxs{}\hole{\prgs{}}}
		\\
		\text{ or equivalently: }
			&\
			\forall\prgs{},\ctxt{} \ldotp\exists\ctxs{}\ldotp \forall \prefix \ldotp
				\trg{\ctxt{}\hole{\comp{\prgs{}}}}\semt\prefix 
				\piff
				\src{\ctxs{}\hole{\prgs{}}}\sems\prefix
	\end{align*}
\end{definition}
By looking at the statement of \rhpref it is not clear that amounts to the preservation of all \hp.
Such a result has been provided by \citet{rhc}, as they prove that \rhpref is equivalent to another statement that more clearly amounts to the preservation of all \hp.

\bigskip

Having defined the two frameworks of interest, we now describe the connection between them.

\section{Connecting the Frameworks}\label{sec:connecting}

This section first discusses our connection visually and informally (\Cref{sec:informal}) %
before formally proving it (\Cref{sec:proof-connection}).

\subsection{Informal Connection}\label{sec:informal}

\begin{figure}[!htb]
\centering
\begin{tikzpicture}
	\node[](att){\patt{}};
	\node[right =1 of att](pi){\prot{}};
	\node[above right =.7 and .3 of att](env){\env{}{}};
	\node[above =.2 of env.north, font=\footnotesize](tout){$\comuc{0/1}$};

	\draw[<->](att.east) to (pi.west);
	\draw[<->](att.70) to (env.south west);
	\draw[<->](pi.110) to (env.south east);
	\draw[myimpl](env.north) to (tout.south);

	\node[right = 2 of att](si){\simu{}};
	\node[right =1 of si](id){\idfu{}};
	\node[above right =.7 and .3 of si](env2){\env{}{}};
	\node[above =.2 of env2.north, font=\footnotesize](sout){$\comuc{0/1}$};

	\draw[<->](si.east) to (id.west);
	\draw[<->](si.70) to (env2.south west);
	\draw[<->](id.110) to (env2.south east);
	\draw[myimpl](env2.north) to (sout.south);

	\path (pi.east) -- (si.west) node[midway,sloped,yshift=2em] (eq) {$\indistp$};
	\path (env.east) -- (env2.west) node[midway,yshift=1em] (elli1) {};

	\node[right = 1 of id](trgp){\comp{\prgs{}}};
	\node[left =.3 of trgp](trgctx){\ctxt{}};
	\path (trgp.west) -- (trgctx.east) node[midway,sloped] (tlnk) {\trg{\linksymbol}};
	\node[above =.5 of tlnk.north](ttr){\prefix};
	\path (tlnk.north) -- (ttr.south) node[midway,sloped] (tsem) {$\semt$};

	\node[right = 1.5 of trgp](srcp){\prgs{}};
	\node[left =.3 of srcp](srcctx){\ctxs{}};
	\path (srcp.west) -- (srcctx.east) node[midway,sloped] (slnk) {\src{\linksymbol}};
	\node[above =.5 of slnk.north](str){\prefix};
	\path (slnk.north) -- (str.south) node[midway,sloped] (ssem) {$\sems$};

	\path (trgp.east) -- (srcctx.west) node[midway,sloped] (eq2) {$\piff$};
	\path (ttr.east) -- (str.west) node[midway,] (elli2) {};

	\draw[dashed,] (att.south) to ([yshift=-.5em]att.south) to ([yshift=-.5em]att.south -| trgctx.south) to (trgctx.south);
	\draw[dashed,] (si.south) to ([yshift=-1.5em]si.south) to ([yshift=-1.5em]si.south -| srcctx.south) to (srcctx.south);

	\draw[dotted] (pi.south) to ([yshift = -1em]pi.south) to ([yshift = -1em]pi.south -| trgp.south) to (trgp.south);
	\draw[dotted] (id.south) to ([yshift = -2em]id.south) to ([yshift = -2em]id.south -| srcp.south) to (srcp.south);

	\draw[loosely dotted, very thick] (tout.north) to ([yshift=.7em]tout.north) to ([yshift=.7em]tout.north -| str.north) to (str.north);
	\draw[loosely dotted, very thick] ([yshift=.7em]tout.north -| sout.north) to (sout.north);
	\draw[loosely dotted, very thick] ([yshift=.7em]tout.north -| ttr.north) to (ttr.north);

	\node[left = .5 of att](ucdef){$\prot{}\vdashuc\idfu{} \isdef$};
	\node[right = .5 of srcp](rcdef){$\isdef\  \vdash\comp{\cdot}:\rhpref$};
\end{tikzpicture}
\caption{\label{fig:diagram}Connecting \UC and \rhpref, visually.} 
\end{figure}

\Cref{fig:diagram} presents a visual depiction of our connection.
There, (almost) each element in a framework has a corresponding counterpart in the other.
Protocols \prot{} correspond to compiled programs \comp{\prgs{}}, ideal functionalities \idfu{} correspond to source programs \prgs{} and the environment \env{} corresponds to the prefix \prefix.
Concrete adversaries \patt{} correspont to target attackers \ctxt{} and existentially-quantified simulators \simu{} correspond to existentially-quantified source attackers \ctxs{}.
However, the connection between certain elements is non-trivial, and we discuss these cases below.
 
In the following, we use the word \emph{system} to mean one of the four pairs at the bottom of each triangle in the figure (i.e., what in \RC we called a complete program).
As such, the system is always composed of an adversary (or an attacker, or a simulator) and another entity: the protocol, the functionality, the source program or its compiled counterpart.

\paragraph{The Role of Environments}\label{sec:env-role}
The connection between the environment \env{} and the prefix \prefix{} is most apparent when thinking about their role, which is observing the system and providing input to it.

A difference between environments and prefixes is their security role---at least at the intuitive level.
In \UC, the environment is a \emph{benign} entity, so it is not supposed to provide messages that deviate from the protocol and its observation is assumed to be fair.
In \RC, a prefix is just an objective indication that some reduction has happened in the system, so a prefix cannot be benign or malicious.

Fortunately, from the formal perspective, the environment in \UC is an \ITM that has no way to be restricted to just benign behaviour.
Instead, since the environment is universally quantified, it can behave both in a benign and in a malicious way.
For this reason, formally, the (malicious) adversary role is subsumed by the environment, and in \UC it is therefore possible to replace the adversary with a dummy forwarder (this is the dummy attacker theorem, as we discuss in \Cref{sec:dummy-for-rc}).

On the other hand, there are languages in which prefixes are not an objective indication of a reduction happening, but they are generated by a computing entity that the system is reacting to: these are called \emph{reactive languages}.
Our connection works just both reactive and non-reactive languages, and we discuss how using reactive ones simplifies certain results (such as porting the dummy attacker theorem from \UC to \RC) in \Cref{sec:react-vs-nonreact-langs}.

\paragraph{Traces VS Single Bit}
A seeming discrepancy in the connection is the single-bit differentiation output of \UC contrasted with a full prefix in \RC.
We argue this is merely a cosmetic differente: in \UC all parties exchange messages -- whose concatenation forms indeed a prefix -- and the bit is a funcion of those messages (as captured by \Cref{ax:final-bit}).
It is therefore just for convenience (and a technicality due to dealing with probability distributions) that the environment outputs a single bit in order to tell whether it is interacting with the real or ideal world.

\paragraph{What are Compilers in \UC?}\label{sec:compilers-in-uc}
In the \UC framework there is no element that corresponds to compilers.
Instead, the translation of high-level \ide{functionalities} into \con{protocols} is human made.
As such, the typical \UC translation risks being imprecise and error prone, and it cannot be automated for large functionalities, or between functionalities with recurring parts.

On the other hand, in \RC the compilers we consider are not what one expects: they do not traverse the program in input and translate its subparts.
Instead, they perform a syntactic check on the whole source program: if it is a specific one (i.e., the ideal functionality), then the related protocol is produced by the compiler.
In turn, this means that the compilers we consider are not total functions on their input, but partial ones, defined only on those source programs that are ideal functionalities.
Notation-wise, we identify such partial functions as $\comp{\prgs{}}\partialmapsto\prgt{}$.
\subsection{Formally Proving the Connection}\label{sec:proof-connection}

\begin{figure}[!htb]
\centering
\begin{tikzpicture} [
    smallfont/.style = {font=\tiny}
    ]
    \node[](idf){\idfu{}};
    \node[below = 1 of idf](prot){\prot{}};
    \path(prot) to node[midway, sloped](vdash){$\vdashuc$} (idf);

    \node[below right = 1 and 2 of idf.center](srcuc){\src{\idealfuncletter}};
    \node[right = -.1 of srcuc](inucls){}; %
    \node[below = 1 of srcuc](trguc){\trg{\protocolletter}};
    \node[right = -.1 of trguc](inuclt){}; %
    \draw[->](srcuc.south) to node[midway, right](vdashruc){} (trguc.north);

    \node[right = 6 of idf.center](src){\prgs{}};
    \node[]at(prot -| src)(trg){\comp{\prgs{}}};
    \draw[->](src.south) to node[midway, right](vdashr){$\vdash\compgen{\cdot}:\rhpref$} (trg.north);
    \path[](vdashruc.north) to node[pos=.35, sloped](compuc){$\vdash\compucl{\cdot}:\rhpref$} (vdashr.north west);

    \draw[myiff, thick, dashed] (idf.north east) to node[midway, above, font=\small ]{\Cref{thm:uc-is-sc}: our main result} (src.north west);
    \draw[myiff, thick] (trguc.south west) to node[midway, below, sloped, font=\small, align=left ]{\Cref{thm:rghsp-iff-uc} \\ \UC-\RC connection \\ (same language)} (prot.south east);
    \draw[myiff, thick] (trg.south west) to node[midway, below, sloped, font=\small, align=left ]{\Cref{thm:uc-lift} \\ \RC composition } (trguc.south east);

    \draw[color=gray] (idf.east) to node[midway, sloped, above, font=\footnotesize](same1){same} (srcuc.west);
    \draw[color=gray] (prot.east) to node[midway, sloped, above, font=\footnotesize](same2){same} (trguc.west);

    \draw[color=gray] (srcuc.east) to node[midway, sloped, above, font=\footnotesize](same3){natural translation} (src.west);
    \draw[color=gray] (trguc.east) to node[midway, sloped, above, font=\footnotesize, align=center](same4){natural translation} (trg.west);
\end{tikzpicture}
\caption{\label{fig:proof-diagram} Visual depiction of the proof of \Cref{thm:uc-is-sc}.}
\end{figure}
The proof strategy we adopt to formally prove the connection is depicted in \Cref{fig:proof-diagram}.
Our main result is the dashed co-implication on top, which represents the equivalence of \UC and \RC (\Cref{thm:uc-is-sc} in \Cref{sec:main-proof}).
We break down that equivalence into two other equivalences, the sloped, full co-implications on the bottom (\Cref{thm:rghsp-iff-uc} and \Cref{thm:uc-lift}).

First we must establish the formal semantics of \UC.
Despite there being a number of flavours of \UC, we define an abstract language \UCLang that captures the essence of all these flavours.
We define the semantics of \UCLang via a set of axioms (\Cref{sec:axioms-connection}); later (\Cref{sec:diff-uc}), we show that many \UC flavours respect these axioms.

With \UCLang, we can prove the connection between \UC and \RC provided that we are using \UCLang in the \RC statements (\Cref{thm:rghsp-iff-uc} in \Cref{sec:connection-uclang}).
That is, the compiler we consider must have the same source and target language: \UCLang, and we indicate this compiler as \compucl{\cdot}.
Since the semantics of \ITMs (in \UC) and of \UCLang (in \RC) is the same, in the figure we indicate that \idfu{}\ (resp. \prot{}) and \src{\idealfuncletter}\ (resp. \trg{\protocolletter}) are the `same'.

Once this connection is proven, we lift the restriction on the languages of the compiler and consider any compiler between an arbitrary source and an arbitrary target language (\Cref{sec:connection-anylang}).
We assume a ``natural translation'', i.e., way to translate programs between \S and \UCLang and between \T and \UCLang that preserves the same behaviour.
We argue that such natural translations already exists, since real-world protocols are not written in \ITMs but in programming languages.
So finally, by composing these natural translations with \compucl{\cdot} we obtain \compgen{\cdot}.
By relying on existing results about the composition of translations~\cite{csc22} we obtain that \compgen{\cdot} is also \rhpref (\Cref{thm:uc-lift}).

\subsubsection{Axioms for \UC Semantics}\label{sec:axioms-connection}

The first axiom relates the \UC semantics of \ITMs to the one of \UCLang, a programming language semantics in the style of \RC, and it defines what we mean when an adversary and a program produce a trace (\Cref{ax:relation-uc}).%
\footnote{
    This axiom is typeset in \comuc{\commoncol} since it is valid for connecting source and ideal functionality semantics as well as trace and real protocol semantics.
    Technically both pairs talk about the same semantics since source and target languages here are \UCLang and real and ideal world talk about \ITMs.
} 
In this axiom (as well as in subsequent ones) we assume a function $\canZfun{\cdot} : \prefix\to\env{}$ that assigns each possible trace a \emph{canonical environment}.
A canonical environment for $\prefix$ produces exactly this prefix and then halts the execution with the final bit $1$. 
Such a canonical environment is a known concept in the programming languages research literature, it is the environment obtained from the backtranslation of a trace~\cite{rsc,rhc,scoo-j,catalinRSC,llfatr-j,javajr}.
We will discuss this in more detail in \Cref{sec:howto-rc-proofs}.

\begin{axiom}[\UC and \UCLang Semantics]\label{ax:relation-uc}
    \begin{align*}
            \comsc{\ctxc{}\hole{\prgc{}}}\semc \prefix
            &\piff 
            \Pr[\ExecTfun{ \canZfun{\prefix},\comuc{\contextletter},\comuc{\programletter} } = \inprefix] = \prob > 0 
            &
            \text{ where }
            \prefix = (\inprefix,\prob)
    \end{align*}
\end{axiom}

Each execution in \UC requires an environment to be defined.
To be sure that the canonical environment does not introduce undesired behaviour, we require that it is correct (\Cref{ax:canonical}), i.e., the canonical environment outputs $1$ only if it the execution produces exactly the prefix the environment took in input.
The side conditions required for this axiom ensure that the canonical environment exists for a non-probabilistic environment $\env{}$ that can produce the prefix at hand.

\begin{axiom}[Canonical Environment Correctness]\label{ax:canonical}
    \begin{align*}
    \Pr[\ExecTfun{\env{},\patt{},\prot{}} = \inprefix ]  =
    \Pr[\ExecTfun{\canZfun{\prefix},\patt{},\prot{}} = \inprefix ]  =
    \Pr[\Execfun{\canZfun{\prefix},\patt{},\prot{}} = 1 ]
        \\
        \text{ where } \prefix = \inprefix,\prob \pand \text{\env{} is non-probabilistic}
        \pand \exists \patt{e}, \prot{e}\ldotp \Pr[\ExecTfun{\env{},\patt{e},\prot{e}} = \inprefix ] > 0
    \end{align*}
\end{axiom}

The construction of this canonical environment is straightforward in most \UC frameworks: it matches each incoming input against the expected input in the prefix and hard-codes its output. 
Thus, intuitively, the canonical environment reduces the environment to its input/output behaviour.

In the axiom we assume $\env{}$ is non-probabilistic, meaning that the probability of that prefix depends only on the randomness used in $\patt{}$ and $\prot{}$.
This is not a real limitation, since non-probabilistic environments are complete (\Cref{ax:nonprob}).

\begin{axiom}[Non-Probabilistic Environments are Complete]\label{ax:nonprob}
    \begin{align*}
        \text{ if }
            &\
            \Execfun{\env{},\patt{},\prot{}}
            \nindistp 
            \Execfun{\env{},\simu{},\idfu{}}
        \text{ then }
            \exists \text{ non-probabilistic } \env{n}\ldotp
            \Execfun{\env{n},\patt{},\prot{}}
            \nindistp
            \Execfun{\env{n},\simu{},\idfu{}}
    \end{align*}
\end{axiom}

For \UC, \citet[p.~47]{canettiUniversallyComposableSecurity2001} states this is the case. 
Intuitively, if there is a distinguishing environment, then there is at least one set of random choices that we can condition the environment's randomness on such that the environment still succeeds in distinguishing both systems. 
Since the environment is quantified after the adversary and protocol (\Cref{def:uc}), we can hard-code these choices into the environment, making it deterministic.

Finally, prefixes must hold enough information to extract whether the environment has decided to output a final bit $b$ or not ($\bot$).
In \UC~\cite{canettiUniversallyComposableSecurity2001}, the environment terminates the trace with the final bit, hence any prefix must have it.
Abstracting away from the encoding, we assume 
an extraction function $\beta : \prefix \to \{0,1,\bot\}$.
\begin{axiom}[Finite Traces Contain the Final Bit]\label{ax:final-bit}
    \[
        \Pr[\Execfun{\env{},\patt{},\prot{}}= b ] 
        = 
        \sum_{\beta(\prefix)=b} \Pr[\ExecTfun{\env{},\patt{},\prot{}} = \prefix]
    \]
\end{axiom}

\subsubsection{Connecting \UC and \rhpref Between \UCLang}\label{sec:connection-uclang}

With the semantics of \UC defined via the axioms, we can prove a simplified version of our connection.
If a protocol \UC-realises a functionality, then the \UCLang-compiler from that functionality to that protocol is \rhpref (\Cref{thm:rghsp-iff-uc}).

\begin{theorem}[\UC and \rhpref Coincide for \UCLang]\label{thm:rghsp-iff-uc}
    \begin{align*}
        \compucl{\prgs{}} \vdashucp \prgs{}
        \piff
        \vdash\compucl{\cdot}:\rhpref
    \end{align*}
\end{theorem}
\begin{proof}
    \theoryProof{\theoryTRef{RHPtoUC} and \theoryTRef{UCtoRHP}.}

    In both directions, we proceed by contradiction, we only discuss the $\rhpref\Rightarrow\UC$ one since the structure of the two directions is the same.
    At the intuitive level, the proof relies on the axioms of \Cref{sec:axioms-connection} in order to switch between the $\ExecT$ representation of \ITMs semantics (of the \UC framework) to the $\sem$ representation of the same semantics (of the \RC framework).
    \begin{description}
        \item[$\rhpref\Rightarrow\UC$]
        By contradiction, we assume the negation of \UC (\Cref{def:uc}), so program $\prgs{}$ is not emulated by its compiled counterpart \comp{\prgs{}}.

        This gives us an adversary $\patt{}$ such that for any simulator $\simu{}$, there exists an environment $\env{}$ that can distinguish real and ideal interactions (negation of \Cref{def:uc}, point 1).

        Thus, without loss of generality, we have a prefix $\prefix$ whose generating execution is more probable in the real world than in the ideal one (\Cref{ax:final-bit}).

        This execution can also be generated by feeding $\prefix$ to the canonical environment (\Cref{ax:canonical}), so we eliminate the environment in favour of the canonical one.
        
        We now use \Cref{ax:relation-uc} twice, on the real and on the ideal world, in order to translate point 1 into \RC.
        Thus, we obtain the \RC counterparts of the \UC elements: $\ctxt{}=\patt{}$ and $\ctxs{}=\simu{}$ such that this holds:
        \[
            \trg{\ctxt{} \hole{\comp{\prgs{}}}} \semt \prefix
        \]
        but at the same this does not hold:
        \[
            \ctxs{}\holes{\prgs{}} \sems \prefix
        \]
        This is in direct contrast with the \rhpref assumption (\Cref{def:rhp}): contradiction.
        \qedhere
    \end{description}
\end{proof}

\subsubsection{Connecting \UC and \rhpref Between Any Language}\label{sec:connection-anylang}
\label{sec:proof-lifting}
\UC and \rhpref coincide for programs written in \UCLang, but what about other languages?
To answer this, we need to take a protocol in \UCLang and essentially translate it into any other language without changing its meaning: we call this a \emph{natural translation}.

We argue this natural translations are already done in practice, at an abstract level, and in the following, we formalise the meaning of natural translations.
In fact, while \UC proofs are carried out between \ITMs, real protocol implementations are done in real programming languages.
Thus, cryptographers already know how to naturally translate between \UCLang and any other programming language.

We say that it is possible to have a natural translation from programs in a language \Lo to programs in \Lt, as witnessed by a compiler $\compll{\cdot}$ if \Lo is in pre-agreement with \Lt (denoted with $\extprecompat{\compll{\cdot}}{\Lo}{\Lt}$).
Intuitively, a language is in pre-agreement with another one if there is a way to go from programs of the former to programs of the latter preserving the behaviour.
That is, there must be a \rhpref compiler from the former to the latter.
Formally, language pre-agreement is captured in \Cref{def:lang-pre-agr}, and if two languages pre-agree in both directions, then they agree (\Cref{def:lang-agr}).
\begin{definition}[Language Pre-Agreement]\label{def:lang-pre-agr}
    \begin{align*}
        \extprecompat{\compll{\cdot}}{\Lo}{\Lt} \isdef
            &\
            \vdash\compll{\cdot}:\rhpref
    \end{align*}
\end{definition}
\begin{definition}[Language Agreement]\label{def:lang-agr}
    \begin{align*}
        \extcompat{\compo\compll{\cdot}}{\compt\compll{\cdot}}{\Lo}{\Lt} \isdef
            &\
            \extprecompat{\compo\compll{\cdot}}{\Lo}{\Lt}
            \pand
            \extprecompat{\compt\compll{\cdot}}{\Lt}{\Lo}
    \end{align*}
\end{definition}

We say that two programs agree if their languages agree, and one is a natural translation of other, as captured by the compiler yielded by the language agreement.%
\footnote{
    So, the agreement of two languages is witnessed by the existence of a galois insertion between components of each language whose abstraction and concretisation functions are the two \rhpref compilers.
    Moreover, this insertion must induce equivalence between the behaviours of the insertion-related programs.
    Since it is not the focus of this paper, we leave exploring the interesting ramifications of this fact for future work.
}
\begin{definition}[Program Agreement]\label{def:prog-pre-agr}
    \begin{align*}
        \prgs{}\compat\prgt{} \isdef
            &\
            \prgs{}\in\S \text{ and } \prgt{}\in\T
        \pand  
            \extcompat{\compgen{\cdot}}{\compgents{\cdot}}{\S}{\T}
        \pand
            \compgen{\prgs{}}\partialmapsto\prgt{}
            \pand
            \compgents{\prgt{}}\partialmapsto\prgs{}
    \end{align*}
\end{definition}

Besides language (and program) agreement, we rely on a property of the composition of \rhpref compilers: sequentially-composing \rhpref compilers yields a \rhpref compiler (\Cref{thm:rschp-trans}). 
Sequentially composing two compilers $\compo\compgen{\cdot}$ and $\compt{\compgento{\cdot}}$ means running the latter on the output of the former (see~\cite{csc22} for a more general treatment of compiler composition).%
\footnote{
    Here and later, we typeset a third language used in the \RC framework in a \oth{pink}, \oth{italics} font.
}
\begin{theorem}[\rhpref is Transitive]\label{thm:rschp-trans}
    \begin{align*}
            \left(
            \vdash\compo\compgen{\cdot}:\rhpref
        \pand
            \vdash\compt\compgento{\cdot}:\rhpref
            \right)
        \piff
            &\
            \vdash\compt\compgento{\compo\compgen{\cdot}}:\rhpref
    \end{align*}
\end{theorem}
\begin{proof}
    \theoryProof{\theoryTRef{RHPRel}.}

    The $\Ra$ direction follows directly from the composition of \rhpref compilers and it is rather intuitive: if each pass preserves \hp, so should the composition.

    The $\La$ direction is more subtle.
    In fact, composition of \RC compilers tells us that the composition preserves the least upper bound of the properties of the composing compilers.
    Thus, $\compo\compgen{\cdot}$ or $\compt\compgento{\cdot}$ may preserve \emph{more} than just \hp.
    However, if they preserve more, they also preserve \hp, so we can lower their guarantees to just \rhpref.
\end{proof}

We now have everything to connect \rhpref for compilers between arbitrary languages to \rhpref for compilers between \UCLang (\Cref{thm:uc-lift}).
For this, assume arbitrary languages \S and \T agree with \UCLango, so there are natural translations between them.
Then, we can perform a natural translation from a program in \S to the same program in \UCLango, apply the \UCLango-to-\UCLango \rhpref compiler and then do another natural translation back from \UCLango to \T.
Transitivity of \rhpref ensures the resulting translation is also \rhpref.
\begin{theorem}[\rhpref for Arbitrary Languages]\label{thm:uc-lift}
    \begin{align*}
        \text{ let }
            &\
            \extcompat{\compsucl{\cdot}}{\compucls{\cdot}}{\S}{\UCLango}
        \\
        \pand 
            &\
            \extcompat{\compuclt{\cdot}}{\comptucl{\cdot}}{\UCLango}\T
        \\
        \text{ we have }
            &\
            \vdash\compuclo{\cdot}:\rhpref
        \piff
            \vdash \compuclt{\compuclo{\compsucl{\cdot}}} : \rhpref
        \\
        \pand
            &\
            \vdash\compuclo{\cdot}:\rhpref
        \piff
            \vdash \compucls{\compuclo{\comptucl{\cdot}}} : \rhpref
    \end{align*}
\end{theorem}
\begin{proof}
    \theoryProof{\theoryTRef{implementUC}.}
    Follows from \Thmref{thm:rghsp-iff-uc}, \Thmref{thm:rschp-trans} (twice) and the \rhpref translations of \Cref{def:lang-agr}.
\end{proof}

\subsubsection{Our Main Result, Formally}\label{sec:main-proof}
Joining all these results toghether yields the formal proof of our main result: \UC and \rhpref coincide (\Cref{thm:uc-is-sc}).
\begin{theorem}[\UC and \rhpref Coincide]\label{thm:uc-is-sc}
    \begin{align*}
        \pif
            &\
            \prgs{}\compat\idfu{}
            \pand
            \prot{}\compat\prgt{}
        \pand
            \compgen{\cdot} \isdef
                \compgen{\prgs{}} \partialmapsto \prgt{}
        \\
        \pthen
            &\
            \prot{}\vdashuc\idfu{}
            \piff
            \vdash\compgen{\cdot}:\rhpref
    \end{align*}
\end{theorem}
\begin{proof}
    \theoryProof{\theoryTRef{UCcompiler1} and \theoryTRef{UCcompiler2}.}
    From the agreement between \idfu{} and \prgs{} (resp. between \prot{} and \prgt{}), we know that \S and \UCLango (resp. \T and \UCLango) agree.
    The co-implication of this theorem is split into the two-coimplications depicted in \Cref{fig:proof-diagram}.
    From left to right, we obtain the result from \Thmref{thm:uc-lift} i.e., from \Thmref{thm:rghsp-iff-uc} and \Thmref{thm:rschp-trans}.
    From right to left, we proceed by contradiction. Given an unsimulatable UC attacker against $\prot{}$, we use the agreements 
            $\prgs{}\compat\idfu{}$ and
            $\prot{}\compat\prgt{}$ to construct a counterexample 
    against the assumption that
            $\vdash\compgen{\cdot}:\rhpref$.
\end{proof}
\medskip

\paragraph{Discussion}
A way to summarise our result is thus:
\begin{center}
    A security proof between \ITMs is a \UC proof, 
    \\
    a security proof between arbitrary languages is an \rhpref one.
\end{center}
However, oftentimes (mainly for simplicity) we are interested in setting up the connection for the same source and target languages, which we now refer to as just \S (i.e., $\T=\S$).
For example, this is what we do in \Cref{sec:sec-comp-proofs}.
This specialisation requires \S and \UCLango to agree (as per \Cref{def:lang-agr}), so the easiest candidates for \S are those that are shown to behave like \ITMs.
Since \ILC has been proven to model \ITMs faithfully, that is why we will choose it and extend it in order to showcase an instance of our connection.

Overall, we believe the connection between \S (and \T) and \UCLango is not interesting: cryptographers have been modelling real-world code as \ITMs for decades.
Thus, we believe the most important interpretation of \Cref{thm:uc-is-sc} is that one should really focus on the \S to \T translation and its security proof.

At this point, however, one may wonder whether our connection is as precise as it gets, or whether there are other \RC notions (or programming languages research notions, as conjectured by \citet{conjecture}) that connect with \UC.
We will prove that our connection is the most precise one later on, in \Cref{sec:wrong-conn}.

\paragraph{Using our Connection}
We now have a good intuition that the connection holds, and of what it connects.
What are we to make of this connection then, and how can we reap its benefits?

In a nutshell, the first step now is to instantiate abstract languages \S and \T.
The second step is devising a compiler from some ideal functionality written in \S to some concrete protocol written in \T, and prove that compiler is \rhpref.

In this paper, we set \S and \T to be the same language: \rilc (\Cref{sec:rilc}).
This is the simplest solution that lets us showcase the goodness of our connection by proving \UC of a commitment protocol as a \rhpref proof (\Cref{sec:sec-comp-proofs}).
Building on this language also lets us mechanise this proof (\Cref{sec:mechanisation}).
 
A different approach would be to define a specification language (for \S) and model the language in which some protocol is implemented (i.e., \T) and prove \rhpref at the level of the compiler from \S to \T.
While mechanisation is more complex in this case -- we conjecture such proofs could be mechanised in Coq%
\footnote{
    We remark that mechanising proofs of \RC (and more in general of secure compilation) is not an easy task, and at the time of writing, very few such examples exist~\cite{catalinRSC,akram22}.
}
-- its applicability to real-world protocols is clearer.
We leave exploring this avenue of results for future work.

\medskip

Before we showcase our connection, however, we need to unravel a few more assumptions on the semantics of programming languages.
In fact, for now we have imposed very few constraints on the semantics of programming languages involved in the connection.
Thus, one may wonder whether, in practice, the connection can be instantiated with realistic (formal) languages.
This is what we investigate next in \Cref{sec:composition}.

\section{Generalising \UC Corollaries in \RC}\label{sec:composition}

In this section, we import two important corollaries from \UC that are essential to its use for composing secure protocols: the composition theorem and the dummy attacker one.
By proving these corollaries in the \RC setting, we identify additional conditions for languages that can be used with our connection.
Thus this section first proves the composition theorem in the \RC setting (\Cref{sec:compos-for-rc}) and formulates the dummy attacker theorem (\Cref{sec:dummy-for-rc}), providing us with concrete requirements for secure composition in programming languages.

\subsection{The Composition Theorem in \RC}\label{sec:compos-for-rc}

As discussed in \Cref{sec:uc}, \UC results always talk about the same language: \ITMs.
As such, the different types of composition (e.g., \emph{protocol composition} between two
protocols and \emph{protocol execution} composing  adversary, protocol and environment to
an executable system~\cite{uc}) are always between \ITMs.
Conversely, \RC results talk about different languages, so when one tries to derive a composition result, one has to deal with composition of programs in different languages.
Thus, porting the composition theorem in \RC lets us identify additional conditions that the composition of programs in different languages must fulfil.

We state these conditions as a set of axioms that regulate the behaviour of \RC composition in order to be able to port a key \UC result in \RC.
Specifically, the \UC framework comes with a key free theorem called the \emph{Composition Theorem} (\Cref{thm:comp-thm-uc}), meaning that any protocol that \UC-realises a functionality also enjoys \Cref{thm:comp-thm-uc}.
Intuitively, the composition theorem lets one replace a sub-protocol that is part of a larger protocol with its ideal functionality, which is typically much simpler to reason about.
Thus, the composition theorem is the basis for the modular analysis of protocols.

\Cref{thm:comp-thm-uc} contains the formal definition of the composition theorem.
There, and in the following, we indicate a larger protocol \prot{large} composed with a sub-protocol \prot{small} (which could also be an ideal functionality) as \compos{\prot{large}}{\prot{small}}.
\begin{theorem}[Composition Theorem in \UC~\cite{uc}]\label{thm:comp-thm-uc}
    \begin{align*}
        \pif
            &\
            \prot{s}\vdashuc\idfu{s}
        \pthen
            \compos{\prot{l}}{\prot{s}} \vdashuc \compos{\prot{l}}{\idfu{s}}
    \end{align*}
\end{theorem}
The theorem states that if a protocol \prot{s} \UC-realises a functionality \idfu{s} and \prot{s} is used within a larger protocol \prot{l}, then protocol \prot{l} with \prot{s} \UC-realises \prot{l} with \idfu{s} in place of \prot{small}.

Why \Cref{thm:comp-thm-uc} is a corollary of \UC-realisation can be explained by reasoning about the role of the environment.
Ideally, the environment represents all possible larger protocols \prot{l}, so \UC-emulation guarantees that no \prot{l} cannot distinguish between its sub-part being $\prot{s}$ or $\idfu{s}$.

The key notion for this result is the composition operator, so we first discuss this operator in \UC and in \RC (\Cref{sec:compos-op}).
Then, we present the set of axioms that regulate the behaviour of composition (\Cref{sec:compo-axioms}).
As we demonstrate later, providing a concrete instantiation of the \RC composition is fairly simple, and there exists plenty of such instances in the real world; moreover, equivalently easy is showing that these instances respect the axioms.
Finally, we state (and prove) the analogous of \Cref{thm:comp-thm-uc} in \RC (\Cref{sec:compos-thm-rc}).

\subsubsection{Composition Operators}\label{sec:compos-op}

\paragraph{\UC Composition}
Depending on the \UC framework, composition is expressed as a program transformation~\cite{uc,Hofheinz2011GNUC:-A-New-Uni}, where two protocols are inlined into a wrapper/sandbox ITM,  or as
a rebinding of input and output channels~\cite{iuc}.
In both cases, 
the composition of \prot{l} with \prot{s} into a single protocol \compos{\prot{l}}{\prot{s}} restricts the communication interfaces of both protocols. 
For example, the sub-protocol \prot{s} cannot communicate directly with the environment, but any such communication is routed through \prot{l}.
Additionally, \prot{s} and \prot{l} have different interfaces to the adversary, that is, there is an adversary for \prot{s} and one for \prot{l}.
Finally, adversaries for \prot{s} and \prot{l} do not communicate directly, but only via the environment.

\paragraph{\RC Compostition}
When defining \RC composition we rely on canonical programming language semantics notions and do not model the communication restriction mentioned above for \UC.
We do this for the sake of simplicity and generality,
and in order to reuse much of the existing semantics theory developed for \RC.
We leave investigating the semantics of more intricate composition mechanisms for future work.
Essentially, given the visibility qualifiers of a function defined in a program (e.g., \mtt{public}, \mtt{private}, etc), composition does not change those qualifiers.
In \UC, it composition can change these qualifiers, and composing \compos{\prot{l}}{\prot{s}} can turn \mtt{public} interfaces of \prot{s} into \mtt{private} ones.
Recall from \Cref{sec:rc} that our languages have a notion of programs \prgc{} and of attackers \ctxc{} that can be linked together (\comsc{\ctx{\contextletter}\hole{\prgc{}}}) into a complete program.
Since the way we define for \RC can talk about multiple languages, we define multiple composition operators in \RC, each differing in its signature, as we present below.
\begin{description}
    \item[linking: $\linksymbol$] 
        is the already-presented linking. It models intra-language composition and its signature is $\ctxc{}\to\prgc{}\to\wprgc{}$.
        In \UC, linking corresponds to the combination of a protocol with an adversary.
	
    \item[program FFI:  $\linkprog$]
        models cross-language composition of programs, so its signature is $\prgs{}\to\prgt{}\to\prgs{}$.
        As a notational convention, cross-language composition symbols are annotated with languages: the top-most is the language of the first argument and of the result, while the bottom-most is the second argument's.

        Intuitively, program FFI composition yields a \S program made of a \T and an \S sub-part (resp, \prgt{} and \prgs{}), that can communicate with each other.
        When real-world languages let programs of different languages \S and \T interoperate with each other, they rely on an actual foreign function interface between \S and \T.
        As such, any concrete implementation of this operator must include one such foreign function interface that marshals and unmarshals (i.e., converts) values of one language into the other and vice-versa.
        In language semantics, this can be modelled as a multilanguage system~\cite{matthews_operational_2009,perconti,amal2}.
        We need not use this model in our instantiation of the composition operators (later in this paper, in \Cref{sec:rilc,sec:sec-comp-proofs}), because for simplicity, we resort to using a single language for \S and \T.

        This composition operator also exists in \UC, where it is dubbed the protocol composition operator, and it has the same signature.
        In fact, program FFI composition combines two programs into one program while protocol composition combines two protocols into one protocol.

    \item[attacker FFI: $\linkatk$] 
        models a different cross-language parallel composition, since its signature is $\ctxs{}\to\ctxt{}\to\ctxs{}$.

        Although many works let attackers and programs have the same
        structure~\cite{uc,abadiMobileValuesNew2001,
        swamyDependentTypesMultimonadic2016},
        this is not always the case~\cite{rhc,rsc,aizatulinExtractingVerifyingCryptographic2011, backesComputationalSoundnessDalvik2016, guancialeInSpectreBreakingFixing2020}.
        For this reason, we need a composition operator for attackers that may differ from the program one.
        This operator is not always given a name in \UC, but appears as
        a construction in the composition proof (e.g.,
        \cite[Fig.~8]{uc}).

    \item[complete FFI: $\linkwhole$] models a cross-language composition for complete programs, so its signature is $\wprgs{}\to\wprgt{}\to\wprgs{}$.

            In order to link any syntactic category of our abstract languages, program and attacker FFI are not sufficient. Therefore we define the complete FFI, too.
            It defines the composition of systems (pairs of protocol
            plus adversaries), where the former becomes the
            environment to the latter.

            In \UC, the environment is an \ITM that provides inputs to
            both the adversary and protocol. It `represents whatever
            is external to the current protocol execution. This
            includes other protocol executions and their adversaries,
            human users, etc.'~\cite{uc}. 
            \UC's environment (this \ITM) thus serves as a proxy for
            the higher-level system and the \emph{protocol execution},
            composing environment and a pair of adversary and
            protocol, can be seen as a system composition.

\end{description}

\subsubsection{Conditions for \RC Composition}\label{sec:compo-axioms}
To understand what can be a valid or an invalid instance of these composition operators, we define axioms that regulate their behaviour.
These axioms are eventually used to derive \rhpref, and thus they are stated in terms of behavioural equivalence ($\beheq$ from \Cref{sec:rc}).%
\footnote{
    We conjecture that proving this connection for computational \UC may result in a connection with a different \RC criterion that may not be \rhpref.
    As such, we believe the following axioms would need to replace $\beheq$ with whatever is appropriate in that connection.
    For the sake of generality, our Isabelle formalisation is stated in terms of an arbitrary transitive relation.
}

\Cref{ax:par-decomp} requires that any attacker communicating with a program composed of two sub-programs can be decomposed into two attackers, each in the language of the respective program.
The reason we need the two programs is to know into what kind of languages to split the attacker.
Had we only a single program, we would not know the language of the second attacker.
\begin{axiom}[Attacker Decomposition]\label{ax:par-decomp}
    $\forall \ctxs{},\prgs{},\prgt{}\ldotp$ $\exists\ctxs{'},\ctxt{'}\ldotp$
    \[
        \ctxs{}  \links \left( \prgs{}\linkprog\prgt{} \right)
        \relateAbs
        \left(\ctxs{'}\linkatk\ctxt{'}\right) \links \left( \prgs{}\linkprog\prgt{} \right)
    \] 
\end{axiom}

\Cref{ax:inter-comp} states that linking an attacker and a program that are themselves composed using their FFI compositions is equivalent to linking each attacker and program and then use complete FFI composition for the result.
\begin{axiom}[FFI Recombination]\label{ax:inter-comp}
\[
    \left(\ctxs{}\linkatk\ctxt{}\right)\links
        \left( \prgs{}\linkprog\prgt{} \right)
    \relateAbs
    \left( \ctxs{} \links \prgs{} \right) \linkwhole
        \left( \ctxt{} \linktt \prgt{} \right)
\]
\end{axiom}

Finally, we formalise the relation between the behaviour of a program and the composition of this program with a new one. 
Essentially, we require that any two programs that have the same behaviour can be FFI-composed with \emph{the same} program, and still have the same behaviour.

\begin{axiom}[Constant Addition]\label{ax:const-elim}
    \[
        \pif 
        \prgs{}\relateAbs\prgt{}
        \pthen
        \prgo{} \linkwholeos \prgs{} \relateAbs \prgo{} \linkwholeot \prgt{}
    \]
\end{axiom}
Technically, \Cref{ax:const-elim} can be proven as a co-implication instead of as an implication, since our \RC composition operators do not change the visibility qualifiers within programs and attackers.
Since the implication is the required direction for the general composition theorem, we state the weaker axiom.

\subsubsection{Deriving the Composition Theorem}\label{sec:compos-thm-rc}

We can now state the \RC analogue to \Cref{thm:comp-thm-uc} as \Cref{thm:composition}.
Here, in the conclusion, we identify the simulator (or, source-level attacker) with \oth{S} in order to differentiate it from the adversary (or target-level attacker) \ctxo{}.
Apart from this cosmetic change, the theorem states that if a compiler is \rhpref when linking its target and source with a target attacker and a source simulator, then it is still \rhpref when linking against an external program (\prgo{}) and then against an external attacker \ctxo{} and simulator \oth{S}.
\begin{theorem}[Composition in \RC]\label{thm:composition}
    \[
     \pif
     \forall \ctxt{}.\exists\ctxs{}\ldotp
				\trg{\ctxt{}\hole{\comp{\prgs{}}}}
                                \relateAbs
				\src{\ctxs{}\hole{\prgs{}}}
    \pthen
    \forall \ctxo{}.\exists \oth{S}\ldotp
    \oth{\ctxo{}\hole{\prgo{}\ffiot{}\comp{\prgs{}}}}
                                \relateAbs
				\oth{S\hole{\prgo{}\ffios\prgs{}}}
    \]
\end{theorem}
\begin{proof}%
    \theoryProof{\theoryTRef{composition}.}
    \Cref{fig:composition} provides a visual depiction of this proof and how each of the presented axioms contribute to proving it (via transitivity of $\relateAbs$).
    Below, we indicate assumption $\trg{\ctxt{}\hole{\comp{\prgs{}}}}\relateAbs\src{\ctxs{}\hole{\prgs{}}}$ with \mi{HP}.

    \begin{figure}[!ht]
    \centering
    \begin{tikzpicture} [
        smallfont/.style = {font=\tiny}
        ]
        \node[](oatt){\ctxo{}};
        \node[right = .5 of oatt, align = center](ffilinkop){
            \prgo{}
            \\
            \linkprogot
            \\
            \comp{\prgs{}}
        };
        \draw[-] (ffilinkop.north west) -| (ffilinkop.south east) -| (ffilinkop.north west);
        \path[](oatt.center) to node[midway](linko){$\linko$\phantom{ii}} (ffilinkop.center);

        \node[right = .6 of ffilinkop, align = center](parlinkop){
            \ctxo{'}
            \\
            \linkatkot
            \\
            \ctxt{'}
        };
        \draw[-] (parlinkop.north west) -| (parlinkop.south east) -| (parlinkop.north west);
        \path[](ffilinkop.center) to node[midway](rel1){$\relateAbs$} (parlinkop.center);

        \node[right = .5 of parlinkop, align = center](ffilinkopp){
            \prgo{}
            \\
            \linkprogot
            \\
            \comp{\prgs{}}
        };
        \draw[-] (ffilinkopp.north west) -| (ffilinkopp.south east) -| (ffilinkopp.north west);
        \path[](parlinkop.center) to node[midway](linko2){$\linko$} (ffilinkopp.center);

        \node[right = .6 of ffilinkopp, align = center](genlinkop){
            $\ctxo{'}\linko\prgo{}$
            \\[0.3em]
            \quad \linkwholeot%
            \\[0.3em]
            $\ctxt{'}\linktt\comp{\prgs{}}$
        };
        \draw[-] (genlinkop.north west) -| ([yshift= 3.1em]genlinkop.south east) -| (genlinkop.north west);
        \draw[-] (genlinkop.south east) -| ([yshift=-3.1em]genlinkop.north west) -| (genlinkop.south east);
        \path[](ffilinkopp.center) to node[midway](rel2){$\relateAbs$\phantom{i}} (genlinkop.center);

        \node[right = .6 of genlinkop, align = center](genlinkos){
            $\ctxo{'}\linko\prgo{}$
            \\[0.3em]
            \quad \linkwholeos%
            \\[0.3em]
            $\ctxs{'}\links\prgs{}$
        };
        \draw[-] (genlinkos.north west) -| ([yshift= 3.1em]genlinkos.south east) -| (genlinkos.north west);
        \draw[-] (genlinkos.south east) -| ([yshift=-3.1em]genlinkos.north west) -| (genlinkos.south east);
        \path[](genlinkop.center) to node[midway](rel3){$\relateAbs$\phantom{i}} (genlinkos.center);

        \node[right = .6 of genlinkos, align = center](parlinkos){
            \ctxo{'}
            \\
            \linkatkos
            \\
            \ctxs{'}
        };
        \draw[-] (parlinkos.north west) -| (parlinkos.south east) -| (parlinkos.north west);
        \path[](genlinkos.center) to node[midway](rel4){\phantom{i}$\relateAbs$} (parlinkos.center);
        \draw [decorate, decoration = {brace}] ([yshift=-.2em]parlinkos.south east) to node[midway, below ]{$\isdef \oth{S}$} ([yshift=-.2em]parlinkos.south west);

        \node[right = .5 of parlinkos, align = center](ffilinkops){
            \prgo{}
            \\
            \linkprogos
            \\
            \prgs{}
        };
        \draw[-] (ffilinkops.north west) -| (ffilinkops.south east) -| (ffilinkops.north west);
        \path[](parlinkos.center) to node[midway](linko3){$\linko$} (ffilinkops.center);

        \node[right = .5 of ffilinkops, align = center](s){
           $\oth{S}$
        };
        \path[](ffilinkops.center) to node[midway](rel5){\phantom{ii}$\isdef$} (s.center);

        \node[right = .4 of s, align = center](ffilinkopsp){
            \prgo{}
            \\
            \linkprogos
            \\
            \prgs{}
        };
        \draw[-] (ffilinkopsp.north west) -| (ffilinkopsp.south east) -| (ffilinkopsp.north west);
        \path[](s.center) to node[midway](linko4){$\linko$\phantom{i}} (ffilinkopsp.center);

        \draw[decorate, decoration = {brace}] ([xshift = -.2em, yshift = -.2em]ffilinkopp.south east |- genlinkos.south) to node[midway, below, align = center ]{\Cref{ax:par-decomp}} ([xshift = .2em, yshift = -.2em]oatt.south west |- genlinkos.south);

        \draw[decorate, decoration = {brace}] ([xshift = -.2em, yshift = -2em]genlinkop.south east |- genlinkos.south) to node[midway, below, align = center ]{\Cref{ax:inter-comp}} ([xshift = .2em, yshift = -2em]parlinkop.south west |- genlinkos.south);

        \draw[decorate, decoration = {brace}] ([xshift = -.2em, yshift = -.2em]genlinkos.south east |- genlinkos.south) to node[midway, below, align = center ]{\Cref{ax:const-elim} {\small(with \mi{HP})}} ([xshift = .2em, yshift = -.2em]genlinkop.south west |- genlinkos.south); %

        \draw[decorate, decoration = {brace}] ([xshift = -.2em, yshift = -2em]ffilinkops.south east |- genlinkos.south) to node[midway, below, align = center ]{\Cref{ax:inter-comp} \\+ Symmetry of $\beheq$} ([xshift = .2em, yshift = -2em]genlinkos.south west |- genlinkos.south);

    \end{tikzpicture}
    \caption{\label{fig:composition} Visual depiction of the proof of \Cref{thm:composition}.}
    \end{figure}
\end{proof}

The proof to this theorem is remarkably compact (ten lines of Isabelle/HOL) which makes it easier to comprehend and separate the framework-specific insights (which are compartmentalised in the operators and their axioms) from the structure of the proof.
In \Cref{sec:ilc}, we will demonstrate that instantiating these composition operators and proving these axioms for a formal language is easy.
Moreover, it simplifies the proof of the composition theorem, which is usually considered the main result of any work presenting a \UC-like framework.

\subsection{The Dummy Attacker in \RC}\label{sec:dummy-for-rc}

The second key \UC result we want in \RC is the dummy adversary theorem (\Cref{thm:dummy-uc}).
Intuitively, \Cref{thm:dummy-uc} states that instead of considering all possible adversaries, one should consider only a specific one, the \emph{dummy adversary}.
This dummy adversary only forwards messages from the environment to
the protocol (which perceives them as coming from the adversary), and
relays messages from the protocol to the adversary to the environment.
The common \UC frameworks define the dummy
adversary explicitly, e.g., as a stateless \ITM that
implements a `transparent
channel`~\cite{uc,Hofheinz2011GNUC:-A-New-Uni}
or as set of channels between environment and protocol~\cite{iuc}.
In the context of $\RC$, we cannot explicitly define the dummy
adversary, as the concrete formulation is dependent on the programming
language. Instead, we will axiomatically define how it interacts with
the other part of the system.

\begin{definition}[Dummy Adversary (informal)]\label{def:dummy-adv}
    In \UC, the dummy adversary (\pattdummy{})
    is a proxy that relays all messages from the environment to the protocol and vice versa.     
\end{definition}
With the dummy adversary, we can state its theorem:
\begin{theorem}[Dummy Adversary Theorem in \UC~\cite{uc}]\label{thm:dummy-uc}
	\[
        \forall \patt{}\ldotp \exists \simu{}\ldotp \forall \env{}\ldotp
        \Execfun{\env{},\patt{},\prot{}}
        \indistp
        \Execfun{\env{},\simu{},\idfu{}}
        \piff
        \exists \simu{}\ldotp \forall \env{}\ldotp
        \Execfun{\env{},\pattdummy{},\prot{}}
        \indistp
        \Execfun{\env{},\simu{},\idfu{}}
 	\]
\end{theorem} 
From a formal perspective it is simple to understand why this theorem holds, the $\Rightarrow$ direction of the co-implication being trivial and the $\Leftarrow$ direction being the interesting one.
In fact, the distinction that the environment provides honest interactions and the adversary provides malicious ones only exists in theory, but not in any concrete formalisation of the \ITMs semantics.
So, since both the adversary and the environment are \ITMs that are $\forall$-quantified, it is sufficient to keep one entity to provide inputs and observe outputs (in this case the environment) and replace the other (i.e., the adversary) with a proxy.

The de-facto standard for \UC proofs is to prove \UC-emulation w.r.t.\ the dummy adversary and then apply \Cref{thm:dummy-uc} to obtain \UC-emulation in the sense of \Cref{def:uc}.
Importing this theorem into the \RC framework provides one key advantage for the proof of \rhpref.
Once the simulator is defined, proving \rhpref amounts to proving trace equivalence of the compiled program and of the source program linked with the simulator.
The key advantage is that these pieces of code are all defined, and so there is no need to perform any complicated induction and one can just reason about the semantics of the source and target programs.
Even better, proving trace equivalence can be mechanised by using existing program analysis tools (as we do later with Deepsec in \Cref{sec:mechanisation}) that verify precisely that property.

As before, we first need to identify key axioms that regulate the behaviour (and the composition) of the dummy attacker in \RC (\Cref{sec:dummy-axions}) before proving \Cref{thm:dummy-uc} in \RC (\Cref{sec:dummy-rc}).

\subsubsection{Additional Conditions for Dummy Attacker in \RC}\label{sec:dummy-axions}

Borrowing from the proof to the dummy adversary theorem in \UC, we rely on a \emph{dummy program}.
\begin{definition}[Dummy Program (informal)]\label{def:dummy-prot}
    The dummy program is a proxy that relays all messages from the
    environment to the program it is program-FFI composed with, and
    vice versa.
\end{definition}
Intuitively, the dummy program is the dual of the dummy attacker from \Cref{def:dummy-att}.
However, the dummy program only plays a role in the proof of the dummy attacker theorem and it does not appear anywhere else in any other formal statement.
Likewise, in \UC, the analogue notion appears only in the dummy
adversary theorem, as part of the construction of distinguishing
environment, but is never given a name.\footnote{Not to be confused
    with \UC's \emph{dummy parties}, which have an entirely different
purpose.}

From a proof perspective, the dummy program appears as part of the attacker, as captured by \Cref{ax:dummy-prog}.
In fact, the Axiom essentially states that any attacker linked with a program ($\ctxt{}\linktt\prgt{}$) can be separated into an equivalent system with the same program, an attacker, and the dummy protocol ($\ctxt{'} \linktt (\prgtdummy{} \linkprogtt \prgt{})$).
The second statement is essentially the same, but it allows the re-composition of an attacker (\trg{\ctxt{}}) and the dummy program (\trg{\prgtdummy{}}) into a single attacker (\src{\ctxs{'}}).
\begin{axiom}[Dummy Program Decomposition]\label{ax:dummy-prog}
$\forall \ctxt{}, \prgt{}\ldotp \exists \ctxt{'}, \ctxs{'}\ldotp$
    \[
        \ctxt{}\linktt\prgt{} \relateAbs \ctxt{'} \linktt (\prgtdummy{} \linkprogtt \prgt{})
        \text{ and }
        \ctxt{} \linktt (\prgtdummy{} \linkprogts \prgs{}) \relateAbs \ctxs{'}\links\prgs{}
    \]
\end{axiom}

Once the adversary part that is the dummy program has been identified, we can also carve out the part of the adversary that is the dummy attacker, as in \Cref{ax:dummy-att}.

\begin{definition}[Dummy Attacker (informal)]\label{def:dummy-att}
    In \RC, 
    the dummy attacker (\ctxtdummy{}) 
    is a proxy that relays all messages from the environment to the attacker, and vice versa.     
\end{definition}

\begin{axiom}[Dummy Attacker Decomposition]\label{ax:dummy-att}
$\forall \ctxt{}, \prgt{}\ldotp \exists \ctxt{'}\ldotp$
    \[
        \ctxt{}\linktt(\prgtdummy\linkprogtt\prgt{}) \relateAbs (\ctxt{'}\linkatktt\ctxtdummy)\linktt (\prgtdummy\linkprogtt\prgt{})
    \]    
\end{axiom}

Essentially, these two axioms state that it is possible to isolate the dummy program and the dummy attacker from an initial adversary, without altering the behaviour of the system.

\subsubsection{Deriving the Dummy Attacker Theorem}\label{sec:dummy-rc}
\Cref{thm:dummy-rc} presents the dummy attacker theorem in \RC.

\begin{theorem}[Dummy Attacker in \RC]\label{thm:dummy-rc}
	\[
     \forall \ctxt{}.\exists\ctxs{}\ldotp
				\trg{\ctxt{}\hole{\comp{\prgs{}}}}
                                \relateAbs
				\src{\ctxs{}\hole{\prgs{}}}
    \piff
    \exists\ctxs{'}\ldotp
				\trg{\ctxtdummy{}\hole{\comp{\prgs{}}}}
                                \relateAbs
				\src{\ctxs{'}\hole{\prgs{}}}
	\]
\end{theorem} 
\begin{proof}
	\theoryProof{\theoryTRef{dummy}}
    \Cref{fig:dummy} depicts the non-trivial direction of this proof ($\Leftarrow$) and how each of the presented axioms contribute to proving it (via transitivity of $\relateAbs$).
    Below, we indicate assumption $\trg{\ctxtdummy{}\hole{\comp{\prgs{}}}} \relateAbs \src{\ctxs{'}\hole{\prgs{}}}$ with \mi{HP}.
    One interesting bit is that this proof relies not just on the axioms presented in this section, but on two of the axioms presented in the previous one: \Thmref{ax:inter-comp} and \Thmref{ax:const-elim}.
    \begin{figure}[!ht]
    \centering
    \begin{tikzpicture} [
        smallfont/.style = {font=\tiny}
        ]
        \node[](n1){
            \ctxt{}\linktt\comp{\prgs}
        };
        
        \node[below = .5 of n1, align = center](n2){
            \ctxt{'}\linktt(\prgtdummy\linkprogtt\comp{\prgs})
        };
        \path[](n1.south) to node[sloped, midway](r1){$\relateAbs$} (n2.north);

        \node[right = .5 of n2, align = center](n3){
            (\ctxt{''}\linkatktt\ctxtdummy)\linktt(\prgtdummy\linkprogtt\comp{\prgs})
        };
        \path[](n2.east) to node[midway](r2){$\relateAbs$} (n3.west);
        
        \node[below = .5 of n3, align = center](n4){
            (\ctxt{''}\linktt\prgtdummy)\linkwholett(\ctxtdummy\linktt\comp{\prgs})
        };
        \path[](n3.south) to node[midway,sloped](r3){$\relateAbs$} (n4.north);

        \node[right = .5 of n4, align = center](n5){
            $(\ctxt{''}\linktt\prgtdummy)\linkwholets(\ctxs{'}\links\prgs{})$
        };
        \path[](n4.east) to node[midway](r4){$\relateAbs$} (n5.west);

        \node[below = .5 of n5, align = center](n7){
            $(\ctxt{''}\linkatkts\ctxs{'})\linktt(\prgtdummy\linkprogts\prgs{})$
        };
        \path[](n5.south) to node[midway,sloped](r6){$\relateAbs$} (n7.north);

        \node[right = .5 of n7, align = center](n8){
            $(\ctxs{}\links\prgs{})$
        };
        \path[](n7.east) to node[midway](r8){$\relateAbs$} (n8.west);

        \draw[decorate, decoration = {brace, aspect=0.2}] ([xshift = -.2em]n2.south east |- n1.north east) to node[pos=.2, right, align = center ]{\Cref{ax:dummy-prog}} ([xshift = -.2em]n2.south east);

        \draw[decorate, decoration = {brace, aspect=0.8}] ([xshift = -.2em]n2.west |- n2.north east) to node[pos=.8, above, align = center ]{\Cref{ax:dummy-att}} ([xshift = -.2em]n3.east |- n2.north east);

        \draw[decorate, decoration = {brace, aspect=0.2}] ([xshift = -.2em]n3.west |- n4.south west) to node[pos=.2, left, align = center ]{\Cref{ax:inter-comp}} ([xshift = -.2em]n3.west |- n3.north west);
        
        \draw[decorate, decoration = {brace, aspect=0.7}] ([xshift = .2em]n5.east |- n5.south east) to node[pos=.7, below, align = center ]{\Cref{ax:const-elim} {\small(with \mi{HP})}} ([xshift = -.2em]n4.west |- n4.south west);

        \draw[decorate, decoration = {brace, aspect=0.2}] ([xshift = .2em]n5.east |- n5.north) to node[pos=.2, right, align = center ]{\Cref{ax:inter-comp}} ([xshift = .2em]n5.east |- n7.south);

        \draw[decorate, decoration = {brace}]([yshift = -.2em]n8.east |- n7.south) to node[pos=.5, below, align = center ]{\Cref{ax:dummy-prog}} ([yshift = -.2em]n7.west |- n7.south) ;

    \end{tikzpicture}
    \caption{\label{fig:dummy} Visual depiction of the proof of \Cref{thm:dummy-rc}.}
    \end{figure}
\end{proof}

We have now identified the set of axioms that allow a \RC result to enjoy key \UC properties such as the composition theorem and the dummy adversary one.
Thus, the next section defines a language that fulfills those axioms (\Cref{sec:rilc}), in order to later on derive \UC results as \RC ones (\Cref{sec:sec-comp-proofs}).

We believe both the composition theorem and the dummy attacker ones are crucial, and thus one should prove all of the presented axioms in whatever language one chooses for \RC.
However, note that these axioms are not necessary for the connection to hold, they merely provide additional useful results.
In particular, the dummy attacker one simplifies the proofs to such a degree that mechanising them becomes possible using existing tools, as we show in \Cref{sec:mechanisation}.

\section{The Reactive, Interactive $\lambda$-Calculus (\rilc)}\label{sec:rilc}

This section presents the Reactive, Interactive $\lambda$-Calculus (\rilc), a reactive language that we use to carry out \UC proofs as \RC ones.
\rilc extends the Interactive $\lambda$-calculus (\ilc), so this section first presents the syntax, typing and operational semantics of \ilc, alongside the key properties of the language (\Cref{sec:ilc}).
Then this section presents the \rilc-proper extensions: a trace model as required by \RC and a module system to understand where each logical entity (e.g., protocol, attacker) ends (\Cref{sec:reactive}).
Finally, this section showcases the properties of \rilc, including the fact that it satisfies the axioms of \Cref{sec:composition} and thus it is ok to use it for our connection (\Cref{sec:rilc-props}).

\smallskip

As the name suggests, \rilc is a \emph{reactive} language.
Reactive languages differ from non-reactive ones in one key element: they do not have a notion of whole program.
In non-reactive languages, a program can run only if it is whole: all of its dependencies are resolved and there are no unknown function symbols.
Instead, reactive languages have a built-in notion of a black-box environment that a program can both `call' and `be called' (or send a message to and receive a message from).
The black box is therefore a catch-all element for all those symbols that are not resolved within the program.

Reactive languages are widely used in formal models for cryptographic protocol verification~\cite{acetoReactiveSystemsModelling2007},
and in our case they are the easiest kind of program for satisfying the axioms of \Cref{sec:composition}.
This is why we use a reactive language to instantiate our connection.
However, we believe it is possible to instantiate the connection with non-reactive languages too, as we discuss in \Cref{sec:react-vs-nonreact-langs}.

\subsection{The Interactive $\lambda$-Calculus (\ILC)}\label{sec:ilc}

In a nutshell, \ILC~\cite{ilc} is a process calculus that adapts \ITMs to a subset of the $\pi$-calculus~\cite{sangioPiBook} through an affine typing discipline.
To ensure that only one process is active (i.e., it can write) at any given time, processes implicitly pass around an affine ``write token''.
When a process \comsc{A} writes to another process \comsc{B}, process \comsc{A} must have the write token, and by writing it passes the toke to process \comsc{B}, which now has the write token. 
Moreover, to maintain that the order of activations is fully determined, the read endpoints of channels are (non-duplicable) affine resources, and so each write operation corresponds to a single, unique read operation. 
Together, these give \ILC its central metatheoretic property of confluence.

This section recaps the elements of \ILC we borrow from \citet{ilc}, namely its syntax (\Cref{sec:rilc-syn}), its typing (\Cref{sec:rilc-typ}) and its operational semantics (\Cref{sec:rilc-sem}).
Then, in order to familiarise the reader with \ILC processes, it presents some shorthands for \ILC notation that we use throughout this paper as well as an example \ILC process (\Cref{sec:short-ilc}).

\subsubsection{Syntax}\label{sec:ilc-syn}\label{sec:rilc-syn}
The syntax of \ILC is best described quoting from \citet{ilc}, starting from the types of the language.

\begin{align*}
	&
	\text{All Types } 	
		&
		\comsc{U}, \comsc{V}
		\bnfdef&\
			\comsc{A} \mid \comsc{X}
	\\
	&
	\text{Sendable Types } 	
		&
		\comsc{S}, \comsc{T}
		\bnfdef&\
			\comsc{1} \mid \comsc{S\times S} \mid \comsc{S+ S}
	\\
	&
	\text{Unrestricted Types } 	
		&
		\comsc{A}, \comsc{B}
		\bnfdef&\
			\comsc{S} \mid \comsc{A\times A} \mid \comsc{A+ A} \mid \comsc{\wrty{S}} \mid \comsc{A\myto{\infty} U} \mid \comsc{A\myto{w} U}
	\\
	&
	\text{Affine Types } 	
		&
		\comsc{X}, \comsc{Y}
		\bnfdef&\
			\comsc{\bang{A}} \mid \comsc{\rdty{S}} \mid \comsc{X\otimes X} \mid \comsc{X\oplus X} \mid \comsc{X\myto{1} U}
\end{align*}
 
Types (written \comsc{U}, \comsc{V} ) are bifurcated into unrestricted types (written \comsc{A}, \comsc{B}) and affine types (written \comsc{X}, \comsc{Y}).
A subset of the unrestricted types are sendable types (written \comsc{S}, \comsc{T}), i.e., the types of values that can be sent over channels. 
This restriction ensures that channels model network channels, which send only data. 
The sendable types include unit, products, and sums.
The unrestricted types include the sendable types, products, sums, write endpoint types (\comsc{\wrty{S}}), arrows, and write arrows (\comsc{A\myto{w} U}).
Write arrows specify unrestricted abstractions for which the write token can be moved into the affine context of the abstraction body during reduction.
The affine types include bang types, read endpoint types (\comsc{\rdty{S}}), products, sums, and arrows.

\begin{gather*}
\begin{aligned}
	&\text{Labels }
		&
		\comsc{\ell}
		\bnfdef&\
			\comsc{\pi} \mid \comsc{w}
	&
	&\text{Multiplicity Labels }
		&
		\comsc{\pi}
		\bnfdef&\
			\comsc{1} \mid \comsc{\infty}
	\\
	&\text{Channel Endpoint }
		&
		\comsc{c}
		\bnfdef&\
			\comsc{\rdchan{d}} \mid \comsc{\wrchan{d}}
	&
	&\text{Channel Names }
		&
		\comsc{d}
		\in&\
			\mathcal{D}
\end{aligned}
\\
\begin{aligned}
	&\text{Values }
		&
		\comsc{v}
		\bnfdef&\
			\comsc{\unit} \mid \comsc{\lamlb{\ell}{x:U}{e}} \mid \comsc{\pairlb{\ell}{v,v}} \mid \comsc{\inllb{\ell}{v}} \mid \comsc{\inrlb{\ell}{v}} \mid \comsc{c} \mid \comsc{\bang{v}}
	\\
	&\text{Expressions }
		&
		\comsc{e}
		\bnfdef&\
			\comsc{x} \mid \comsc{v} \mid \comsc{\pairlb{\ell}{e,e}} \mid \comsc{\inllb{\ell}{e}} \mid \comsc{\inrlb{\ell}{e}} 
		\mid
			\comsc{\applb{\ell}{e}{e}} \mid \comsc{\fixlb{\ell}{x.e}} \mid \comsc{\letinlb{\pi}{x}{e}{e}} 
	\\
		&&
		\mid&\ 
			\comsc{\splitlb{\ell}{e, x_1.x_2.e}}
		\mid
			\comsc{\caseoflb{\ell}{e}{e}{e}} 
	\\
		&&
		\mid&\
			\comsc{\bang{e}} \mid \comsc{\gnab{e}} \mid \comsc{\newch{x_1,x_2}{e}} \mid \comsc{\wrex{e}{e}} \mid \comsc{\rdex{e}{x.e}}
		\mid
			\comsc{\chanex{e}{x.e}{e}{x.e}} \mid \comsc{\forkex{e}{e}}
\end{aligned}
\end{gather*}
For concision, certain syntactic forms are parameterized by a multiplicity \comsc{\pi} to distinguish between the unrestricted (\comsc{\infty}) and affine (\comsc{1}) counterparts.
Other syntactic forms are parameterized by a syntax label \comsc{\ell}, which includes the multiplicity labels and the write label \comsc{w} (related to write effects). 
On introduction and elimination forms for functions (abstraction, application, and fixed points), the label \comsc{w} denotes variants that move around the write token as explained above. 
On introduction and elimination forms for products and sums, the label w denotes the sendable variants.

Values in \ILC include unit, lambda expressions, pairs, sums, channel endpoints (written \comsc{c}), and banged values. 
We distinguish between the names of channel endpoints (\comsc{\rdchan{d}} and \comsc{\wrchan{d}}) and the channel \comsc{d} itself that binds them. 
\ILC supports a fairly standard feature set of expressions:
pairs creation, left and right tagging, application, fixpoints, let-bindings, pair destruction, tag destruction, replication and un-replication, channel creation, writing and reading to channels, choice over multiple channel reads, and forking of a process.

\subsubsection{Typing}\label{sec:ilc-typ}\label{sec:rilc-typ}
\begin{align*}
	&\text{Typing Environment }
		&
		\comsc{\Gamma}
		\bnfdef&\
			\come \mid \comsc{\Gamma,(x:A)}
	\\
	&\text{Affine Environment }
		&
		\comsc{\Delta}
		\bnfdef&\
			\come \mid \comsc{\Delta, (x:X)} \mid \comsc{\Delta, \wrtk}
	\\
	&\text{Channel Environment }
		&
		\comsc{\Psi}
		\bnfdef&\
			\come \mid \comsc{\Psi, (d:S)}
\end{align*}
Typing relies on three environments: a classical typing environment that binds variables to unrestricted types, an affine environment that binds variables to affine types, and a channel enviornment binding channel names to sendable types.
Notice that the write token \comsc{\wrtk} lives in the affine context, though it cannot be bound to any variable. 
Instead, it flows around implicitly by virtue of where read and write effects are performed, as captured by the typing rules.

The typing judgement 
	\[
		\comsc{\Psi;\Delta;\Gamma}\vdash\comsc{e}:\comsc{U}
	\]
reads: \emph{``under channel environment \comsc{\Psi}, affine environment \comsc{\Delta}, and typing environment \comsc{\Gamma}, expression \comsc{e} has type \comsc{U}''}.

Most typing rules are standard for process calculi and affine types, we report them below and refer the reader to \citet{ilc} for an in-depth discussion of them.
\begin{center}
	\typerule{\ILC-Type-rdend}{
		\comsc{\Psi}(\comsc{d}) = \comsc{S}
	}{
		\comsc{\Psi;\Delta;\Gamma}\vdash \comsc{\rdchan{d}} : \comsc{\rdty{S}}
	}{ilc-ty-rdend}
	\typerule{\ILC-Type-wrend}{
		\comsc{\Psi}(\comsc{d}) = \comsc{S}
	}{
		\comsc{\Psi;\Delta;\Gamma}\vdash \comsc{\wrchan{d}} : \comsc{\wrty{S}}
	}{ilc-ty-wrend}
	\typerule{\ILC-Type-unit}{}{
		\comsc{\Psi;\Delta;\Gamma}\vdash \comsc{\unit} : \comsc{1}
	}{ilc-ty-unit}
	\typerule{\ILC-Type-var-u}{
		\comsc{\Gamma}(\comsc{x}) = \comsc{A}
	}{
		\comsc{\Psi;\Delta;\Gamma}\vdash \comsc{x} : \comsc{A}
	}{ilc-ty-var-u}
	\typerule{\ILC-Type-var-a}{
		\comsc{\Delta}(\comsc{x}) = \comsc{X}
	}{
		\comsc{\Psi;\Delta;\Gamma}\vdash \comsc{x} : \comsc{X}
	}{ilc-ty-var-a}
	\typerule{\ILC-Type-pair-u}{
		\comsc{\Psi;\Delta_1;\Gamma}\vdash \comsc{e_1} : \comsc{A_1}
		&
		\comsc{\Psi;\Delta_2;\Gamma}\vdash \comsc{e_2} : \comsc{A_2}
	}{
		\comsc{\Psi;\Delta_1,\Delta_2;\Gamma}\vdash \comsc{\pairlb{\infty}{e_1,e_2}} : \comsc{A_1\times A_2}
	}{ilc-ty-pair-u}
	\typerule{\ILC-Type-pair-s}{
		\comsc{\Psi;\Delta_1;\Gamma}\vdash \comsc{e_1} : \comsc{S_1}
		&
		\comsc{\Psi;\Delta_2;\Gamma}\vdash \comsc{e_2} : \comsc{S_2}
	}{
		\comsc{\Psi;\Delta_1,\Delta_2;\Gamma}\vdash \comsc{\pairlb{w}{e_1,e_2}} : \comsc{S_1\times S_2}
	}{ilc-ty-pair-s}
	\typerule{\ILC-Type-pair-a}{
		\comsc{\Psi;\Delta_1;\Gamma}\vdash \comsc{e_1} : \comsc{X_1}
		&
		\comsc{\Psi;\Delta_2;\Gamma}\vdash \comsc{e_2} : \comsc{X_2}
	}{
		\comsc{\Psi;\Delta_1,\Delta_2;\Gamma}\vdash \comsc{\pairlb{1}{e_1,e_2}} : \comsc{X_1\otimes X_2}
	}{ilc-ty-pair-a}
	\typerule{\ILC-Type-inl-u}{
		\comsc{\Psi;\Delta;\Gamma}\vdash \comsc{e} : \comsc{A_1}
	}{
		\comsc{\Psi;\Delta;\Gamma}\vdash \comsc{\inllb{\infty}{e}} : \comsc{A_1 + A_2}
	}{ilc-ty-inl-u}
	\typerule{\ILC-Type-inl-s}{
		\comsc{\Psi;\Delta;\Gamma}\vdash \comsc{e} : \comsc{S_1}
	}{
		\comsc{\Psi;\Delta;\Gamma}\vdash \comsc{\inllb{w}{e}} : \comsc{S_1 + A_2}
	}{ilc-ty-inl-s}
	\typerule{\ILC-Type-inl-a}{
		\comsc{\Psi;\Delta;\Gamma}\vdash \comsc{e} : \comsc{X_1}
	}{
		\comsc{\Psi;\Delta;\Gamma}\vdash \comsc{\inllb{1}{e}} : \comsc{X_1 \oplus X_2}
	}{ilc-ty-inl-a}
	\typerule{\ILC-Type-inr-u}{
		\comsc{\Psi;\Delta;\Gamma}\vdash \comsc{e} : \comsc{A_2}
	}{
		\comsc{\Psi;\Delta;\Gamma}\vdash \comsc{\inrlb{\infty}{e}} : \comsc{A_1 + A_2}
	}{ilc-ty-inr-u}
	\typerule{\ILC-Type-inr-s}{
		\comsc{\Psi;\Delta;\Gamma}\vdash \comsc{e} : \comsc{S_2}
	}{
		\comsc{\Psi;\Delta;\Gamma}\vdash \comsc{\inrlb{w}{e}} : \comsc{S_1 + S_2}
	}{ilc-ty-inr-s}
	\typerule{\ILC-Type-inr-a}{
		\comsc{\Psi;\Delta;\Gamma}\vdash \comsc{e} : \comsc{X_2}
	}{
		\comsc{\Psi;\Delta;\Gamma}\vdash \comsc{\inrlb{1}{e}} : \comsc{X_1 \oplus X_2}
	}{ilc-ty-inr-a}
	\typerule{\ILC-Type-split-u}{
		\comsc{\Psi;\Delta_1;\Gamma}\vdash \comsc{e} : \comsc{A_1\times A_2}
		&
		\comsc{\Psi;\Delta_2;\Gamma,x_1:A_1,x_2:A_2}\vdash \comsc{e'} : \comsc{U}
	}{
		\comsc{\Psi;\Delta_1,\Delta_2;\Gamma}\vdash \comsc{\splitlb{\infty}{e, x_1.x_2.e'}} : \comsc{U}
	}{ilc-ty-split-u}
	\typerule{\ILC-Type-split-s}{
		\comsc{\Psi;\Delta_1;\Gamma}\vdash \comsc{e} : \comsc{S_1\times S_2}
		&
		\comsc{\Psi;\Delta_2;\Gamma,x_1:S_1,x_2:S_2}\vdash \comsc{e'} : \comsc{U}
	}{
		\comsc{\Psi;\Delta_1,\Delta_2;\Gamma}\vdash \comsc{\splitlb{w}{e, x_1.x_2.e'}} : \comsc{U}
	}{ilc-ty-split-s}
	\typerule{\ILC-Type-split-a}{
		\comsc{\Psi;\Delta_1;\Gamma}\vdash \comsc{e} : \comsc{X_1\times X_2}
		&
		\comsc{\Psi;\Delta_2,x_1:X_1,x_2:X_2;\Gamma}\vdash \comsc{e'} : \comsc{U}
	}{
		\comsc{\Psi;\Delta_1,\Delta_2;\Gamma}\vdash \comsc{\splitlb{1}{e, x_1.x_2.e'}} : \comsc{U}
	}{ilc-ty-split-a}
	\typerule{\ILC-Type-case-u}{
		\comsc{\Psi;\Delta_1;\Gamma}\vdash \comsc{e} : \comsc{A_1 + A_2}
		&
		\comsc{\Psi;\Delta_2;\Gamma,x_1:A_1}\vdash \comsc{e_1} : \comsc{U}
		&
		\comsc{\Psi;\Delta_2;\Gamma,x_2:A_2}\vdash \comsc{e_2} : \comsc{U}
	}{
		\comsc{\Psi;\Delta_1,\Delta_2;\Gamma}\vdash \comsc{\caseoflb{\ell}{e}{e_1}{e_2}} : \comsc{U}
	}{ilc-ty-case-u}
	\typerule{\ILC-Type-case-s}{
		\comsc{\Psi;\Delta_1;\Gamma}\vdash \comsc{e} : \comsc{S_1 + S_2}
		&
		\comsc{\Psi;\Delta_2;\Gamma,x_1:S_1}\vdash \comsc{e_1} : \comsc{U}
		&
		\comsc{\Psi;\Delta_2;\Gamma,x_2:S_2}\vdash \comsc{e_2} : \comsc{U}
	}{
		\comsc{\Psi;\Delta_1,\Delta_2;\Gamma}\vdash \comsc{\caseoflb{w}{e}{e_1}{e_2}} : \comsc{U}
	}{ilc-ty-case-s}
	\typerule{\ILC-Type-case-a}{
		\comsc{\Psi;\Delta_1;\Gamma}\vdash \comsc{e} : \comsc{X_1 \oplus X_2}
		&
		\comsc{\Psi;\Delta_2,x_1:X_1;\Gamma}\vdash \comsc{e_1} : \comsc{U}
		&
		\comsc{\Psi;\Delta_2,x_2:X_2;\Gamma}\vdash \comsc{e_2} : \comsc{U}
	}{
		\comsc{\Psi;\Delta_1,\Delta_2;\Gamma}\vdash \comsc{\caseoflb{1}{e}{e_1}{e_2}} : \comsc{U}
	}{ilc-ty-case-a}
	\typerule{\ILC-Type-lam-u}{
		\comsc{\Psi;\come;\Gamma,x:A}\vdash \comsc{e} : \comsc{U}
	}{
		\comsc{\Psi;\Delta;\Gamma}\vdash \comsc{\lamlb{\infty}{x:A}{e}} : \comsc{A\myto{\infty}U}
	}{ilc-ty-lam-u}
	\typerule{\ILC-Type-lam-s}{
		\comsc{\Psi;\come,\wrty;\Gamma,x:A}\vdash \comsc{e} : \comsc{U}
	}{
		\comsc{\Psi;\Delta;\Gamma}\vdash \comsc{\lamlb{w}{x:A}{e}} : \comsc{A\myto{w}U}
	}{ilc-ty-lam-s}
	\typerule{\ILC-Type-lam-a}{
		\comsc{\Psi;\Delta,x:X;\Gamma}\vdash \comsc{e} : \comsc{U}
	}{
		\comsc{\Psi;\Delta;\Gamma}\vdash \comsc{\lamlb{1}{x:X}{e}} : \comsc{X\myto{1}U}
	}{ilc-ty-lam-a}
	\typerule{\ILC-Type-app-u}{
		\comsc{\Psi;\Delta_1;\Gamma}\vdash \comsc{e} : \comsc{A\myto{\infty}U}
		&
		\comsc{\Psi;\Delta_2;\Gamma}\vdash \comsc{e'} : \comsc{A}
	}{
		\comsc{\Psi;\Delta_1,\Delta_2;\Gamma}\vdash \comsc{\applb{\infty}{e}{e'}} : \comsc{U}
	}{ilc-ty-app-u}
	\typerule{\ILC-Type-app-s}{
		\comsc{\Psi;\Delta_1;\Gamma}\vdash \comsc{e} : \comsc{A\myto{w}U}
		&
		\comsc{\Psi;\Delta_2;\Gamma}\vdash \comsc{e'} : \comsc{A}
	}{
		\comsc{\Psi;\Delta_1,\Delta_2,\wrty;\Gamma}\vdash \comsc{\applb{w}{e}{e'}} : \comsc{U}
	}{ilc-ty-app-s}
	\typerule{\ILC-Type-app-a}{
		\comsc{\Psi;\Delta_1;\Gamma}\vdash \comsc{e} : \comsc{X\myto{1}U}
		&
		\comsc{\Psi;\Delta_2;\Gamma}\vdash \comsc{e'} : \comsc{X}
	}{
		\comsc{\Psi;\Delta_1,\Delta_2;\Gamma}\vdash \comsc{\applb{1}{e}{e'}} : \comsc{U}
	}{ilc-ty-app-a}
	\typerule{\ILC-Type-fix-u}{
		\comsc{\Psi;\come;\Gamma,x:A\myto{\infty}U}\vdash \comsc{e} : \comsc{A\myto{\infty}U}
	}{
		\comsc{\Psi;\Delta;\Gamma}\vdash \comsc{\fixlb{\infty}{x.e}} : \comsc{A\myto{\infty}U}
	}{ilc-ty-fix-u}
	\typerule{\ILC-Type-fix-s}{
		\comsc{\Psi;\come;\Gamma,x:A\myto{w}U}\vdash \comsc{e} : \comsc{A\myto{w}U}
	}{
		\comsc{\Psi;\Delta;\Gamma}\vdash \comsc{\fixlb{w}{x.e}} : \comsc{A\myto{w}U}
	}{ilc-ty-fix-s}
	\typerule{\ILC-Type-fix-a}{
		\comsc{\Psi;\come,x:X\myto{1}U;\Gamma}\vdash \comsc{e} : \comsc{X\myto{1}U}
	}{
		\comsc{\Psi;\Delta;\Gamma}\vdash \comsc{\fixlb{1}{x.e}} : \comsc{X\myto{1}U}
	}{ilc-ty-fix-a}
	\typerule{\ILC-Type-let-u}{
		\comsc{\Psi;\Delta_1;\Gamma}\vdash \comsc{e} : \comsc{A}
		&
		\comsc{\Psi;\Delta_2;\Gamma,x:A}\vdash \comsc{e} : \comsc{U}
	}{
		\comsc{\Psi;\Delta_1,\Delta_2;\Gamma}\vdash \comsc{\letinlb{\infty}{x}{e}{e'}} : \comsc{U}
	}{ilc-ty-let-u}
	\typerule{\ILC-Type-let-a}{
		\comsc{\Psi;\Delta_1;\Gamma}\vdash \comsc{e} : \comsc{X}
		&
		\comsc{\Psi;\Delta_2,x:A;\Gamma}\vdash \comsc{e} : \comsc{U}
	}{
		\comsc{\Psi;\Delta_1,\Delta_2;\Gamma}\vdash \comsc{\letinlb{1}{x}{e}{e'}} : \comsc{U}
	}{ilc-ty-let-u}
	\typerule{\ILC-Type-bang}{
		\comsc{\Psi;\Delta;\Gamma}\vdash \comsc{{e}} : \comsc{{A}}
	}{
		\comsc{\Psi;\Delta;\Gamma}\vdash \comsc{\bang{e}} : \comsc{\bang{A}}
	}{ilc-ty-bang}
	\typerule{\ILC-Type-gnab}{
		\comsc{\Psi;\Delta;\Gamma}\vdash \comsc{{e}} : \comsc{\bang{A}}
	}{
		\comsc{\Psi;\Delta;\Gamma}\vdash \comsc{\gnab{e}} : \comsc{A}
	}{ilc-ty-gnab}
	\typerule{\ILC-Type-nu}{
		\comsc{\Psi;\Delta,x_1:\rdty{S};\Gamma,x_2:\wrty{S}}\vdash \comsc{e} : \comsc{U}	
	}{
		\comsc{\Psi;\Delta;\Gamma}\vdash \comsc{\newch{x_1,x_2}{e}} : \comsc{U}
	}{ilc-ty-nu}
	\typerule{\ILC-Type-wr}{
		\comsc{\Psi;\Delta_1;\Gamma}\vdash \comsc{e} : \comsc{S}
		&
		\comsc{\Psi;\Delta_2;\Gamma}\vdash \comsc{e'} : \comsc{\wrty{S}}
	}{
		\comsc{\Psi;\Delta_1,\Delta_2,\wrtk;\Gamma}\vdash \comsc{\wrex{e}{e'}} : \comsc{1}
	}{ilc-ty-wr}
	\typerule{\ILC-Type-rd}{
		\comsc{\Psi;\Delta_1;\Gamma}\vdash \comsc{e} : \comsc{\rdty{S}}
		&
		\comsc{\Psi;\Delta_2,\wrtk,x:\bang{S}\otimes\rdty{S};\Gamma}\vdash \comsc{e'} : \comsc{U}
		&
		\comsc{\wrtk}\notin\comsc{\Delta_2}
	}{
		\comsc{\Psi;\Delta_1,\Delta_2;\Gamma}\vdash \comsc{\rdex{e}{x.e'}} : \comsc{U}
	}{ilc-ty-rd}
	\typerule{\ILC-Type-choice}{
		\comsc{\Psi;\Delta_1;\Gamma}\vdash \comsc{e_1} : \comsc{\rdty{S}}
		&
		\comsc{\Psi;\Delta_2;\Gamma}\vdash \comsc{e_2} : \comsc{\rdty{T}}
		&
		\comsc{\Psi;\Delta_3,\wrtk,x:\bang{S}\otimes\rdty{S}\otimes\rdty{T};\Gamma}\vdash \comsc{e'_1} : \comsc{U}
		\\
		\comsc{\Psi;\Delta_3,\wrtk,x:\bang{T}\otimes\rdty{S}\otimes\rdty{T};\Gamma}\vdash \comsc{e'_2} : \comsc{U}
		&
		\comsc{\wrtk}\notin\comsc{\Delta_3}
	}{
		\comsc{\Psi;\Delta_1,\Delta_2,\Delta_3;\Gamma}\vdash \comsc{\chanex{e_1}{x.e'_1}{e_2}{x.e'_2}} : \comsc{U}
	}{ilc-ty-choice}
	\typerule{\ILC-Type-fork}{
		\comsc{\Psi;\Delta_1;\Gamma}\vdash \comsc{e} : \comsc{U}
		&
		\comsc{\Psi;\Delta_2;\Gamma}\vdash \comsc{e'} : \comsc{V}
	}{
		\comsc{\Psi;\Delta_1,\Delta_2;\Gamma}\vdash \comsc{\forkex{e}{e'}} : \comsc{U}
	}{ilc-ty-fork}
\end{center}

\subsubsection{Operational Semantics}\label{sec:ilc-sem}\label{sec:rilc-sem}
\begin{gather*}
\begin{aligned}
	&
	\text{Process Names }
		&
		\comsc{p}, \comsc{q}
		\in&\
			\mathcal{P}
	&
	&\text{Names Sets }
		&
		\comsc{\Sigma}
		\bnfdef&\
			\come \mid \comsc{\Sigma, d} \mid \comsc{\Sigma, p}
	\\
	&\text{Process Pool }
		&
		\comsc{\Pi}
		\bnfdef&\
			\come \mid \comsc{\Pi, p:e}
	&
	&\text{Configurations }
		&
		\comsc{C}
		\bnfdef&\
			\comsc{\pair{\Sigma;\Pi}}
\end{aligned}
	\\
\begin{aligned}
	&\text{Evaluation Ctx. }
		&
		\comsc{\evalctx}
		\bnfdef&\
			\comsc{\holev{\cdot}} \mid \comsc{\pairlb{\ell}{\evalctx,e}} \mid \comsc{\pairlb{\ell}{v,\evalctx}} \mid \comsc{\inllb{\ell}{\evalctx}} \mid \comsc{\inrlb{\ell}{\evalctx}}
		\mid
			\comsc{\applb{\ell}{\evalctx}{e}} \mid \comsc{\applb{\ell}{v}{\evalctx}} \mid \comsc{\letinlb{\pi}{x}{\evalctx}{e}}
	\\
		&&
		\mid&\ 
			\comsc{\splitlb{\ell}{\evalctx, x_1.x_2.e}}
		\mid
			\comsc{\caseoflb{\ell}{\evalctx}{e}{e}} 
			\mid
			\comsc{\bang{\evalctx}} \mid \comsc{\gnab{\evalctx}}
	\\
		&&
		\mid&\
			\comsc{\wrex{\evalctx}{e}} \mid \comsc{\wrex{v}{\evalctx}} \mid \comsc{\rdex{\evalctx}{x.e}}
		\mid
			\comsc{\chanex{\evalctx}{x.e}{e}{x.e}} \mid \comsc{\chanex{e}{x.e}{\evalctx}{x.e}}
\end{aligned}
\end{gather*}
The operational semantics of \ILC relies on the notion of configuration \comsc{C}, which is a tuple of (existing) channel and process names \comsc{\Sigma}, and a pool of running and terminated processes \comsc{\Pi}.
The operational semantics of \ILC is a small-step concurrent semantics that relies on a contextual semantics for the reduction of expressions.
The following judgements comprise the operational semantics of \ILC:
\begin{align*}
	\comsc{C\equiv C'}
		&&&
		\text{ congruence rules between configurations }
	\\
	\comsc{c \leadsto c'}
		&&&
		\text{ channel endpoint communication: \comsc{c} communicates to \comsc{c'} }
	\\
	\comsc{C \myred C'}
		&&&
		\text{ configuration reduction: \comsc{C} takes a step and becomes \comsc{C'} }
	\\
	\comsc{e \redp e'}
		&&&
		\text{ expression primitive reduction: \comsc{e} takes a primitive step and becomes \comsc{e'} }
\end{align*}

The rules of the operational semantics are standard for process calculi, we report them below and refer the reader to \citet{ilc} for an in-depth discussion of them.
\begin{center}
	\typerule{\ILC-eq-conf}{
		\comsc{\Pi}\equiv_{\mi{perm}}\comsc{\Pi'}
	}{
		\comsc{\pair{\Sigma;\Pi}}\equiv\comsc{\pair{\Sigma;\Pi'}}
	}{ilc-sem-eq-conf}

	\typerule{\ILC-connect}{}{
		\comsc{\wrty{d}\leadsto\rdty{d}}
	}{ilc-sem-connect}

	\typerule{\ILC-Sem-local}{
		\comsc{e \redp e'}
	}{
		\comsc{\pair{\Sigma;\Pi,p:\evalctx\holev{e}} \myred \pair{\Sigma;\Pi,p:\evalctx\holev{e'}}}
	}{ilc-sem-local}
	\typerule{\ILC-Sem-fork}{
		\comsc{q}\notin\comsc{\Sigma}
	}{
		\comsc{\pair{\Sigma;\Pi,p:\evalctx\holev{\forkex{e}{e'}}} \myred \pair{\Sigma,q;\Pi,p:\evalctx\holev{e},q:e'}}
	}{ilc-sem-fork}
	\typerule{\ILC-Sem-nu}{
		\comsc{d}\notin\comsc{\Sigma}
	}{
		\comsc{\pair{\Sigma;\Pi,p:\evalctx\holev{\newch{x_1,x_2}{e}}} \myred \pair{\Sigma,d;\Pi,p:\evalctx\holev{e\subst{\rdchan{d}}{x_1}\subst{\wrchan{d}}{x_2}}}}
	}{ilc-sem-nu}
	\typerule{\ILC-Sem-read-write}{
		\comsc{c_2 \leadsto c_1}
	}{
		\begin{multlined}
			\comsc{\pair{\Sigma;\Pi,p:\evalctx\holev{\rdex{c_1}{x.e}},q:\evalctx'\holev{\wrex{v}{c_2}}} \myred}
			\\
			\comsc{\pair{\Sigma;\Pi,p:\evalctx\holev{e\subst{\pairlb{1}{!v,c_1}}{x}},q:\evalctx'\holev{\unit} }}
		\end{multlined}
	}{ilc-sem-rw}
	\typerule{\ILC-Sem-choice}{
		\comsc{c \leadsto c_i} \text{ for } \comsc{i}\in\comsc{1..2}
	}{
		\begin{multlined}
			\comsc{\pair{\Sigma;\Pi,p:\evalctx\holev{\chanex{c_1}{x_1.e_1}{c_2}{x_2.e_2}},q:\evalctx'\holev{\wrex{v}{c}}} \myred }
			\\
			\comsc{\pair{\Sigma;\Pi,p:\evalctx\holev{e_i\subst{\pairlb{1}{!v,c_1,c_2}}{x_i}},q:\evalctx'\holev{\unit} }}
		\end{multlined}
	}{ilc-sem-ch}
	\typerule{\ILC-Sem-cong}{
		\comsc{C_1}\equiv\comsc{C'_1}
		&
		\comsc{C_1' \myred C_2'}
		&
		\comsc{C_2'}\equiv\comsc{C_2}
	}{
		\comsc{C_1 \myred C_2}
	}{ilc-sem-cong}

	\typerule{\ILC-Sem-ex-letin}{
	}{
		\comsc{\letinlb{\pi}{x}{v}{e} \redp e\subst{v}{x}}
	}{ilc-sem-ex-letin}
	\typerule{\ILC-Sem-ex-beta}{
	}{
		\comsc{\applb{\ell}{\lamlb{\ell}{x}{e}}{v} \redp e\subst{v}{x}}
	}{ilc-sem-ex-beta}
	\typerule{\ILC-Sem-ex-split}{
	}{
		\comsc{\splitlb{\ell}{\pairlb{\ell}{v_1,v_2}, x_1.x_2.e} \redp e\subst{v_1}{x_1}\subst{v_2}{x_2}}
	}{ilc-sem-ex-split}
	\typerule{\ILC-Sem-ex-case-l}{
	}{
		\comsc{\caseoflb{\ell}{\inllb{\ell}{v}}{e}{e'} \redp e\subst{v}{x_1}}
	}{ilc-sem-ex-case-l}
	\typerule{\ILC-Sem-ex-case-r}{
	}{
		\comsc{\caseoflb{\ell}{\inrlb{\ell}{v}}{e}{e'} \redp e'\subst{v}{x_2}}
	}{ilc-sem-ex-case-r}
	\typerule{\ILC-Sem-ex-fix}{
	}{
		\comsc{\fixlb{\ell}{x.e} \redp e\subst{\fixlb{\ell}{x.e}}{x}}
	}{ilc-sem-ex-fix}
	\typerule{\ILC-Sem-ex-gnabang}{
	}{
		\comsc{\gnab{\bang{v}} \redp v}
	}{ilc-sem-ex-gb}
\end{center}

\subsubsection{Shorthands and \ILC Example}\label{sec:short-ilc}

The syntax of \ILC lacks many useful real-world shorthands.
In this section we present the semantics for the following such useful shorthands:
\begin{itemize}
	\item \comsc{fwd({\rdchan{d}},{\wrchan{d'}})} defines a proxy that forwards all messages read on channel \comsc{d} to channel \comsc{d'} (\Cref{tr:ilc-sem-eq-fwd});
	\item \comsc{\bop} models the usual binary operations ($+$,$-$,$==$, etc.) on values (\Cref{tr:ilc-sem-ex-bop});
	\item \Cref{tr:ilc-sem-ex-iftet,tr:ilc-sem-ex-iftef} show if-then-else reductions that can be obtained from case destruction of tagged values;
	\item \comsc{loop(e)} models an unbound loop (\Cref{tr:ilc-sem-ex-loop});
	\item \comsc{loop(n)(e)} models an bounded loop that runs for \comsc{n} iterations (\Cref{tr:ilc-sem-ex-loopn,tr:ilc-sem-ex-loop0}).
\end{itemize}

\begin{center}
	\typerule{\ILC-eq-fwd}{
	}{
		\comsc{\pair{\Sigma;\Pi, p:\evalctx\holev{fwd({\rdchan{d}},{\wrchan{d'}})} }}\equiv\comsc{\pair{\Sigma;\Pi, p: \evalctx\holev{ \rdex{\rdchan{d}}{x.\wrex{x}{\wrchan{d'}}} } }}
	}{ilc-sem-eq-fwd}

	\typerule{\ILC-Sem-ex-bop}{
		\comsc{v''} = \bop (\comsc{v},\comsc{v'})
	}{
		\comsc{ v \bop v' \redp v''}
	}{ilc-sem-ex-bop}
	\typerule{\ILC-Sem-ex-ifte-t}{
	}{
		\comsc{ \ifte{\true}{e}{e'} \redp e}
	}{ilc-sem-ex-iftet}
	\typerule{\ILC-Sem-ex-ifte-f}{
	}{
		\comsc{ \ifte{\false}{e}{e'} \redp e'}
	}{ilc-sem-ex-iftef}

	\typerule{\ILC-Sem-ex-loop}{
	}{
		\comsc{ loop(e) \redp e;loop(e)}
	}{ilc-sem-ex-loop}
	\hspace{-1em}
	\typerule{\ILC-Sem-ex-loopN}{
		n > 1
	}{
		\comsc{ loop(n)(e) \redp e;loop(n-1)(e)}
	}{ilc-sem-ex-loopn}
	\typerule{\ILC-Sem-ex-loop1}{
	}{
		\comsc{ loop(1)(e) \redp e}
	}{ilc-sem-ex-loop0}
\end{center}

We defer presenting an example of \ILC code (and its semantics) to \Cref{sec:sec-comp-proofs}.
Note that, in the examples, we often massage the \ILC syntax to enhance readability.
For example, we will write code in an imperative form (more similar to other similar languages used by tools that perform protocol verification).
Moreover, we will write \lstinline{wr (Msg v) from Write(PQ) } instead of writing \comsc{\wrex{Msg~v}{\wrchan{PQ}}}, assuming that message carry a defining header (in this case, \comsc{Msg}).
Finally, we will write \lstinline{rd (Msg x) from Read(PQ)} instead of writing a more complex read expression that reads on channel end \comsc{\rdchan{PQ}} and then checks that the read value matches some header \comsc{Msg} and binds the read value in \comsc{x}.

\subsection{\rilc: Modules, Traces and Cryptographic Additions to \ILC}\label{sec:reactive}
This section presents our addition to \ILC that define \rilc.
The first such addition is a series of cryptographic primitives that are used to describe the hash-based commitment protocol of \Cref{sec:sec-comp-proofs} (\Cref{sec:rilc-crypto}).

The second one is a module system that is used to clearly identify the different modules (read, protocols and sub-protocols) that comprise a \rilc program (\Cref{sec:rilc-module-sys}).

The final addition is a reactive trace semantics whose purpose is twofold (\Cref{sec:rilc-traces}).
First, the reactive semantics lifts the language into a reactive setting, where \rilc programs (now called modules) interact with an unspecified environment.
Second, the traces capture any interaction that happens on the interface between modules and the environment.

\subsubsection{Cryptographic Additions}\label{sec:rilc-crypto}
Vanilla \ILC does not contain cryptographic primitives, but the authors provide some additions to \ILC~\cite[Appendix]{ilc} in order to model a commitment protocol (the same protocol we also model later).
In this version of \rilc, we add cryptographic primitives for symbolic pseudorandom generation with a trapdoor and xor-ing of symbolic values.
Our cryptographic additions are handled in a separate part of the program state, so that they can be easily extended and replaced with minimal effort for ensuring their additions do not violate any metatheoretical property.

\begin{gather*}
\begin{aligned}
	&\text{Security Parameter }& \comsc{\secpam} \in
		&\
		\mb{N}
	&&&
	&\text{Random Seed }& \comsc{\nrnd} \bnfdef
		&\
		\comsc{b_1\cdots b_j}
	\\
	&\text{Seals }& \comsc{\sigma}
		\in&\ 
		\comsc{\mc{S}}
	&&&
	&\text{Values }& \comsc{v}
		\bnfdef&\ 
		\cdots \mid \comsc{\sigma}
\end{aligned}
	\\
\begin{aligned}
	&\text{ Expressions }& \comsc{e}
		\bnfdef&\ 
		\cdots \mid \comsc{\takerand} \mid \comsc{\keygen{e}} \mid \comsc{\prgen{e}{e}} \mid \comsc{\invert{e}{e}} \mid \comsc{xors(e, e)}
	\\
	&\text{ Evaluation Contexts }& \comsc{E}
		\bnfdef&\ 
		\cdots \mid \comsc{\keygen{E}} \mid \comsc{\prgen{E}{e}} \mid \comsc{\prgen{v}{E}} \mid \comsc{\invert{E}{e}} 
	\\
	&&\mid&\
		\comsc{\invert{v}{E}} \mid \comsc{xors(E, e)} \mid \comsc{xors(v, E)}
\end{aligned}
\end{gather*}
To model the output of a symbolic prg, we introduce the notion of seals \comsc{\sigma}, which are values and opaque tokens (which have been used to model other cryptographic primitives such as additions in other work~\cite{articlespi,sumii_logical_2003,Sumii:2004:BDS:964001.964015}.
There are a number of expressions that deal with cryptographic primitives: taking a $\secpam$-long random bitstring, generating a key, prg-hashing a value with a key, inverting the result of a hash via a trapdoor, and xor-ing values.
In general, we assume a security parameter \comsc{\secpam} which is an integer and a randomness parameter \comsc{\nrnd}, which is a string of bits whose length is a polynomial function of \comsc{\secpam}.

\paragraph{Typing of Cryptographic Additions}

Typing cryptographic additions is straightforward (and therefore omitted), since they do not pass the write token around.
The only meaningful change to the typing is that \comsc{\takerand} needs to know the length of the security parameter in order to tell the type of the bitstring it returns.
Thus, typing rules need to be extended to carry around $\secpam$.
Since this is a minor, tedious addition that mostly clutters the notation, we elide that annotation in subsequent judgements.

\paragraph{Operational Semantics of Cryptographic Additions}
\begin{align*}
	&\text{Cryptographic Stores }& \comsc{G} \bnfdef
		&\
		\come \mid \comsc{G, \xorbinding{\sigma}{v}{v}} 
		\mid \comsc{G,\com{\cryptobinder{v}{v'}{\pair{\sigma,\bot,\bot}}}}
		\mid \comsc{G,\com{\cryptobinder{v}{v'}{\pair{\sigma,v'',\sigma'}}}}
	\\
	&\text{Leaked Knowledge }& \comsc{H} \bnfdef
		&\
		\come \mid \comsc{H , v}
	\\
	&\text{Cryptographic Configuration}& \comsc{K} \bnfdef
		&\
		\comsc{\pair{H ; G ; \secpam ; \nrnd}}
\end{align*}
In order to present the semantics of the cryptographic expressions, we need to define additional elements of the runtime state that the operational semantics rules rely on.

Cryptographic stores \comsc{G} are a store needed to collect any information required by the cryptographic additions.
In this work, the store contains: the result of \comsc{xors} operations, bindings for a random number \comsc{v} and a fresh key \comsc{v'} bound to a trapdoor \comsc{\sigma} that has not been used (thus the two \comsc{\bot}), and bindings for a key \comsc{v} used to create the prg \comsc{\sigma'} for a value \comsc{v''}.
Leaked knowledge represent all values that have been leaked, the specific way in which leakage takes place (i.e., how can modules leak data to the attacker) is presented later, in \Cref{sec:rilc-traces}.
Cryptographic configurations \comsc{K} are the part of a runtime state that contain all cryptography-related additions.

The operational semantics that handles cryptographic additions follows these judgements
\begin{align*}
	\comsc{K \triangleright \exc{}\redp {K}' \triangleright \exc{'}}
		&&&
		\text{ Primitive reduction relying on the cryptographic configuration }
	\\
	\comsc{\pair{K,C}\myred \pair{K',C'}}
		&&&
		\text{ Cryptographic configuration takes a step  }
\end{align*}

Below are the semantics rules of the cryptographic additions.
\begin{center}
	\typerule{\rilc-Sem-take}{
		\comsc{K} = \comsc{\pair{H ; G ; \secpam ; \nrnd}}
		&
		\comsc{\nrnd} = \comsc{b_0\cdots b_{\secpam-1}\cdots b_{j}}
		&
		\comsc{\secpam-1\leq j}
		\\
		\comsc{n'} = \comsc{b_0\cdots b_{\secpam-1}}
		&
		\comsc{\nrnd_c} = \comsc{b_{\secpam}\cdots b_{j}}
		&
		\comsc{K'} = \comsc{\pair{H ; G ; \secpam ; \nrnd_c}}
	}{
		\comsc{K \triangleright \takerand\ \redp K' \triangleright n'}
	}{rrilc-e-take}
	\typerule{\rrilc-Sem-keygen}{
		\comsc{K} = \comsc{\pair{H ; G ; \secpam ; \nrnd}}
		&
		\com{G'} = \com{G},\com{\cryptobinder{v_{rand}}{v_{pubkey}}{\pair{\sigma_{trap},\bot,\bot}}}
		&
		\fun{fresh}{v_{rand}, G}
		\\
		\fun{fresh}{v_{pubkey}, G}
		&
		\fun{fresh}{\sigma_{trap}, G}
		&
		\comsc{K'} = \comsc{\pair{H ; G' ; \secpam ; \nrnd}}
	}{
		\com{K \triangleright \keygen{v_{rand}} \redp K' \triangleright \pairlb{\infty}{v_{pubkey}, \sigma_{trap}} }
	}{rrilc-sem-ex-key2}
	\typerule{\rrilc-Sem-keygen}{
		\comsc{K} = \comsc{\pair{H ; G ; \secpam ; \nrnd}}
		&
		\com{\cryptobinder{v_{rand}}{v_{pubkey}}{\pair{\sigma_{trap},\bot,\bot}}} \in \com{G}
	}{
		\com{K \triangleright \keygen{v_{rand}} \redp K \triangleright \pairlb{\infty}{v_{pubkey}, \sigma_{trap}} }
	}{rrilc-sem-ex-key-old}
	\typerule{\rrilc-Sem-trapdoor}{
		\comsc{K} = \comsc{\pair{H ; G ; \secpam ; \nrnd}}
		&
		\fun{fresh}{\sigma, G}
		&
		\com{\cryptobinder{v_{rand}}{v_{pubkey}}{\pair{\sigma_{trap},\bot,\bot}}} \in \com{G}
		\\
		\com{G'} = \com{G},\com{\cryptobinder{v_{rand}}{v_{pubkey}}{\pair{\sigma_{trap},v_{in},\sigma}}} \setminus \com{\cryptobinder{v_{rand}}{v_{pubkey}}{\pair{\sigma_{trap},\bot,\bot}}}
		&
		\comsc{K'} = \comsc{\pair{H ; G' ; \secpam ; \nrnd}}
	}{
		\com{K \triangleright \prgen{v_{pubkey}}{v_{in}} \redp K' \triangleright \sigma }
	}{rrilc-sem-ex-prg}
	\typerule{\rrilc-Sem-trapdoor-exists}{
		\comsc{K} = \comsc{\pair{H ; G ; \secpam ; \nrnd}}
		&
		\com{\cryptobinder{v_{rand}}{v_{pubkey}}{\pair{\sigma_{trap},v_{in},\sigma}}} \in \com{G}
	}{
		\com{K \triangleright \prgen{v_{pubkey}}{v_{in}} \redp K \triangleright \sigma }
	}{rrilc-sem-ex-prg-ex}
	\typerule{\rrilc-Sem-invert-true}{
		\comsc{K} = \comsc{\pair{H ; G ; \secpam ; \nrnd}}
		&
		\exists \com{v_{in}} \ldotp \com{\cryptobinder{v_{rand}}{v_{pubkey}}{\pair{\sigma_{trap},v_{in},\sigma}}} \in \com{G}
	}{
		\com{K \triangleright\invert{\pairlb{\infty}{v_{pubkey}, \sigma_{trap}}}{\sigma} \redp K \triangleright \true }
	}{rrilc-sem-ex-inv2-t}
	\typerule{\rrilc-Sem-invert-false}{
		\comsc{K} = \comsc{\pair{H ; G ; \secpam ; \nrnd}}
		&
		\nexists \com{v_{in}} \ldotp \com{\cryptobinder{v_{rand}}{v_{pubkey}}{\pair{\sigma_{trap},v_{in},\sigma}}} \in \com{G}
	}{
		\com{K \triangleright\invert{\pairlb{\infty}{v_{pubkey}, \sigma_{trap}}}{\sigma} \redp K \triangleright \false }
	}{rrilc-sem-ex-inv2-f}
	\typerule{\rrilc-Sem-xor-seal}{
		\comsc{K} = \comsc{\pair{H ; G ; \secpam ; \nrnd}}
		&
		\comsc{K'} = \comsc{\pair{H ; G' ; \secpam ; \nrnd}}
		\\
		\text{ if } \xorbinding{v}{\_}{\_}\notin{G} \text{ then } G' = G, \xorbinding{v}{v'}{\sigma} \text{ and } \fun{fresh}{\sigma, G}
		\\
		\text{ if } \xorbinding{v}{v'}{\sigma}\in G \text{ then } G'=G
		\\
		\text{ if } \xorbinding{v}{v''}{\sigma''}\in G \text{ and } v'\neq v'' \text{ then } G'=G, \xorbinding{v}{v'}{\sigma} \text{ and } \fun{fresh}{\sigma, G}
	}{
		\com{K \triangleright xors(v, v') \redp K'\triangleright \sigma }
	}{rrilc-sem-xor-seal}

	\typerule{\rilc-Sem-crypto}{
		\comsc{K \triangleright \exc{}\redp {K}' \triangleright \exc{'}}
	}{
		\comsc{\pair{K;\Sigma;\Pi,p:\evalctx\holev{e}} \myred \pair{K';\Sigma;\Pi,p:\evalctx\holev{e'}}}
	}{rilc-sem-crypto}
	\typerule{\rilc-Sem-local}{
		\comsc{\pair{C} \myred \pair{C}}	
	}{
		\comsc{\pair{K;C} \myred \pair{K;C}}
	}{rilc-sem-local}
\end{center}

\Cref{tr:rrilc-e-take} takes the first \comsc{\secpam}-long bitstring from the randomness bits \comsc{\nrnd} and updates that accordingly.
\Cref{tr:rrilc-sem-ex-key2} takes a random value \comsc{v_{rand}} and generates a fresh binding for a key \comsc{v_{pubkey}} with a known trapdoor \comsc{\sigma_{trap}}.
\Cref{tr:rrilc-sem-ex-key-old} returns the existing binding in case the key generation is called with a known input.
\Cref{tr:rrilc-sem-ex-prg} calculates the fresh prg \comsc{\sigma} of a value \comsc{v_{in}} with some key \comsc{v_{pubkey}}.
\Cref{tr:rrilc-sem-ex-prg-ex} is used in case the prg of some value and some key has already been calculated, in which case, the result is fetched from the cryptographic store \comsc{G}.
\Cref{tr:rrilc-sem-ex-inv2-t} acknowledges a correct inversion of a prg \comsc{\sigma} given its key \comsc{v_{pubkey}} and the trapdoor \comsc{\sigma_{trap}}, while \Cref{tr:rrilc-sem-ex-inv2-f} is used when the inversion is used on a prg with either the wrong key or trapdoor.
\Cref{tr:rrilc-sem-xor-seal} models a simple, one-directional exclusive or.

\Cref{tr:rilc-sem-crypto} lifts the cryptographic primitive reductions to the cryptographic state while \Cref{tr:rilc-sem-local} lifts any previous reduction from \Cref{sec:ilc-sem} that does not use cryptographic primitives to the cryptographic state.

In the following, we define the initial cryptographic state via function \comsc{K_0(\cdot)}, which takes a security parameter and a random bitstring in input and returns the empty cryptographic state.
\begin{align*}
	\comsc{K_0(\secpam,\nrnd)} \isdef&\ \comsc{\pair{\come,\come,\secpam,\nrnd}}
\end{align*}

\paragraph{Cryptography for the Environment}
As we have hinted (but will only show in \Cref{sec:rilc-traces}), \rilc is a reactive language, whose model contains a black-box environment that can send messages to the reactive program being run.
The cryptographic additions should also let the semantics calculate what is the environment allowed to compute, and this is what the cryptographic environment semantics describes.
The judgement for this semantics is:
\begin{align*}
	\comsc{
		\envredkk{K}{K'}{v}
	}
	&&&
	\text{
		From knowledge \comsc{K}, the environment produces value \comsc{v}
	}
	\\
	&&&
	\text{
		 and augments its knowledge to \comsc{K'}
	}
\end{align*}
Below are the rules for this judgement, which we keep to a minimum for simplicity.
\begin{center}
	\typerule{\rilc-Sem-EC-base}{
		\comsc{v} \neq \comsc{\sigma}
	}{
		\envredkk{K}{K}{v}
	}{env-cry-base}
	\typerule{\rilc-Sem-EC-history}{
		\comsc{K} = \comsc{\pair{H ; G ; \secpam ; \nrnd}}
		&
		\comsc{v} \in \comsc{H}
	}{
		\envredkk{K}{K}{v}
	}{env-cry-his}
	\typerule{\rilc-Sem-EC-take}{
		\comsc{K \triangleright \takerand\ \redp K' \triangleright n'}
	}{
		\envredkk{K}{K'}{n'}
	}{env-cry-rand}
	\typerule{\rilc-Sem-EC-xor}{
		\envredkk{K}{K'}{v}
		&
		\envredkk{K'}{K''}{v'}
		&
		\com{K'' \triangleright xors(v, v') \redp K'''\triangleright \sigma }
		\\
		\comsc{K'''} = \comsc{\pair{H ; G ; \secpam ; \nrnd}}
		&
		\comsc{K_f} = \comsc{\pair{H,\sigma ; G ; \secpam ; \nrnd}}
	}{
		\envredkk{K}{K_f}{\sigma}
	}{env-cry-xor}
	\typerule{\rilc-Sem-EC-keygen}{
		\envredkk{K}{K'}{v}
		&
		\com{K' \triangleright \keygen{v} \redp K'' \triangleright \pairlb{\infty}{v_{pubkey}, \sigma_{trap}} }
		\\
		\comsc{K''} = \comsc{\pair{H ; G ; \secpam ; \nrnd}}
		&
		\comsc{K_f} = \comsc{\pair{H,\sigma_{trap},v_{pubkey} ; G ; \secpam ; \nrnd}}
	}{
		\envredkk{K}{K_f}{ \pairlb{\infty}{v_{pubkey}, \sigma_{trap}} }
	}{env-cry-key}
	\typerule{\rilc-Sem-EC-prg}{
		\envredkk{K}{K'}{v}
		&
		\envredkk{K'}{K''}{v'}
		&
		\com{K'' \triangleright \prgen{v}{v'} \redp K''' \triangleright \sigma }
		\\
		\comsc{K'''} = \comsc{\pair{H ; G ; \secpam ; \nrnd}}
		&
		\comsc{K_f} = \comsc{\pair{H,\sigma ; G ; \secpam ; \nrnd}}
	}{
		\envredkk{K}{K_f}{\sigma}
	}{env-cry-prg}
\end{center}
An attacker can generate any value that is a ground value (\Cref{tr:env-cry-base}), any value that has been sent to it in the past (\Cref{tr:env-cry-his}), or any random value from the randomness bitstring (\Cref{tr:env-cry-rand}).
An attacker can also compute the xor of any value it could generate (\Cref{tr:env-cry-xor}) and generate a fresh key (\Cref{tr:env-cry-key}), an act that leaks the key trapdoor to the attacker knowledge.
Finally, an attacker can create a fresh prg, so long as the values used in its computation were known to it (\Cref{tr:env-cry-prg}).
Given that our treatment of cryptographic elements is symbolic, we do not concern ourselves with modelling a semantics with a computational bound for the attacker.
We believe lifting this simplification can be done by relying on existing work, e.g.,~\cite{ipdl}.

\subsubsection{A Module System}\label{sec:rilc-module-sys}

\begin{align*}
	&\text{Modules }& 
		\comsc{\modc{}} \bnfdef
		&\
		\comsc{( {\chans} ; {\procs} ; {\intfs} ; {\expos})}
	\\
	&\text{Endpoint Types }& 
		\comsc{\tau}\bnfdef
		&\
		\comsc{\rdty{S}}
		\mid
		\comsc{\wrty{S}}
	\\
	&\text{Imports, Exports, Definitions }& 
		\comsc{\intfs}, \comsc{\expos}, \comsc{\chans} \bnfdef
		&\
		\come \mid \comsc{ I , {\chan{} : \tau} }
	\\
	&\text{Processes }& 
		\comsc{\procs} \bnfdef
		&\
		\come \mid \comsc{ {\procs} , {\com{p : \rdex{\rdchan{d}}{x.\exc{'}}}}}
\end{align*}

The key element of a module system is the definition of modules (\comsc{M}), which are a tuple consisting of defined channel names, process definitions, imports and exports.
First, any top-level process starts by reading from a channel, and thus we change the top-level definition of processes accordingly.
For well-typedness reasons, that top-level read channel, alongside any other channel used in the process of the body form the definitions \comsc{\chans} of a module.
Imports define the dependencies of the module: those channel names are not defined by the module and thus external modules (or the environment) will provide an implementation for them.
Exports define what a module supplies for external modules to use, essentially any channel name that is defined but not exported is local to the module.
Any channel end \comsc{c} that appears in definitions, imports or exports carries its endpoint type, meaning, it describes what kind of channel end is and how it is supposed to be used.

In the following, for simplicity, with a slight abuse of notation, we assume that these types mention sessions, i.e., the full sequence of the type of messages that are to be exchanged on the channel.
Note that it would be possible to simply use a different channel after each message is communicated, but this would make many an example hard to follow.
We leave a full formalisation of session types for \rilc for future work.

The operation that happens on modules is linking, denoted with $\comsc{\modc{_1}\linksymbol\modc{_2}}$ (\Cref{tr:rrilc-linking}).
\begin{center}
	\typerule{\rilc-Linking}{
		\comsc{\modc{_1}} = \comsc{{\chans_1} ; {\procs_1} ; {\intfs_1} ; {\expos_1}}
		&
		\comsc{\modc{_2}} = \comsc{{\chans_2} ; {\procs_2} ; {\intfs_2} ; {\expos_2}}
		\\
		\comsc{\intfs} = \comsc{\intfs_1}\setminus\comsc{\expos_2} \cup \comsc{\intfs_2}\setminus\comsc{\expos_1}
		&
		\comsc{\expos} = \comsc{\expos_1}\setminus\comsc{\intfs_2} \cup \comsc{\expos_2}\setminus\comsc{\intfs_1}
		&
		\comsc{\chans} = \comsc{\chans_1}\cup\comsc{\chans_2}
		&
		\comsc{\procs} = \comsc{\procs_1}\cup\comsc{\procs_2}
	}{
		\comsc{\modc{_1}\linksymbol\modc{_2}} = \comsc{\chans ; \procs ; \intfs ; \expos}
	}{rrilc-linking}
\end{center}
Module linking does what one expects, it generates another module whose processes and definitions are the union of the processes and definitions of its sub-modules.
Then, the resulting module has all the imports and exports of its sub-modules, minus those that are fulfilled by the other module.
That is, the resulting module has all imports of the two sub-modules, minus their exports, and all the exports of the two sub-modules, minus their imports.
When performing the set difference (as in $\comsc{\intfs_1}\setminus\comsc{\expos_2}$), we assume that subtracting a write (resp. read) end for a channel remove the respective read (resp. write) end from the set, i.e., removing \comsc{\wrchan{d}} from \comsc{\{\rdchan{d}, \rdchan{e}\}} results in \comsc{\{\rdchan{e}\}}.

\textbf{Note on Notation:}
From the type of linking ($\linksymbol : \modc{}\times\modc{}\to\modc{}$), it is evident that the top-level notion of partial programs that is of interest is indeed modules.
In the previous part of the paper, we used metavariable \prgc{} to range over such partial programs.
In the case of \rilc specifically, we use \modc{} to indicate what was previously denoted with \prgc{}.
We do not rebind the metavariable \prgc{} for \rilc, which can be then thought as an alternative to \modc{}.

\paragraph{Typing of Modules}
Modules are typed according to these new judgements.
\begin{align*}
	\comsc{\vdash \modc{}}
		&&&
		\text{ Well-typed module}
	\\
	\comsc{ {\chans} , {\intfs}  \vdash \procs{} }
		&&&
		\text{ Well-typed process with its defined channels and imports}
\end{align*}

The typing of modules is straightforward (\Cref{tr:rrilc-e-wt-module}).
A module is well-typed if its processes are, according to the process typing (\Cref{tr:rrilc-e-wt-proc}), which in turns relies on the expression typing of \Cref{sec:rilc-typ}.
Note that module and process typing do not tamper with the write token, nor with the affinity of resources, and thus their addition easily preserve the properties of \ILC.

\begin{center}
	\typerule{\rilc-Type-Module}{
		\comsc{\modc{}} = \comsc{( {\chans} ; {\procs} ; {\intfs} ; {\expos} )}
		&
		\comsc{\procs} = \comsc{ {\procs{_1}}, {{\ldots}, {\procs{_n}}}}
		&
		\underset{i\in1..n}{\bigcap} \comsc{\Pi_i.p} = \come
		\\
		\comsc{\chans{}} = \underset{i\in1..n}{\bigcup} \comsc{\chans{_i}}
		&
		\comsc{\intfs{}} = \underset{i\in1..n}{\bigcup} \comsc{\intfs{_i}}
		&
		\comsc{\expos{}} \subseteq \comsc{\chans{}}
		&
		\forall i \in 1..n\ldotp  
		\comsc{\chans{}} ; \comsc{\intfs{}}  \vdash \comsc{\procs{_i} }
	}{
		\comsc{\vdash \modc{}}
	}{rrilc-e-wt-module}	
	\typerule{\rilc-Type-Process}{
		\comsc{\procs{}} = \comsc{p : \rdex{\rdchan{d}}{x.\exc{'}}}
		&
		\comsc{ \rdchan{d}:\rdty{S}{}} \in \comsc{ {\chans{}} }
		\\
		\comsc{ {\chans{}};\come;\come\vdash \rdchan{d} : \rdty{S} }
		&
		\comsc{ {\chans{}}, {\intfs{}};\come;\come \vdash {\rdex{\rdchan{d}}{x.\exc{}}} : U }
	}{
		\comsc{\chans{}} ; \comsc{\intfs{}}  \vdash \comsc{\procs{} }
	}{rrilc-e-wt-proc}
\end{center}
With a slight abuse of notation we write \chans\ and \intfs\ to mean their channel and types that are used to create the $\Psi$ used for typing expressions in the premise of \Cref{tr:rrilc-e-wt-proc}.

\subsubsection{A Reactive Trace Semantics}\label{sec:rilc-traces}
The reactive trace semantics that makes \rilc a reactive language relies on two key notions: reactive programs and traces.
A reactive program contains some internal state, which is not directly observable from an external observer and reacts to a sequence of inputs from an environment by producing a sequence of observable outputs. 
After each input, the program may update its internal state, allowing all past inputs to influence an output.
By definition, a reactive program is really a component of a larger system that provides it inputs.
In \rilc, the reactive program is a module (and we will use the two words interchangeably in this section) and the larger system is an unspecified black-box that represents the classical \UC environment.
The input and output sequences of messages exchanged between the reactive program and the environment are what constitute an interaction trace (from \Cref{sec:bg}).

\begin{align*}
	&\text{Reactive Configurations }& \confc{} \bnfdef
		&\
		\comsc{\pair{\eta ; K ; \Xi ; C }}
	\\
	&\text{ Token Tag }& \comsc{\eta} \bnfdef
		&\
		\comsc{\wrtk} \mid \comsc{\bot}
	\\
	&\text{ External Names }& \comsc{\Xi} \bnfdef
		&\
		\come \mid \comsc{{\Xi},  \chan{}:S }
\end{align*}
From an operational semantics perspective, dealing with reactive programs changes the notion of configuration, which now becomes a reactive configuration \confc{}.
A reactive configuration keeps track of the write token in the token tag \comsc{\eta}, in order to know whether the reactive program (\comsc{\bot}) or the environment (\comsc{\wrtk}) has the right to write.
Then, it also keeps track of the cryptographic store \comsc{K}, as presented in \Cref{sec:rilc-crypto}.
The reactive configuration then contains a list of external names \comsc{\Xi}, that is, a list of those names that the reactive program relies on and that the environment is supposed to handle the other end of.
Finally, the reactive configuration contains the elements of an \ILC configuration from \Cref{sec:ilc-sem}, namely the existing names \comsc{\Sigma} and the process pool \comsc{\Pi}.

\begin{align*}
	&\text{ Actions }& \comsc{\ac} \bnfdef
		&\
		\comsc{\snd{v}{\chan{}}}
		\mid
		\comsc{\sndb{v}{\chan{}}}
		\mid
		\comsc{\noact}
	&
	&\text{ Interaction Traces }& \comsc{\intrace} \bnfdef
		&\
		\come \mid \comsc{\stackht{\ac}{\intrace}}
	\\
	&\text{ Probabilities }& \comsc{\rho} \in
		&\
		0..1
	&
	&\text{ Traces }& \comsc{\trace } \bnfdef
		&\
		(\rho, \intrace)
\end{align*}
Traces \comsc{\trace} are pairs of a probability \comsc{\rho} and an interaction trace \comsc{\intrace}.
An interaction trace is a sequence of actions \comsc{\ac}.
Actions include: the environment sending a message \comsc{v} on channel \comsc{\chan{}} to the reactive program (\comsc{\snd{v}{\chan{}}}), the reactive program sending a message \comsc{v} on channel \comsc{\chan{}} to the environment (\comsc{\sndb{v}{\chan{}}}), %
and a silent action \comsc{\noact}.
The actions use \comsc{?} and \comsc{!} decoration borrowed from the process calculi literature~\cite{sangioPiBook} to indicate the ``direction'' of the message.
Note that the semantics will insist that interaction traces start with a \comsc{?} action and then alternate between \comsc{!} and \comsc{?} ones.

\paragraph{Operational Reactive Trace Semantics}
The semantics relies on the following judgements, where we recap all previous judgements too, for the sake of completeness.
\begin{align*}
	\comsc{\confc{}\equiv\confc{'}}
		&&&
		\text{ Configurations are equivalent, as in \Cref{sec:ilc-sem} }
	\\
	\comsc{\chan{} \leadsto \chan{'}}
		&&&
		\text{ \chan{} writes where \chan{'} reads, as in \Cref{sec:ilc-sem} }
	\\
	\comsc{\exc{} \redp \exc{'}}
		&&&
		\text{ Expression reduction, as in \Cref{sec:ilc-sem} }
	\\
	\comsc{C\myred C'}
		&&&
		\text{ Configuration takes a step, as in \Cref{sec:ilc-sem}  }
	\\
	\comsc{\pair{K,C}\myred \pair{K,C'}}
		&&&
		\text{ Crypto configuration takes a step, as in \Cref{sec:rilc-crypto}  }
	\\
	\comsc{\confc{}\myredx{\ac} \confc{'}}
		&&&
		\text{ Reactive configuration takes a step with action \comsc{\ac} }
	\\
	\comsc{\confc{} \Xto{\intrace} \confc{'}}
		&&&
		\text{ Reactive configuration big-steps with interaction trace \intrace }
\end{align*}

Many of the judgements of the semantics we rely on are untouched from previous definitions.
Below are those rules of the semantics that deal with the robustness and the traces.
\begin{center}
	\typerule{\rrilc-Sem-Write}{
		\comsc{\chan{}}\in\comsc{\Xi}
		&
		\comsc{K} = \comsc{\pair{H ; G ; \secpam ; \nrnd}}
		&
		\comsc{K'} = \comsc{\pair{{H},v ; G ; \secpam ; \nrnd}}
	}{
		\comsc{
			\pair{\bot ; K ; \Xi ; \Sigma ; \Pi, p:\evalctx\holev{\wrex{v}{\chan{}}} } \xto{\sndb{v}{\chan{}}} \pair{\wrtk ; K' ; \Xi ; \Sigma ; \Pi, p:\evalctx\holev{\unit} }
		}
	}{rrilc-e-write}
	\typerule{\rrilc-Sem-Read}{
		\comsc{\chan{}}\in\comsc{\Xi}
		&
		\comsc{\envredkk{K}{K'}{v}}
	}{
		\comsc{
			\pair{\wrtk ; K ; \Xi ; \Sigma ; \Pi, p:\evalctx\holev{\rdex{\chan{}}{x.\exc{}}} } \xto{\snd{v}{\chan{}}} \pair{\bot ; K' ; \Xi ; \Sigma ; \Pi, p:\evalctx\holev{\exc{}\subst{\pair{v,\chan{}}}{x}} }
		}
	}{rrilc-e-read}
	\typerule{\rrilc-Sem-choice}{
		\comsc{\chan{_i}}\in\comsc{\Xi}
		&
		\comsc{\envredkk{K}{K'}{v}}
	}{
		\begin{multlined}
			\comsc{
				\pair{\wrtk ; K ; \Xi ; \Sigma;\Pi,p:\evalctx\holev{\chanex{\chan{_1}}{x_1.\exc{_1}}{\chan{_2}}{x_2.\exc{_2}}}} \myred 
			}
			\\
			\comsc{
				\pair{\bot ; K' ; \Xi ; \Sigma;\Pi,p:\evalctx\holev{\exc{_i}\subst{\pairlb{1}{!v,\chan{_1},\chan{_2}}}{x_i}} }
			}
		\end{multlined}
	}{rrilc-e-ch}
	\typerule{\rrilc-Sem-Other}{
		\comsc{\pair{ K; \Sigma ; \Pi} \myred \pair{K';\Sigma ; \Pi' }}
	}{
		\comsc{
			\pair{\bot ; K ; \Xi ; \Sigma ; \Pi } \xto{\noact} \pair{\bot ; K' ; \Xi ; \Sigma ; \Pi'}
		}
	}{rrilc-e-rest}
	\typerule{\rrilc-Sem-Error}{
	}{
		\comsc{
			\pair{\bot ; K ; \Xi ; \Sigma ; \Pi, p:\evalctx\holev{ error } } \xto{\errt} \pair{\wrtk ; K ; \Xi ; \Sigma ; \come }
		}
	}{rrilc-e-error}
\end{center}
\Cref{tr:rrilc-e-write} states that when the reactive program has the write token (\comsc{\eta}=\comsc{\bot}), and it is writing on a channel \comsc{\chan{}} that belongs to the environment ($\comsc{\chan{}}\in\comsc{\Xi}$), then the environment can read the value that is being written, and that value gets added to the attacker knowledge \comsc{K'}.
This is the way leakage (as left hanging from \Cref{sec:rilc-crypto}) is generated: values that are written into channels that belong to external names are effectively leaked to the environment (that is, to the attacker).

Dually, \Cref{tr:rrilc-e-read} states that when the environment has the write token (\comsc{\eta} = \comsc{\wrtk}), and the reactive program is reading on a channel \comsc{\chan{}} that belongs to the environment, then the environment can send a value \comsc{v} that the program reads.
The value is generated according to what the environment can compute according to the environment cryptographic reductions of \Cref{sec:rilc-crypto} (\comsc{\envredkk{K}{K'}{v}}).

\Cref{tr:rrilc-e-rest} relies on the previously-defined semantics to execute a reduction within the reactive program, note that this rule can only be triggered when the program has the write token.

Finally, \Cref{tr:rrilc-e-error} halts the computation whenever the \comsc{error} expression is executed.

\begin{center}
	\typerule{Trace-Refl}{}{
		\com{ \confc{} \Xto{\noact} \confc{}}
	}{rrilc-t-refl}
	\typerule{Trace-Action}{
		\com{\confc{}\xto{\ac} \confc{''}}
		&
		\com{\confc{''} \Xto{\intrace} \confc{'}}
	}{
		\com{ \confc{} \Xto{\stackht{\ac}{\intrace}} \confc{'}}
	}{rrilc-t-act}
	\typerule{Trace-No-Action}{
		\com{\confc{}\xto{\noact} \confc{''}}
		&
		\com{\confc{''} \Xto{\intrace} \confc{'}}
	}{
		\com{ \confc{} \Xto{\intrace} \confc{'}}
	}{rrilc-t-no-act}
\end{center}

\paragraph{Reactive Program Traces}

The following judgements define when does a reactive program generate a (set of) traces.
\begin{align*}
	\comsc{\wprgc{};\secpam;\nrnd \sem \{ \intrace \}}
		&&&
		\text{ Reactive program \comsc{\wprgc{}} produces a set of interactive traces \{ \comsc{\intrace} \} }
	\\
		&&&
		\text{ with security parameter \comsc{\secpam} and randomness \comsc{\nrnd}}
	\\
	\wprgc{};\secpam;\nrnd \semfat \{ \trace \}
		&&&
		\text{ Reactive program \comsc{\wprgc{}} produces a set of traces \{ \comsc{\trace} \} }
	\\
		&&&
		\text{ with security parameter \comsc{\secpam} and randomness \comsc{\nrnd}}
	\\
	\comsc{\behavc{\wprgc{},\secpam}} = \{\trace\}
		&&&
		\text{ Behaviours (i.e., set of traces \trace) of a reactive program \comsc{\wprgc{}}}
	\\
		&&&
		\text{ with a security parameter \comsc{\secpam} } 
\end{align*}

Before we explain each judgement and their rules, we describe some auxiliary functions.
\begin{align*}
	\comsc{\totrand} = 
		&\
		\myset{ \comsc{n} }{ 
			\begin{aligned}
				&
				\comsc{n}:\myarray{\Bits} \text{ and } \mylength{\myarray{\Bits}} = f(\secpam) \text{ and } 
				f:\mb{N}\to\mb{N} \text{ and } \vdash \fun{poly}{f}  
			\end{aligned}
		}
\end{align*}
\comsc{\totrand} generates all the random bitstrings $\{\comsc{n}\}$ that a reactive program can use, and the length of these bitstrings is a polynomial function ($\vdash\fun{poly}{f}$) of the security parameter \comsc{\secpam}.
In the following, we use \mylength{\cdot} to indicate the length of a list, or the cardinality of a set.

\begin{center}
	\typerule{\rilc-Initial State}{
		\comsc{\wprgc{}} = \comsc{{\chans} ; {\procs} ; {\intfs} ; {\expos}}
		&
		\comsc{\vdash\wprgc{}}
		&
		\comsc{\Xi} = \comsc{\intfs}\cup\comsc{\expos}
		&
		\comsc{\Sigma} = \comsc{\dom{\Pi}}\cup\comsc{\dom{\Xi}}
	}{
		\comsc{\SInit{\wprgc{}; \secpam; \nrnd }} = \comsc{\pair{\wrtk ; K_0(\secpam,\nrnd) ; \Xi ; \Sigma ; \Pi}}
	}{rrilc-init}
	\typerule{\rilc-Terminal State}{
		\nexists \confc{'},\ac \ldotp \comsc{\confc{}\myredx{\ac} \confc{'}}
	}{
		\isterm{\confc{}}
	}{rrilc-term}
\end{center}
Given a complete (as per the definition of ``complete'' given in \Cref{sec:rc}) reactive program \comsc{\wprgc{}} (and the security parameter, and a randomness bitstring \comsc{\nrnd}), \Cref{tr:rrilc-init} defines the starting state for that program.
Dually, \Cref{tr:rrilc-term} tells that a configuration is terminal, i.e., it cannot step any more.

\begin{center}
	\typerule{\rilc-Interaction Traces}{}{
		\wprgc{};\secpam;\nrnd \sem
			\myset{ \intrace }{ \exists \confc{}\ldotp \SInit{\wprgc{};\secpam;\nrnd}\Xto{\intrace}\confc{} \text{ and } \isterm{\confc{}} }
	}{rrilc-interactiontrace}
\end{center}
\Cref{tr:rrilc-interactiontrace} calculates the set of interaction traces that a complete reactive program generates starting from a starting state until a terminal state.

\begin{center}
	\typerule{\rilc-Interaction Traces}{}{
		\wprgc{};\secpam;\nrnd \semfat
			\myset{ (\intrace,\rho) }{ 
				\begin{aligned}
					&
					\wprgc{};\secpam;\nrnd \sem \{\intrace\} \text{ and } \intrace\in \{\intrace\} \text{ and } 
					\rho = 1/(\mylength{\{\intrace\}} \times \mylength{\totrand})
				\end{aligned} 
			}
	}{rrilc-traces}
\end{center}
\Cref{tr:rrilc-traces} calculates the set of traces of a complete reactive program.
The probability of each trace is calculated as the probability of obtaining the trace over all possible traces (and all randomnesses) that exist.

\begin{center}
	\typerule{\rilc-Interaction Traces}{}{
	\behavc{\wprgc{},\secpam} = 
		\myset{ (\intrace, \rho) }{ 
			\wprgc{};\secpam;\nrnd \semfat (\intrace,\rho_0)
			\text{ and }
			\rho = \sum_{\rho_0} 
			\text{ and } \nrnd \in \totrand 
		}
	}{rrilc-beh}
\end{center}
\Cref{tr:rrilc-beh} calculates the behaviour of a complete reactive program.
The probability \comsc{\rho} of each trace is calculated as the sum of all the probabilities \comsc{\rho_0} to generate that trace calculated with any possible randomness \comsc{\nrnd} taken from the space of randomnesses \comsc{\totrand}.

\subsection{Properties of \rilc}\label{sec:rilc-props}
\rilc enjoys a number of properties, first that its additions from \Cref{sec:reactive} still preserve well-typedness and confluence from the original \ILC work (\Cref{sec:rilc-lang-props}).
Then, \rilc fulfils the axioms of \Cref{sec:composition}, so it can be used in our connection and it enjoys the composition theorem and the dummy adversary one (\Cref{sec:rilc-axioms}).

\subsubsection{Language Properties}\label{sec:rilc-lang-props}

In order to state confluence for \rilc we define two projection functions on traces that extract the inputs ($\inps{\cdot}$) and outputs ($\outs{\cdot}$) actions of a trace:
\begin{center}
	\typerule{Inputs - empty}{}{
		\inps{\come} = \noact
	}{aux-inps-e}
	\typerule{Inputs - ?}{
		\ac = \snd{v}{c}
	}{
		\inps{\stackht{\ac}{\intrace}} = \stackht{\ac}{\inps{\intrace}}
	}{aux-inps-i}
	\typerule{Inputs - !}{
		\ac = \sndb{v}{c}
	}{
		\inps{\stackht{\ac}{\intrace}} = \inps{\intrace}
	}{aux-inps-o}

	\typerule{Optputs - empty}{}{
		\outs{\come} = \noact
	}{aux-outs-e}
	\typerule{Outputs - ?}{
		\ac = \snd{v}{c}
	}{
		\outs{\stackht{\ac}{\intrace}} = \outs{\intrace}
	}{aux-outs-i}
	\typerule{Outputs - !}{
		\ac = \sndb{v}{c}
	}{
		\outs{\stackht{\ac}{\intrace}} = \stackht{\ac}{\outs{\intrace}}
	}{aux-outs-o}
\end{center}
Additionally, we formalise the fact that during two executions the environment changes the cryptographic state in the same manner.
We say that two such executions are consistent wrt cryptographic inputs, denote this fact as \cic{\cdot}{\cdot} and formalise it as follows:
\begin{center}
	\typerule{Cryptographic input Consistency - Base}{}{
		\cic{\confc{_1}\Xto{\epsilon}\confc{_1}}{\confc{_2}\Xto{\epsilon}\confc{_2}}
	}{kic-base}
	\typerule{Cryptographic input Consistency - Inductive}{
		\cic{\confc{_1''}\Xto{\trace}\confc{_1'}}{\confc{_2''}\Xto{\trace}\confc{_2'}}
		&
		\cic{\confc{_1}\xto{\ac}\confc{_1'}}{\confc{_2}\Xto{\ac}\confc{_2''}}
	}{
		\cic{\confc{_1}\xto{\ac}\confc{_1''}\Xto{\trace}\confc{_1'}}{\confc{_2}\xto{\ac}\confc{_2''}\Xto{\trace}\confc{_2'}}
	}{kic-ind}
	\typerule{Cryptographic input Consistency - Single-nop}{
		\comsc{\ac} = \comsc{\epsilon} \vee \comsc{\ac} = \comsc{\ac!}
	}{
		\cic{\confc{_1}\xto{\ac}\confc{_1'}}{\confc{_2}\xto{\ac}\confc{_2'}}
	}{kic-sing-no}
	\typerule{Cryptographic input Consistency - Single-in}{
		\comsc{\confc{_1'}.K} = \comsc{\confc{_2'}.K}
	}{
		\cic{\confc{_1}\xto{\ac?}\confc{_1'}}{\confc{_2}\xto{\ac?}\confc{_2'}}
	}{kic-sing-in}
\end{center}

\rilc has a full confluence property (\Cref{thm:f-conf}) telling that if a program state \confc{} performs two different traces up to two terminal states \confc{_1} and \confc{_2} with the same environment-supplied inputs, and with the same changes to the cryptographic state from the environment, then the terminal states are renamings of each other (with renaming function $g(\cdot)$), and the outputs of the traces are also the same.

\begin{theorem}[Full Confluence]\label{thm:f-conf}
	\begin{align*}
		\pif
			&
			\confc{} \Xto{\trace} \confc{_1}
		\pand
			\confc{} \Xto{\trace'} \confc{_2}
		\pand
			\isterm{ \confc{_1} }
		\pand
			\isterm{ \confc{_2} }
		\\
		\pand
			&
			\inps{\trace}=\inps{\trace'}
		\pand
			\cic{\confc{} \Xto{\trace} \confc{_1}}{\confc{} \Xto{\trace'} \confc{_2}}
		\\
		\pthen
			&
			\confc{_1} = g(\confc{_2})
		\pand
			\isterm{ g(\confc{_2}) }
		\pand 
			\outs{\trace} = \outs{\trace'}
	\end{align*}
\end{theorem}

The proof of \Cref{thm:f-conf} is an unsurprising induction over the generated traces, with the silent actions case being inherited from \citet{ilc}.
The only interesting case is the single visible action lemma (\Cref{thm:conf-sing}).
\begin{lemma}[Generalised Confluence Single]\label{thm:conf-sing}
	\begin{align*}
		\pif
			&
			\confc{} \xto{\ac} \confc{_1}
		\pand
			\comsc{f(\confc{})} \xto{\ac'} \confc{_2}
		\pand
			\inps{\ac}=\inps{\ac'}
		\pand
			\comsc{\confc{_1}.K}  = \comsc{\confc{_2}.K}
		\\
		\pthen
			&
			{\confc{_1}} = \comsc{g(\confc{_2})}
		\pand
			\outs{\ac}=\outs{\ac'}
	\end{align*}
\end{lemma}
In the proof of \Cref{thm:conf-sing}, we have to reason about the rules generating the single actions, which are the key additions to \rilc.
Thus, this proof essentially shows that our additions to \rilc still keep the language confluent, assuming that the environment provides the same inputs and the same changes to the cryptographic state.

\subsubsection{Fulfilling the Axioms of \Cref{sec:composition}}\label{sec:rilc-axioms}

The Axioms of \Cref{sec:composition} are split in two groups, those needed for the composition theorem and those needed for the dummy adversary one.

\paragraph{Axioms for Composition}
In order to state the Axioms of \Cref{sec:compos-for-rc} in \rilc, we need to instantiate the different abstract composition operators ($\linksymbol, \linkprog, \linkatk \linkwhole$) in \rilc.
Since \rilc has a single composition operator ($\linkc$ from \Cref{tr:rrilc-linking}), we instantiate all the operators with $ \linkc $.
We believe in other languages where the distinction between attacker and program needs to be carried around (e.g., as in the work of \citet{akram}), we could not instantiate all the operators with a single one, but we would have to define different ones.

The key property that we rely on for proving the axioms is linking commutativity (\Cref{thm:link-comm}), whose proof is straightforward and thus left in the technical report~\cite{ucissc-prisc}.
\begin{lemma}[Linking Commutativity]\label{thm:link-comm}
		${\ctxc{}\linkc\modc{}} \relateAbs {\modc{}\linkc\ctxc{}}$
\end{lemma}

Below are the statements of the axioms instantiated for \rilc, as their proofs are tedious and not insightful, we leave them in the technical report~\cite{ucissc-prisc}.
First we present the ones related to composition:
\begin{description}
	\item[\Cref{ax:par-decomp}] $\forall\ctxc{_0}, \modc{}\ldotp\exists\ctxc{_1},\ctxc{_2}\ldotp {\ctxc{_0}\linksymbol\modc{}} \relateAbs {(\ctxc{_1}\linksymbol\ctxc{_2})\linksymbol\modc{}}$
	\item[\Cref{ax:inter-comp}] ${(\ctxc{_1}\linksymbol\ctxc{_2})\linksymbol(\modc{1}\linksymbol\modc{2})} \relateAbs {(\ctxc{_1}\linksymbol\modc{1})\linksymbol(\ctxc{_2}\linksymbol\modc)}$
	\item[\Cref{ax:const-elim}] $\pif \modc{1}\relateAbs\modc{2} \pthen \modc{3} \linkc \modc{1} \relateAbs \modc{3} \linkc \modc{2}$
\end{description}
From these Axioms, we can derive \Cref{thm:compos-rilc} below, which is \Thmref{thm:composition} specified for \rilc:
\begin{corollary}[Composition for \rilc]\label{thm:compos-rilc}
	\[
     \pif
     \forall \ctxc{}.\exists\ctxc{'}\ldotp
				\comsc{\ctxc{}\hole{\comp{\prgc{}}}}
                                \relateAbs
				\comsc{\ctxc{'}\hole{\prgc{}}}
    \pthen
    \forall \ctxc{''}.\exists \comsc{S}\ldotp
    \comsc{\ctxc{''}\hole{\prgc{s}\linksymbol\comp{\prgc{}}}}
                                \relateAbs
				\comsc{S\hole{\prgc{s}\linksymbol\prgc{}}}
	\]
\end{corollary}

\paragraph{Axioms for the Dummy Attacker}
The Axioms related to the dummy attacker (\Cref{sec:dummy-for-rc}) cannot possibly be stated (let alone proved) in a general fashion.
In fact, different modules (which model different protocols) will provide different signatures, and thus there is not really a single dummy attacker that fits all protocols.%
\footnote{
	We note that there may be some simplifying assumptions that fix the channels in a signature (e.g., one channel per protocol with a unit that is responsible to dispatch the messages to the protocol parties), but that is beyond the scope of this paper.
}
However, we notice that once the signature of a module is fixed, and the meaning of each imported and exported channels is known, it is possible to devise a dummy attacker for that protocol.
Even more, we conjecture that for every protocol it is possible to devise a dummy attacker (though we leave stating and proving such a conjecture for future work).
Thus, we do not provide a general proof of \Cref{ax:dummy-prog} and \Cref{ax:dummy-att}, but leave them to be proved on a per-protocol basis.

Specifically, we will report on the proofs of those Axioms for the specific protocol we consider in this paper in \Cref{sec:comm-static-proof}.

\medskip

With our candidate formal language for instantiating the connection, we now turn to reaping its benefits and show how to derive \UC proofs as \RC ones.

\section{Rigorous, Scalable \UC Proofs as \RC Proofs}\label{sec:sec-comp-proofs}
This section first provides some background on \UC and \RC proofs and it clarifies how to use our connection to provide \UC proofs via \RC ones (\Cref{sec:bg-proofs}).
Then it provides two instances of \UC proofs done via \RC ones.
The first one is for a known result: \UC of a single-bit commitment in the static corruption setting~\cite{uc-comm,ilc} (\Cref{sec:commitment-static}).
The second one is instead for a simple but (to our knowledge) novel result: \UC of the same protocol, but with adaptive corruptions (\Cref{sec:commitment-adaptive}).

\subsection{Background on \UC and \RC Proofs}\label{sec:bg-proofs}
We now recount \UC (\Cref{sec:howto-uc-proofs}) and \RC proofs (\Cref{sec:howto-rc-proofs}), before drawing conclusions on the benefits to reap from each approach (\Cref{sec:betterproofs}).

\subsubsection{\UC Proofs}\label{sec:howto-uc-proofs}

As mentioned earlier, the vast majority of \UC proofs exploit the dummy-adversary theorem (\Cref{thm:dummy-uc}) and thus consists of two steps. 
First, authors establish the existence of the simulator by stating it explicitly. 
There is usually little explanation about why it is defined as it is, since the simulator is the result of a fair amount of trial and error.
Second, the authors prove the indistinguishability property at the heart of \UC:
$ \Execfun{\env{},\pattdummy{},\prot{}}
        \indistp
        \Execfun{\env{},\simu{},\idfu{}}
$. 
This is usually an induction over the length of the trace from the perspective of the environment, or more precisely, over the number of outgoing messages from the environment, which mark the beginning of what is often called an \emph{epoch}. 
The epoch ends when the environment regains control by receiving a message from the attacker or protocol. 
For each epoch, these proofs show (via case distinction over the message) that some state invariant holds; that will ensure that the message the environment receives at the end of the epoch is indistinguishable in the ideal and the real world.
Consequently, the state invariant links the state of the parties in the ideal world ($\simu{},\idfu{}$) and in the real world ($\pattdummy{},\prot{}$) to each other. 
A recurring pattern in \UC proofs is that the simulator $\simu{}$ tracks a large part of the  state of the protocol $\prot{}$ to correctly simulate behaviour that is independent from $\idfu{}$. 

The main work of the proof is hence in the aforementioned case distinction: no matter what the environment sends to the protocol/ideal functionality, the state invariant holds. 
The majority of these cases is about the simulator correctly tracking the protocol state.
For one, this is tedious:  state invariants can become very large, since interactions in the ideal world may involve multiple messages between $\simu{}$ and $\idfu{}$. 
For two, and somewhat tragically, this is mechanical but not easily mechanizable (as admitted by the authors of one mechanised \UC proof~\cite{canettiEasyUCUsingEasyCrypt2019}).
The mathematician is essentially performing a symbolic execution in a deterministic system, which would be automatable, if the language used to describe protocol, simulator and functionality were to have a formal semantics. 
Some proofs omit these tedious steps entirely and concentrate on the few cases where a cryptographic argument is necessary, establishing a reduction of this form: if, given the state invariant, the environment can compute a message, then the same environment can be transformed into an attacker for some cryptographic game.

Summarizing, there is a lot of potential for automation in these proofs, would they be conducted in a formal language. 
We believe some of the low-hanging fruits here are (i) the symbolic execution to prove state equivalence in the majority of cases and (ii) the automated synthesis of the state equivalence relation. 
A more challenging tasks is the partial synthesis of the simulator. 
Given a partial simulator and a partial state equivalence that fixes the cryptographically interesting parts of the interaction, provide a simulator and state equivalence relation that produces equivalent traces for the interactions that are not defined by the partial simulator.

While the informal description above captures the majority of \UC proofs, we note that some follow a more structured approach. 
The game-playing technique~\cite{bellareGamePlayingTechnique2004} consists of structuring cryptographic proofs as a series of small syntactic refinement steps in an informal pseudocode language. 
The technique is also sometimes used in \UC proofs (e.g.,~\cite{abdallaUniversallyComposableRelaxed2020, canettiUniversallyComposableEndtoEnd}) and it has been mechanized in EasyUC but, quoting \citet{canettiEasyUCUsingEasyCrypt2019}: 
	``despite the relative simplicity [of their case studies, proving UC] took an immense amount of work``~\cite{canettiEasyUCUsingEasyCrypt2019}. 
The authors point out the need for a domain specific language that is coroutine-based, as opposed to the underlying EasyCrypt~\cite{pwhile} language they used, as well as the lack of symbolic evaluation.
Moreover, proving composition inside EasyCrypt was only possible for specific instances, but not in general. 
The authors state that a general proof must be performed at the level of EasyCrypt's metatheory, i.e., without automation. 
Our composition axiom can help this task.

\subsubsection{\RC Proofs}\label{sec:howto-rc-proofs}
Most \RC proofs follow a proof technique inherited from secure compilation works~\cite{scsurvey} called \emph{Backtranslation}~\cite{rsc,rhc,catalinRSC,journey-rel,exorcising}.
The goal of a backtranslation is to define how to construct any source attacker (the existentially-quantified \ctxs{} in \Cref{def:rhp}) starting from whatever target information available (i.e., what preceeds said existentially-quantified \ctxs{}).
If the used information is the target attacker \ctxt{}, then the proof is called \emph{Context-based Backtranslation}, if the used information is the target trace, then it is called \emph{Trace-based Backtranslation}.

Once the backtranslation is defined, the \RC proofs essentially proceed by induction over the target trace that needs to be replicated in the source.
Most times those traces follow this inductive principle: they are either empty or the concatenation of a trace with a list of attacker-generated actions ending with a control-transferring action to the program (\comsc{\intrace\cdot \ac?}), followed by a sequence of program-generated actions ending with a control-transferring action to the context (\comsc{\intrace'\cdot \ac!}).

The proof sets up a simulation argument between the target context performing \comsc{\intrace\cdot \ac?} and its backtranslated counterpart, which needs to be proven to produce \comsc{\intrace\cdot \ac?} in the source.
Following estblished compilation-style proofs~\cite{Leroy09b}, this is done by setting up a cross-language state relation ($\src{\Omega}\sim\trg{\bm{\Omega}}$) between the states of the source attacker (\src{\Omega}) and those of its backtranslation-generated counterpart (\trg{\bm{\Omega}}).
Then, after the ?-decorated action, control is transferred to the compiled program.
The proof then proceeds by a classical compiler correctness argument which ensures that the program performs \comsc{\intrace'\cdot \ac!} given that its compilation perform \comsc{\intrace'\cdot \ac!}.
This proof is carried out with a different cross-language state relation relating states of the source program and its compilation ($\src{\Omega}\approx\trg{\bm\Omega}$).
Notably, the relation when control is in the attacker code ($\sim$) tends to enforce a stronger invariant than the other one ($\approx$), and thus there is need to weaken and strengthen the relation.
The weakening and strengthening often happens when the compiler performs some security-related checks that are placed as a wrapper around the program.
It is in fact common for secure compilers to act as ``security wrappers'' around correctly-compiled code~\cite{mfac,popl-journal,nonintfree,sjoin}.

\RC works that use the backtranslation proof technique typically do not use reactive languages, and thus they cannot write the analogous of the dummy attacker theorem.
Thus, backtranslation proofs require quite some effort, with research being active in simplifying said effort~\cite{akram}.
This effort is especially great when these proofs are mechanised, at the time of writing, the only such mechanised effort required 20K lines of Coq just for a part of the backtranslation correctness proof~\cite{akram22}.

\subsubsection{The Best of Both Worlds}\label{sec:betterproofs}
In the following, we will use our connection (\Cref{thm:uc-is-sc}) to provide rigorous \UC proofs, relying on the benefits of \RC proofs.
Specifically, we will consider a protocol and its ideal functionality, coded in \rilc.
By using the \rilc semantics, we will perform the reductions in the real world (for now manually) and compute the real-world traces.
Then, we will perform the ideal world reductions and show that the ideal traces they yield are the same (trace and probability-wise) as the real-world ones.
This will let us conclude that the compiler from the ideal functionality to the protocol is \rhpref, and thus the latter \UC-realises the former.

\subsection{\UC Commiments as \RC Proofs for the Static Corruption Model}\label{sec:commitment-static}
In order to perform this proof we need to define the following \rilc processes:
\begin{enumerate}
	\item the ideal functionality \src{F} (\Cref{sec:comm-static-idf});
	\item the protocol \trg{P}, \trg{Q} that \UC-realises \src{F} (\Cref{sec:comm-static-prot});
	\item the simulator \src{S} used in the proof (\Cref{sec:comm-static-sim}).
\end{enumerate}
Finally, the proof proper is presented in \Cref{sec:comm-static-proof}, followed by a discussion \Cref{sec:protocol-not-looking-nice}.
This proof rests on a few assumptions, which we inherit from the original \UC proofs from \citet{uc-comm}, and that we present below.

\paragraph{Assumptions for the Static Corruption Commitment}
Below are the assumptions that hold for the protocol of this section.
Note that these assumptions are not imposed by us, but they are inherited by the work that defined the protocol in the first place, we simply recap them here for simplicity.
\begin{enumerate}
	\item the protocol assumes authenticated channels, i.e., if a party \comsc{A} receives a message \comsc{m}, it knows the identity of the party \comsc{B} that sent \comsc{m}.
	\item the only party that can be corrupted is the committer (\trg{P}), since corrupting the receiver does not let the attacker break any of the commitment properties.
\end{enumerate}

\subsubsection{Ideal Functionality}\label{sec:comm-static-idf}
The ideal functionality is presented in \Cref{lis:comm-prot-static-if}, with comments explaining the code.
In all our code snippets when we write a channel \comsc{XY}, it is a channel used from process \comsc{X} to send a message read by process \comsc{Y}.
\begin{lstlisting}[language=SRC,label=lis:comm-prot-static-if,caption=Ideal Functionality for the single-bit commitment.]
process F =
	rd Commit b from Read(PF) 		// receive a commitment message for a bit 'b'
	wr Receipt to Write(FS) 			// acknowledge that the commitment has been initiated
	rd Open from Read(PF) 				// receive a message to open the commitment
	wr Opened b to Write(FS) 			// open the commitment to the right bit 'b'
\end{lstlisting}

An interesting bit of this code is its \emph{module signature}, which is in \Cref{lis:comm-prot-static-ideal-idx}.
\begin{lstlisting}[language=SRC,label=lis:comm-prot-static-ideal-idx,caption=Module signature for ideal functionality for the single-bit commitment.,escapechar=|]
imports If = Write(FS)
definitions Df = Read(PF), Write(FS)
exports Xf = Read(PF)
\end{lstlisting}
The reason we deem such signatures interesting is that it describes the channels that will be used.

\subsubsection{Protocol}\label{sec:comm-static-prot}
The protocol consists of two parties (a committer \trg{P} and a receiver \trg{Q}) whose code is in \Cref{lis:comm-prot-static-main} and of a common reference string whose code is in \Cref{lis:comm-prot-static-crs}.
\begin{lstlisting}[language=TRG,label=lis:comm-prot-static-main,caption=Single-bit commitment protocol.,escapechar=|]
process P =
	rd M from Read(ZP)										// receive the static corruption message		|\label{line:comm-prot-static-start}|
	if M == CrptNO then 									// in case no party is corrupted					|\label{line:comm-prot-static-resume}|
		wr StartCrs to Write(PCrs)					// start the CRS functionality |\label{line:comm-prot-static-prea-s1}|
		rd Started from Read(CrsP)					// ack the CRS started
		wr Ok to Write(PZ)									// return control to the environment,
																				//  which will call the CRS
		rd WaitCrs from Read(CrsP)					// synchronise with the environment
		wr Waited to Write (PCrs)						|\label{line:comm-prot-static-prea-e1}|
																				// here starts the commitment protocol, 
																				//  as taken from the literature
		rd Commit b from Read(ZP)						// receive a commitment message for a bit 'b'	|\label{line:comm-prot-static-literature}|
		wr GetCRS to Write(PCrs)						// request the CRS
		rd PublicStrings s pk0 pk1 from Read(CrsP)			// receive the CRS
		let r = takernd								
		let y = (if b == 0 then prg pk0 r else xors(prg(pk1, r), s))	// set y to
																				//  the commitment of bit 'b' using the 'prg'
		wr Commit' y to Write(PQ)						// send the commitment 'y' to the receiver
		rd Open from Read(ZP) 							// receive a message to open the commitment
		wr Open' b r to Write(PQ)						// forward that request to the receiver |\label{line:comm-prot-static-h-end}|
	else 																	// if the corruption message is not 'CrptNO', 
																				//  the committer is corrupted |\label{line:comm-prot-static-else}|
		wr StartCrs to Write(PCrs)					// these synchronisation messages are as above
		rd Started from Read(CrsP)
		wr Ok to Write(PZ)
		rd WaitCrs from Read(CrsP)
		wr Waited to Write (PCrs)
																				// below is the corrupted committer: a proxy 
																				//  from the environment to the receiver
		loop(2)(fwd(Read(ZP))(Write(PQ)))		// we expect 2 iterations, thus the bound |\label{line:comm-prot-static-c-end}|

process Q = 
	rd Commit' y from Read(PQ)			// receive the commitment 'y' from the committer
	wr GetCRS to Write(QCrs)				// request the CRS
	rd PublicStrings s pk0 pk1 from Read(CrsQ)		// receive the CRS
	wr Receipt to Write(QZ)					// acknowledge the commitment has been initiated
	rd Open' b r from Read(PQ)			// receive original commitment bit 'b' 
																	//  and randomness 'r'
																	// below: calculate if the communication 
																	//  of 'b', 'r' and 'y' was done properly
	if (b == 0 && y == prg(pk0, r)) \/ (b == 1 && y == xors(prg(pk1, r), s))	|\label{line:comm-prot-static-if-q}|
		wr Opened b to Write(QZ)
	else error 											// some parameter does not match: abort
\end{lstlisting}
In the static corruption model, corruption is defined at the beginning of the computation with a message from the environment.
That is what happens on \Cref{line:comm-prot-static-start} and that is the reason why the code of the committer \trg{P} is split in two parts: the non-corrupted behaviour (\Cref{line:comm-prot-static-resume} to \Cref{line:comm-prot-static-h-end}) and the corrupted one (\Cref{line:comm-prot-static-else} to \Cref{line:comm-prot-static-c-end}).
Both parts contain effectively the same preamble (\Cref{line:comm-prot-static-prea-s1} to \Cref{line:comm-prot-static-prea-e1}).
Those are additional messages that we inserted \emph{before} the protocol proper to fix the common reference string (CRS)--related interleavings, since this simplifies the proof.
The CRS (presented in \Cref{lis:comm-prot-static-crs} and described below) is a piece of code that \trg{P}, \trg{Q}, and \env{} all interact with.
To both simplify the proof and to establish a security argument (as explained in \Cref{sec:comm-static-proof}), we fix the interaction to be first \env{}, then \trg{P}, then \trg{Q}.
 
Rather than describing the exchange in text, we provide a diagrammatic representation of the expected message exchanges in \Cref{fig:comm-static-msg}.
Apart from the additional code that explicitates the corruption and from the code that simplifies the rigorous reasoning in the proofs, the essence of the protocol code is the same as in its previous definition by \citet{uc-comm}.

\begin{figure}[!ht]
\begin{msc}{Protocol}
	\newinst{p}{\trg{P}}{committer}
	\newinst{f}{\trg{Fcrs}}{crs}
    \newinst{a}{\trg{A}}{attacker}
    \newinst{z}{\comsc{Z}}{env}
    \newinst{q}{\trg{Q}}{receiver}
    \nextlevel

    \mymessalphanl{z}{\comsc{CrptNo}}{p}
    \mymesstau{p}{\trg{(StartCrs)}}{f}
    \mymesstau{f}{\trg{(Started)}}{p}
    \mymessalphanl{p}{\comsc{Ok}}{z}

    \mymessalphanl{z}{\comsc{GetCrs}}{f}
    \mymesstau{f}{\trg{(WaitCrs)}}{p}
    \mymesstau{p}{\trg{(Waited)}}{f}
    \mymessalphanl{f}{\comsc{PublicStrings\ \sigma\ pk0\ pk1}}{z}

   	\mymessalphanl{z}{\comsc{(Commit, b)}}{p}
   	\mymesstau{p}{\trg{(GetCRS)}}{f}
   	\mymesstau{f}{\trg{(PublicStrings\ {\sigma}\ pk0\ pk1)}}{p}
	\mymesstau{p}{\trg{(Commit', y)}}{q}
   	\mymesstau{q}{\trg{(GetCRS)}}{f}
   	\mymesstau{f}{\trg{(PublicStrings\ {\sigma}\ pk0\ pk1)}}{q}
	\mymessalphanl{q}{\comsc{(Receipt)}}{z}

	\mymessalphanl{z}{\comsc{(Open)}}{p}
	\mymesstau{p}{\trg{(Open', b, r)}}{q}
	\mymessalphanl{q}{\comsc{(Opened, b)}}{z}
\end{msc} 
\caption{
	\label{fig:comm-static-msg} 
	Diagram representation of the protocol interaction in the no-corruption case. 
	\trg{Red} actions are \emph{internal} steps while \comsc{black} ones are those that constitute a trace.
}
\end{figure}

\begin{lstlisting}[language=TRG,label=lis:comm-prot-static-crs,caption=Protocol CRS,escapechar=|]
process Crs =
	rd StartCrs from Read(PCrs)				|\label{line:comm-prot-static-crs-b1}|
	wr Started to Write(CrsP)
	rd GetCrs from Read(ZCrs)				|\label{line:comm-prot-static-crs-b2}|
	let s = takernd
	let r0 = takernd
	let r1 = takernd
	let pk0, _ = keygen(r0)
	let pk1, _ = keygen(r1)
	wr WaitCrs to Write(CrsP) 						|\label{line:comm-prot-static-crs-b3}|
	rd Waited from Read(PCrs)						|\label{line:comm-prot-static-crs-b4}|
	wr PublicStrings s pk0 pk1 to Write(CrsZ)
	rd Getcrs from Read(PCrs)
	wr PublicStrings s pk0 pk1 to Write(CrsP)	
	rd GetCrs from Read(QCrs)	
	wr PublicStrings s pk0 pk1 to Write(CrsQ)
\end{lstlisting}
Intuitively, the CRS functionality lets the parties read from a bit string that is sampled from a specified distribution during a one-time setup phase.
For the purposes of the protocol, this setup phase is assumed to be run honestly.
The protocol's use of a CRS sidesteps the impossibility of constructing a simulator for secure commitments in the \emph{plain model} (a setting that makes no assumptions about an honest setup phase); this impossibility result was proved by \citet[\Ss3]{uc-comm}.
As for the protocol, the core intuition behind this code and its original formulation is unchanged from the original paper \cite{uc-comm}.

The signature for the protocol (in \Cref{lis:comm-prot-static-idx}) highlights an interesting difference with the signature of the ideal functionality presented in \Cref{lis:comm-prot-static-ideal-idx}.
\begin{lstlisting}[language=TRG,label=lis:comm-prot-static-idx,caption=Protocol Interface,escapechar=|]
imports	Ip = Write(PCrs), Write(PZ), Write(PQ)
definitions Dp = Write(PCrs), Write(PZ), Write(PQ), Read(ZP), Read(CrsP)
exports Xp = Read(ZP), Read(CrsP)
imports	Iq = Write(QCrs), Write(QZ)
definitions Dq = Write(QCrs), Write(QZ), Read(PQ), Read(CrsQ)
exports Xq = Read(PQ), Read(CrsQ)
imports	Icrs = Write(CrsP), Write(CrsZ), Write(CrsQ)
definitions Dcrs = Write(CrsP), Write(CrsZ), Write(CrsQ), Read(PCrs), Read(ZCrs), Read(QCrs)
exports Xcrs = Read(PCrs), Read(ZCrs), Read(QCrs)

// the resulting signature after linking |\trg{P}|, |\trg{Q}|, and |\trg{Crs}| is
imports I = Write(PZ), Write(QZ), Write(CrsZ)
definitions D = Dp |$\cup$| Dq |$\cup$| Dcrs
exports X = Read(ZP), Read(ZCrs)
\end{lstlisting}
In fact, this signature is \emph{different} from the one of the ideal functionality: \emph{and righteously so}!
In fact, from a module perspective, at the protocol level, this code is all we consider, while at the functionality level, we do not just consider \src{F}, but also the simulator (that we discuss next).
Thus, the signature of the protocol module should be the same as the signature of the module obtained from linking the functionality and the simulator.
As we show below, this is the case, but in order to argue this point, we need to present the simulator first.

\subsubsection{Simulator}\label{sec:comm-static-sim}
The simulator for this proof contains a number of code snippets: the simulator proper, the fake common reference string, and the simulated parties.
The simulator proper is presented in \Cref{lis:s-comm-prot-static-main}.
\begin{lstlisting}[language=SRC,label=lis:s-comm-prot-static-main,caption=Simulator for the single-bit commitment.,escapechar=|] 
process S =
	rd !M from Read(ZP) 													|\label{line:s-comm-prot-static-start}|
	if M == CrptNO then 													|\label{line:s-comm-prot-static-resume}|
		wr CrptNO to Write(SP) 												|\label{line:s-comm-prot-static-pre1}|
		rd !Ack from Read(PS) 												|\label{line:s-comm-prot-static-pre2}|
		wr FakeSetup to Write(SCrs) 								// starts the fake CRS
		rd !Fake pk0 pk1 td0 td1 s from Read(CrsS)  // receives the fake CRS parameters
		wr Ok to Write(SP) 													|\label{line:s-comm-prot-static-pre3}|
		rd !Receipt from Read(FS)
		wr Receipt to Write(SQ)
		fwd(Read(FS))(Write(SQ))  											|\label{line:s-comm-prot-static-pre4}| 	
	else  														|\label{line:s-comm-prot-static-else}|
		wr CrptP to Write(SP)
		rd !Ack from Read(PS)
		wr FakeSetup to Write(SCrs)										// starts the fake CRS
		rd !Fake pk0 pk1 td0 td1 s from Read(CrsS)		// receives the fake CRS parameters
		wr Ok to Write(SP)
		rd !Commit' y from Read(PS)									// receive the Commit' message from P  |\label{line:s-comm-prot-static-commit2}|
		let g0 = (invert(<pk0, td0>, y)) 						// use the trapdoor to calculate if the 
																								//  committed bit 'y'' in the image of 
																								//  the key for 0
		let g1 = (invert(<pk1, td1>, xors(y, s)))		// use the trapdoor to calculate if the 
																								//  committed bit 'y'' in the image of
																								//  the key for 1
		if g0 then 															|\label{line:s-comm-prot-static-g0true}|
			wr Commit 0 to Write(SP) 		// instruct P (and then the F) to commit to 0 |\label{line:s-comm-prot-static-if0}|
		elif g1 then
			wr Commit 1 to Write(SP) 		// instruct P (and then the F) to commit to 1 |\label{line:s-comm-prot-static-if1}|
		else
			wr Commit 0 to Write(SP) 		// fake commitment, just to resume other processes
		fwd(Read(FS)(Write(SQ))) 					 						|\label{line:s-comm-prot-static-resume}|
		rd !Open' b r from Read(PS) 	// receive the Open' message from a (malicious?) P |\label{line:s-comm-prot-static-commit3}|
		if (b == 0 && g0) \/ (b == 1 && g1) then 							|\label{line:s-comm-prot-static-finalif}|
			wr Open to Write SP 				// instruct P (and then the F) to Open
			fwd(Read(FS))(Write(SQ))		// forward the Opened message to Q
		else error
\end{lstlisting}
The simulator receives the static corruption message (\Cref{line:s-comm-prot-static-start}) and behaves differently in case \src{P} is corrupted (\Cref{line:s-comm-prot-static-else}) or not (\Cref{line:s-comm-prot-static-resume}).
In either case, the simulator starts the fake crs and receives the fake parameters.
If \src{P} is not corrupted, the simulator acts as a proxy for the messages it receives \Cref{line:s-comm-prot-static-pre3,line:s-comm-prot-static-pre4}.
If \src{P} is corrupted, the simulator receives the intermediate \src{Commmit' y'} and \src{Open'} messages (\Cref{line:s-comm-prot-static-commit2} and \Cref{line:s-comm-prot-static-commit3} respectively).
Then, it uses its knowledge of the fake CRS to calculate whether the received commitment bit \src{y'} is in the image of the 0 or 1 key.
Once it learns this information, it sends the correct bit (0 or 1) to the ideal functionality (\Cref{line:s-comm-prot-static-if0} and \Cref{line:s-comm-prot-static-if1} respectively).
From here on, the simulator acts as a proxy, and in the end it checks whether the opened bit was correct (\Cref{line:s-comm-prot-static-finalif}).

The fake common reference string is responsible for setting up a common reference string whose backdoor is known to the simulator (\Cref{lis:s-comm-prot-static-fakecrs}).
\begin{lstlisting}[language=SRC,label=lis:s-comm-prot-static-fakecrs,caption=Fake CRS for the single bit commitment.,escapechar=|]
process FAKECRS =
	rd !FakeSetup from Read(SCrs)
	let pk0, td0 = keygen([1..1])		// lambda-long ones
	let pk1, td1 = keygen([1..1])		// lambda-long ones
	let r0 = takernd
	let r1 = takernd
	let s = xors(prg(pk0, r0), prg(pk1, r1))
	wr Fake pk0 pk1 td0 td1 s to Write(CrsS)			|\label{line:s-comm-prot-static-fakecrs-fake}|
	rd GetCrs from Read(ZCrs)
	wr WaitCrs to Write(CrsP)
	rd Waited from Read(PCrs)
	wr PublicStrings s pk0 pk1 to Write(CrsZ)
\end{lstlisting}
The fake CRS is a process we inherit from the original proof of \UC for this protocol~\cite{uc-comm}.
The reason we need this in the ideal world is dictated by some reductions that happen in the real world, which we now discuss.
Essentially, in the case \trg{P} is corrupted, it could behave maliciously or honestly.
In order to detect the honest case, the simulator must know whether its messages are conformant to the data communicated from the CRS (i.e., whether it used the string \trg{s}, and the keys \trg{pk0} and \trg{pk1}).
Thus, the first thing the simulator does is starting the fake CRS process, which computes a known common reference string \src{s} and that communicates the string as well as both the keys and the trapdoors to the simulator (\Cref{line:s-comm-prot-static-fakecrs-fake}).
With the trapdoor, the simulator will be able to undestand if the message received from \src{P} (\Cref{line:s-comm-prot-static-commit2}) is honest or not.

The dummy parties are elements that are present in the literature too, and here we prove that are needed in order to make the signatures of the modules match (\Cref{lis:s-comm-prot-static-pq}).
\begin{lstlisting}[language=SRC,label=lis:s-comm-prot-static-pq,caption=Simulator Parties,escapechar=|]
process P =
	rd !M from Read(ZP)
	wr M to Write(PS) 				|\label{line:comm-prot-static-proxyzs1}|
	rd !M from Read(SP)
	if M == CrptNO then
		wr Ack to Write(PS)
		rd Ok from Read(SP)
		wr Ok to Write(PZ)				|\label{line:comm-prot-static-proxysz1}|
		rd WaitCrs from Read(CrsP)
		wr Waited to Write (PCrs)
		loop(2)(
			fwd(Read(ZP))(Write(PF))	|\label{line:comm-prot-static-proxyzf1}|
		)
	else // M == CrptP
		wr Ack to Write(PS)
		rd WaitCrs from Read(CrsP)
		wr Waited to Write (PCrs)
		loop(2)(
			fwd(Read(ZP))(Write(PS)) 	|\label{line:comm-prot-static-proxyzs2}|
			fwd(Read(SP))(Write(PF))	
		)

process Q =
	fwd(Read(SQ))(Write(QZ))			|\label{line:comm-prot-static-proxysz2}|
\end{lstlisting}
Essentially, the dummy parties act as proxies: they forward environment messages to the functionality (\Cref{line:comm-prot-static-proxyzs1,line:comm-prot-static-proxyzs2}), and simulator messages to the environment (\Cref{line:comm-prot-static-proxysz1,line:comm-prot-static-proxysz2}) (or to the functionality \Cref{line:comm-prot-static-proxyzf1}).

The signature of the simulator is presented in \Cref{lis:s-comm-prot-static-signature}. 
\begin{lstlisting}[language=SRC,label=lis:s-comm-prot-static-signature,caption=Simulator Signature.,escapechar=|]
imports Is = Write(SP), Write(SCrs), Write(SQ)
definitions Ds = Write(SP), Write(SCrs), Write(SQ), Read(ZP), Read(PS), Read(CrsS), Read(FS)
exports Xs = Read(ZP), Read(PS), Read(CrsS), Read(FS)
imports Ifake = Write(CrsS), Write(CrsP), Write(CrsZ)
definitions Dfake = Write(CrsS), Write(CrsP), Write(CrsZ), Read(SCrs), Read(ZCrs), Read(PCrs)
exports Ifake = Read(SCrs), Read(ZCrs), Read(PCrs)
imports Idp = Write(PS), Write(PZ), Write(PCrs), Write(PF)
definitions Ddp = Write(PS), Write(PZ), Write(PCrs), Write(PF), Read(ZP), Read(SP), Read(CrsP)
exports Idp = Read(ZP), Read(SP), Read(CrsP)
imports Idq = Write(QZ)
definitions Ddq = Write(QZ), Read(SQ)
exports Idq = Read(SQ)

// the resulting signature after linking |\src{P}|, |\src{Q}|, |\src{FAKECRS}|, and |\src{S}| is
imports I = Write(CrsZ), Write(PZ), Write(PF), Write(QZ)
definitions D = Ds |$\cup$| Dfake |$\cup$| Ddp |$\cup$| Ddq
exports X = Read(ZP), Read(FS), Read(ZCrs), Read(ZP)


// the resulting signature after linking the result above with |$\src{F}$| is
imports I = Write(CrsZ), Write(PZ), Write(QZ)
definitions D = Ds |$\cup$| Dfake |$\cup$| Ddp |$\cup$| Ddq |$\cup$| Df
exports X = Read(ZP), Read(ZCrs)
\end{lstlisting}

When linking the simulator with the ideal functionality, we obtain a single process whose signature -- in terms of imports and export -- is the same of the protocol signature from \Cref{lis:comm-prot-static-idx}.
This is necessary, since such a (static) difference would already make any \UC (or \rhpref) result impossible.

\subsubsection{\RC Proof}\label{sec:comm-static-proof}
We can now define a compiler that translates \Cref{lis:comm-prot-static-if} into \Cref{lis:comm-prot-static-main} (\Cref{def:comp-comm-static}) and prove it is \rhpref (\Cref{thm:comm-comp-static}).
\begin{definition}[Compiler for the Static Corruption Commitment]\label{def:comp-comm-static}
	\begin{align*}
		\compcomm{\cdot}\isdef&\ \compcomm{ \prgscomm } \partialmapsto \prgtcomm
		&
		\text{ where } &\ 
			\prgscomm \text{ is \Cref{lis:comm-prot-static-if}}
		\\
		&&&\
			\prgtcomm \text{ is \Cref{lis:comm-prot-static-main}}
	\end{align*}
\end{definition}

\begin{theorem}[The \compcomm{\cdot} Compiler is \rhpref]\label{thm:comm-comp-static}
	\begin{align*}
		\vdash \compcomm{\cdot} : \rhpref
	\end{align*}
\end{theorem}
Like most proofs, the one of \Cref{thm:comm-comp-static} is tedious, so we now report the main insights we gained while doing it and leave the full proof in the technical report~\cite{ucissc-prisc}.
We believe these steps may be helpful for those that want to have a paper proof before attempting a mechanisation such as the one presented in the next section (as we did).
\begin{description}
	\item[General structure:] 
		First, we devise a dummy attacker for the commitment and prove \Cref{ax:dummy-att} and \Cref{ax:dummy-prog}, which yield \Thmref{thm:dummy-rc} (again, these uninformative proofs are left in the technical report).
		This lets to get rid of the target attacker and, in turn, ensures that we have all the code (protocol, ideal functionality, simulator) whose behaviour needs to be analysed for the proof.

		Then, intuitively, the proof consists of matching the reductions in the source (\prgscomm plus simulator) with those in the target (\prgtcomm), while keeping track of what actions generate traces.
		While seemingly dull, this exercise alone is useful in getting rid of inconsistencies such as the ideal functionality sending a message with some data, and the protocol sending a message with more data.

	\item[Interleavings:]
		One complication when doing this proof is the amount of interleavings that arise due to concurrency.
		For simplicity, we streamline these interleavings, which, in turn, leaves fewer cases to reason about in the proof.
 
		This is the reason why, for example, we added the code from \Cref{line:comm-prot-static-prea-s1} to \Cref{line:comm-prot-static-prea-e1} to the protocol of \Cref{lis:comm-prot-static-main}.
		Of course, changes in the target behaviour propagate to the source, and thus our simulator in \Cref{lis:s-comm-prot-static-main} also contains code that reduces interleavings.

	\item[\env{} talks to CRS:]
		One of the interleavings that we fixed is the environment talking to the CRS, which is an event that must happen prior to the protocol proper.
		The reason this must happen has to deal with one of the cases of the proof.
		When the sender party is corrupted, it forwards all messages from the environment.
		Thus, we have two cases: either the environment behaves like an honest sender, or it does not.
		In the former case, we need to know that the environment has communicated with the CRS in order to derive the goodness of the values that are being sent.
		Essentially, if we do not force this communication, then we have more cases to consider for the environment to not behave like a honest sender.
		Thus, to simplify the proof, we fixed the communication to happen before the protocol.

	\item[Security argument:] 
		One simplifying assumption of our model is the symbolic crypto model that we have added into \rilc (in \Cref{sec:rilc-crypto}).
		This common assumption~\cite{Sumii2000ParamCrypto,sumii_logical_2003,ABADI200237,articlespi} ensures that the security argument we need to consider is solely about the randomness distributions in traces.

		Specifically, when unravelling the reductions of the protocol, we obtain the following traces.
		Note that for simplicity, we present them in symbolic form, that is, instead of listing all the single-bit combinations, we represent their values abstractly, with a symbol:
		Also, we elide the trace of events between the environment and the \trg{Crs} of \Cref{lis:comm-prot-static-crs} with \trg{\ldots}.
		\begin{itemize}
			\item $(\trg{(CrptNo)? \cdot (Ok)! \cdot \ldots \cdot (Commit~b)? \cdot (Receipt)! \cdot (Open)? \cdot (Opened~b)!}, 1/2)$
			\item $(\trg{(CrptP)? \cdot (Ok)! \cdot \ldots \cdot (Commit'~y)? \cdot (Receipt)! \cdot (Open'~b~r)? \cdot (Opened~b)!}, 1/2\times1/\varphi)$
			\item $(\trg{(CrptP)? \cdot (Ok)! \cdot \ldots \cdot (Commit'~y)? \cdot (Receipt)! \cdot (Open'~b~r)? \cdot (Error)!}, 1/2\times1/(1-\varphi))$
		\end{itemize}
		As we can see, the total of these traces amount to $1$, so they capture all the behaviour of the protocol.
		In these traces, \trg{b} and \trg{y} are single-bit values (i.e., either \trg{0} or \trg{1}), while \trg{r} is a $\lambda$-long sequence of bits.
		The probabilities are the most interesting bit: the first $1/2$ comes from the environment choice on the corruption of \trg{P}, i.e., that is the probability of the environment sending the first message as a \trg{CrptNo} or as a \trg{CrptP}.
		Probability $1/\varphi$ is the probability of the environment guessing both \trg{b} and \trg{r}. 

		Following the steps mentioned above, we follow the source reductions and see that the source modules (simulator, ideal functionality, dummy parties and fake CRS) perform the same traces, at least wrt the emitted actions.
		We then have to argue that the distribution of all target traces matches the source one.
		Wrt the choice of the initial message, they are chosen by the environment using the same message distribution, so the first $1/2$ probability exists in source traces too.
		Wrt the probability of the environment guessing \src{b} and \src{r}, we notice that the environment has the same probability of generating those values, since source and target values come from the same set, and since the environment does not know more in the source than in the target.

		Thus we can conclude that the source does exactly the same traces presented above.
\end{description}

\subsubsection{Excess Code in Protocol}\label{sec:protocol-not-looking-nice}
After seeing the amount of code that is required in order to apply our approach, we see the protocol code having too much boilerplate code as a possible criticism.
We agree with this point, and we believe to be justified in having added said boolerplate, given the simplification they provide.
Moreover, it is worth noting that the additional boilerplate surrounds the original protocol code and thus can be morally separated from it.
 
Some of the boilerplate code have been added as part of the attacker model (e.g., \Cref{line:comm-prot-static-start} in \Cref{lis:comm-prot-static-main} where the corruption message is received) and that could have been handled differently.
For example, given the static corruption setting, we may want to say that in order to prove that a protocol \UC-realises a functionality, one has to provide different implementations: one for the protocol and one for each of the corruption cases considered.
That would simplify the presentation of the protocol, and protocol implementors would benefit from that.

The additional synchronisation messages provide another part of the boilerplate, but we believe that they simplify the pen-and-paper approach significantly, since there would many interleavings to consider.
An alternative would be to fix a scheduling of messages, and prove the protocol \UC-realises a functionality given a certain scheduling.
Finally, as we move to using tools to mechanise these proofs, that boilerplate may not be necessary.
As we report in \Cref{sec:mechanisation}, our encoding of this proof in \DEEPSEC lets us strip the protocol, the CRS, and the simulator of the synchronisation messages.

\subsection{\UC Commiments as \RC Proofs for the Adaptive Corruption Model}\label{sec:commitment-adaptive}

For the adaptive corruption setting, we need to adapt both the ideal functionality and the protocol.
The adaptive setting is a difficult one for commitments: it is known that adaptive security requires either an exotic assumption (e.g., securely erasable memory) or weaker security properties (\citet[\Ss1.2]{hirt21adaptive} discuss in recent work that sidesteps this issue).
For simplicity we take the latter approach, demonstrating a scheme that sacrifices the hiding property.
Thus, we start this section with the new ideal functionality (\Cref{sec:comm-adaptive-idf}).
Then we present the protocol (\Cref{sec:comm-adaptive-prot}), followed by the simulator (\Cref{sec:comm-adaptive-sim}) and my the proof itself (\Cref{sec:comm-adaptive-proof}).
As before, this proof has also be mechanised in \DEEPSEC, which we report in the next section.

\subsubsection{Ideal Functionality}\label{sec:comm-adaptive-idf}

The only difference between the ideal functionality for the adaptive corruption case (\Cref{lis:comm-adaptive-idf}) and the one for the static corruption case (\Cref{lis:comm-prot-static-if}) are the additional messages on \Cref{line:comm-adaptive-idf-start,line:comm-adaptive-idf-end}.
The functionality immediately leaks the commitment bit (\Cref{line:comm-adaptive-idf-start}) and then waits for input to resume the commitment (\Cref{line:comm-adaptive-idf-end}).

\begin{lstlisting}[language=SRC,label=lis:comm-adaptive-idf,caption=Ideal functionality for an only-binding commitment.,escapechar=|]
fob =
	rd Commit b from Read(PF)			// receive the commitment bit
	wr Leak b to Write(FS)				// immediately leak it 					|\label{line:comm-adaptive-idf-start}|
	rd DoCommit from Read(PF)			// resume the functionality as before	|\label{line:comm-adaptive-idf-end}|
	wr Receipt to Write(FS)
	rd Open from Read(PF)
	wr Opened b to Write(FS)
\end{lstlisting}

\subsubsection{Protocol (and CRS)}\label{sec:comm-adaptive-prot}

As before, we enrich all real-world parties (i.e., the committer, the receiver, and the CRS) with additional synchronisation messages, to limit the amount of interleavings we have to deal with in the proof.
The first party that starts is the \trg{CRS}, which awaits a message from the environment and then generates the CRS as before in \Cref{lis:comm-prot-static-main}.
Then it sends the public strings message to the environment first, then to the committer, and finally to the receiver.
The CRS acts as the synchronisation party between committer and receiver, with the additional messages \trg{WaitCrs}, \trg{Waited}, \trg{SynchOpen}, \trg{Synched}, \trg{SynchOpen'}, and \trg{Synched'}.
\begin{lstlisting}[language=TRG,label=lis:comm-prot-adaptive-crs,caption=CRS for adaptive corruption.,escapechar=|]
CRS =
	rd GetCrs from Read(ZCrs) 
	let s = takernd
	let r0 = takernd
	let r1 = takernd
	let pk0, _ = keygen(r0)
	let pk1, _ = keygen(r1)
	wr WaitCrs to Write(CrsP) 								// start the committer
	rd Waited from Read(PCrs)									// ack the committer started
	wr PublicStrings s pk0 pk1 to Write(CrsZ)	// send values to environment
	rd Getcrs from Read(PCrs)									// ack the committer wants the CRS
	wr WaitCrs to Write(CrsQ)									// start the receiver
	rd Waited from Read(QCrs) 								// ack the receiver started
	wr PublicStrings s pk0 pk1 to Write(CrsP)	// send values to the committer
	rd GetCrs from Read(QCrs)									// ack the receiver wants the CRS
	wr SyncOpen to Write(CrsP)								// tell the committer it can open 
																						//  the commitment
	rd Synched from Read(PCrs)								// ack the committer can open
	wr PublicStrings s pk0 pk1 to Write(CrsQ)	// send values to the receiver
	rd SynchOpen' from Read(PCrs)							// ack the committe can forward the open 
																						//  to the receiver
	wr SynchOpen to Write(CrsQ)								// tell the receiver it can open 
																						//  the commitment
	rd Synched from Read(QCrs)								// ack the receiver can open
	wr Synched' to Write(CrsP)								// tell the committer to send 
																						//		the open message
\end{lstlisting}

The committer (\trg{P}) and the receiver (\trg{Q}) protocols are very similar to their static-corruption counterparts save for one key difference.
There is no communication that happens on authenticated channels directly between \trg{P} and \trg{Q}: all protocol-related messages are transmitted on an attacker interface.
In our convention, that is visible in the channel names used in the communication: they are of the form \trg{AP}, \trg{PA}, \trg{AQ} or \trg{QA}.
This indicates that the message sent or received by the protocol party goes through the attacker (and thus possibly the environment) before reaching the other party.
In the protocol description we comment the additional lines that have been added for synchronisation, as well as those lines that have been added to give up the hiding property.
\begin{lstlisting}[language=TRG,label=lis:comm-prot-adaptive-main,caption=Only-binding commitment protocol for adaptive corruption.,escapechar=|]
processes P =
	rd WaitCrs from Read(CrsP)					// added synchronisation
	wr Waited to Write(PCrs)						// added synchronisation
	rd Commit b from Read(ZP)
	wr Leak b to Write(PA) 							// added for no hiding
	rd DoCommit from Read(AP) 					// added for no hiding
	wr GetCRS to Write(PCrs)
	rd PublicStrings s pk0 pk1 from Read(CrsP)
	let r = takernd
	let y = (if b == 0 then prg pk0 r else xors(prg(pk1, r), s))
	wr Commit' y to Write(PA)
	rd SyncOpen from Read(CrsP)					// added synchronisation
	wr Synched to Write(PCrs)						// added synchronisation
	rd Open from Read(AP)
	wr SyncOpen' to Write(PCrs)					// added synchronisation
	rd Synched' from Read(CrsP)					// added synchronisation
	wr Open' b r to Write(PA)
	
Q = 
	rd WaitCrs from Read (CrsQ) 				// added synchronisation
	wr Waited to Write(QCrs)						// added synchronisation
	rd Commit' y from Read(AQ) 
	wr GetCRS to Write(QCrs)
	rd PublicStrings s pk0 pk1 from Read(CrsQ)
	wr Receipt to Write(QA)
	rd SynchOpen from Read(CrsQ)				// added synchronisation
	wr Synched to Write(QCrs)						// added synchronisation
	rd Open' b r from Read(AQ)
	if (b == 0 && y == prg(pk0, r)) \/ (b == 1 && y == xors(prg(pk1, r), s))	|\label{line:comm-prot-adaptive-if-q}|
		wr Opened b to Write(QZ)
	else
		error
\end{lstlisting}

\subsubsection{Simulator}\label{sec:comm-adaptive-sim}

The simulator follows the same structure of the previous one, save for one detail.
Since the value of the committed bit is known, the simulator does not have to rely on inverting the \src{prg} used by the \src{(fake) CRS} in order to know what the commitment bit is.

\begin{lstlisting}[language=SRC,label=lis:s-comm-prot-adaptive-main,caption=Simulator for adaptive corruption.,escapechar=|]
S =
	rd StartS from Read(CrsS)
	wr FakeSetup from Write(SCrs)
	rd Fake pk0 pk1 td0 td1 s from Read(CrsS)
	wr Ok to Write(SCrs)
	rd Leak b from Read(FS)
	wr Leak b to Write(SQ)
	rd Receipt from Read(FS)
	let r = takernd
	let y = if b == 0 then prg pk0 r else xors(prg(pk1, r), s)							|\label{line:sim-adaptive-makey}|
	wr Commit' y to Write(SQ) 										 |\label{line:sim-adaptive-write-yprime}|
	rd Commit' y' from Read(PS)
	wr Receipt to Write(SQ)
	rd Opened b from Read(FS)
	wr Open' b r to Write(SQ)
	rd Open' b' r' from Read(PS)
	if (y ==y') && (b == b') && (r == r')
		wr Opened b to Write(SQ)
	else
		error
\end{lstlisting}

The fake CRS (\Cref{lis:s-comm-prot-adaptive-fakecrs}) is essentially the same as before save for the different synchronisation messages, which we indicate with a comment.
\begin{lstlisting}[language=SRC,label=lis:s-comm-prot-adaptive-fakecrs,caption=Fake CRS for adaptive corruption.,escapechar=|]
FAKECRS =
	rd GetCrs from Read(ZCrs) 				// added synchronisation
	wr StartS to Write(CrsS)					// added synchronisation
	rd !FakeSetup from Read(SCrs)
	let pk0, td0 = keygen([1..1]) 	
	let pk1, td1 = keygen([1..1])	
	let r0 = takernd
	let r1 = takernd
	let s = xors(prg(pk0, r0), prg(pk1, r1))
	wr Fake pk0 pk1 td0 td1 s to Write(CrsS)
	rd Ok from Read(SCrs)					
	wr WaitCrs to Write(CrsP)				// added synchronisation
	rd Waited from Read(PCrs)				// added synchronisation
	wr PublicStrings s pk0 pk1 to Write(CrsZ)
\end{lstlisting}

Since there are more messages being exchanged between the environment and the functionality, the dummy parties (\Cref{lis:s-comm-prot-adaptive-pq}) contain additional proxy messages.
For simplicity, some of these messages are forwarded to the simulator, as in \Cref{line:comm-adaptive-pq-fwstos1,line:comm-adaptive-pq-fwstos2}.
Those are messages that the environment should send to the simulator directly.
To avoid having the simulator listen on the environment channel, and then synchronise with the dummy party, we let the dummy parties be the only entities that talk to the environment.
The dummy parties then forward to the simulator those messages that are not meant for the simulator (in this case, the \src{Commit'} and the \src{Open'} messages).
\begin{lstlisting}[language=SRC,label=lis:s-comm-prot-adaptive-pq,caption=Simulator Parties for adaptive corruption.,escapechar=|]
P =
	rd WaitCrs from Read(CrsP)
	wr Waited to Write(PCrs)
	loop(2)(
		fwd(Read(ZP))(Write(PF))	// forward from environment to functionality
	)
	fwd(Read(ZP))(Write(PS)) 		// forward from environment to simulator	|\label{line:comm-adaptive-pq-fwstos1}|
	fwd(Read(ZP))(Write(PF))		// forward from environment to functionality 
	fwd(Read(ZP))(Write(PS))		// forward from environment to simulator 	|\label{line:comm-adaptive-pq-fwstos2}|

Q =
	loop(5)(
		fwd(Read(SQ))(Write(QZ))
	)
\end{lstlisting}

As before, the interfaces of the linked source parties (dummy \src{P}, dummy \src{Q}, \src{S}, \src{FAKECRS} and \src{fob}) are the same as the interfaces of the linked target ones (\trg{P}, \trg{Q}, and \trg{CRS}).

\subsubsection{\RC Proof}\label{sec:comm-adaptive-proof}

\begin{definition}[Compiler for the Adaptive Corruption Commitment]\label{def:comp-comm-adaptive}
	\begin{align*}
		\compcommad{\cdot}\isdef&\ \compcomm{ \prgscomm } \partialmapsto \prgtcomm
		&
		\text{ where } &\ 
			\prgscomm \text{ is \Cref{lis:comm-adaptive-idf}}
		\\
		&&&\
			\prgtcomm \text{ is \Cref{lis:comm-prot-adaptive-main}}
	\end{align*}
\end{definition}

\begin{theorem}[The \compcommad{\cdot} Compiler is \rhpref]\label{thm:comm-comp-adaptive}
	\begin{align*}
		\vdash \compcommad{\cdot} : \rhpref
	\end{align*}
\end{theorem}

The proof follows the same structure of the one presented in \Cref{sec:comm-static-proof}, so we leave it in the technical report~\cite{ucissc-prisc}.
The traces generated in the adaptive corruption case contain more actions than the previous ones, as presented below.
For simplicity, we only show the action trace, without the probabilities, and abstract the values symbolically.
We also elide the error trace where the last action is replaced by an error message.
This case happens when the environment does not act as a forwarder for the \comsc{Commit'} and the \comsc{Open'} messages, but in the case it acts maliciously and tampers with the communicated \comsc{y}, \comsc{b} and \comsc{r} values.
\begin{itemize}
	\item 
		\(
			\comsc{GetCrs?} \cdot \comsc{PublicStrings\ s\ pk0\ pk1!} \cdot \comsc{Commit\ b?} \cdot \comsc{Receipt\ b!} \cdot \comsc{DoCommit?} \cdot \comsc{Commit'\ y!} \cdot \comsc{Commit'\ y'?} \cdot \comsc{Receipt!} \cdot \comsc{Open?} \cdot \comsc{Open'\ b\ r!} \cdot \comsc{Open'\ b'\ r'?} \cdot \comsc{Opened~b!}
		\)
\end{itemize}
One interesting insight we encountered while doing said proof was the inability to complete it while trying to retain both the binding and hiding properties.
In fact, there comes a time (\Cref{line:sim-adaptive-write-yprime}) when performing the reductions that the simulator must create the commitment value (\src{y} in the code).
If the protocol is trying to retain the hiding property, the simulator has no data to conjure \src{y} from, so the proof gets stuck.
On the other hand, if the protocol gives up hiding, it knows enough data to create such \src{y}, as done in \Cref{line:sim-adaptive-makey}.

\medskip

The proofs provided in this section already showcase how to make good use of our connection in order to obtain rigorous, formal \UC proofs.
However, as we point out, many of the steps of these proofs can be automated, and this insight is what we expand upon in the next section, where we mechanise the proofs presented here.

\section{Mechanising these Proofs}\label{sec:mechanisation}

This section reports on how to mechanise the proofs of the previous section.
First, this section justifies the choice of the tool for this mechanisation: \DEEPSEC (\Cref{sec:tool-choice}).
Then it details how to model protocols in \DEEPSEC, i.e., how to go from the pen-and-paper protocol definitions of the previous section to the internal representation of \DEEPSEC (\Cref{sec:modelling-in-deepsec}).
Finally, this section reports the lessons learned while doing this mechanisation effort (\Cref{sec:lessons-learned}).

\subsection{Picking the Tool}\label{sec:tool-choice}
To explore the potential for mechanisation of the previous proofs, we investigated the applicability of off-the-shelf tools for the analysis of the previous case study. 
Our requirements were twofold. 
First, the tool must provide a high-degree of automation, since the proofs required to manually check the semantics of each program reduction. 
Second, the tool must rely on a language whose communication model is similar to the one of \rilc.
We could not use existing \ilc-based tools since they are currently limited to type-checking processes, with no symbolic execution of processes.
Since we are interested in knowing what is \emph{possible} from a mechanisation perspective, we accept a (small) semantic gap between \rilc and the native language of the tool.

Thus, our tool choice fell on \DEEPSEC~\cite{chevalDEEPSECDecidingEquivalence2018}, which is a decision procedure for process equivalence in the applied-pi calculus from the perspective of a Dolev-Yao attacker (or, environment, in \rilc terms).
\DEEPSEC represents protocols in the applied-$\pi$ calculus, whose semantics is fairly close to the one of \rilc and thus omitted.
The semantics of applied $\pi$ has three notable limitations.
First, applied $\pi$ does not have the cryptographic primitives required by the protocol (i.e., no \src{xors}, no \src{prg}, etc.).
Second, applied $\pi$ lacks an explicit module system like the one of \rilc.
Finally, applied $\pi$ processes are not necessarily confluent.
We need to account for these while modelling protocols and we explain in detail how we do this accounting in the next section.

Another limitation is that the decision procedure of \DEEPSEC is undecidable~\cite{abadiDecidingKnowledgeSecurity2006}, so the tool needs to bound the number of processes running in parallel by some fixed number.
Fortunately, the commit protocol we chose as fixed in \Cref{sec:commitment-static} only covers a single session, so our verification is not affected by this limitation.

\subsection{Modelling Protocols in \DEEPSEC}\label{sec:modelling-in-deepsec}
We now describe the choices behind modelling channels, primitives, and the various entities of the protocol.

\paragraph{Modelling Channels}
\DEEPSEC channels are one-directional, untyped channels, akin to \rilc channel ends, which can be labelled as public or private.
Communication over public channels ends up in \DEEPSEC traces while communication over private ones does not.
Thus, we set up one \DEEPSEC channel for each channel end in \Cref{sec:commitment-static}, that is, one channel between all pairs of protocol, subprotocol, environment and attacker that talk to each other.
The channels that interact with the environment are public, while those that do not are private.

In general, any channel whose read and write ends are both specified in the protocol should be private, to ensure communication on that channel does not generate any trace.
Dually, if at least one of the read and write ends are not specified in the protocol, that channel should be public, since it models communication with the environment.

\paragraph{Modelling Primitives}
The first limitation of \DEEPSEC is the lack of built-in cryptographic primitives that the commitment protocol relies on.
This is a design choice of \DEEPSEC, and the tool lets programmers define arbitrary primitives (via keyword \lstinline{fun}) and their reductions (via keyword \lstinline{reduc}).
Thus we add reductions for the \comsc{prg} and its inversion, as well as for \comsc{xors} (i.e., for those rules added in \Cref{sec:rilc-crypto}). 
\begin{lstlisting}
fun prg/2. 
fun keygen/1.
reduc invert(prg(keygen(trapdoor),seed),trapdoor) -> seed. 

fun xor/2.
reduc dexor1(a,xor(a,b)) -> b.
reduc dexor2(b,xor(a,b)) -> a.
\end{lstlisting}

\paragraph{Modelling Entities as Processes}
The second limitation of applied $\pi$ is its lack of a module system. 
To overcome this, we identify applied $\pi$ processes to have the same role as \rilc modules.
Thus we use process composition ($\mid$) as module composition ($\linksymbol$), since they behave in the same way.

We code a process for each of the entity in \Cref{sec:commitment-static}, in the real world we have \lstinline{committer} (for \trg{P}), \lstinline{receiver} (for \trg{Q}), and \lstinline{CRS}, while in the ideal world we have the ideal functionality \lstinline{COM} (for \src{F}), and then \lstinline{dummy-committer} (for \src{P}), \lstinline{dummy-receiver} (for \src{Q}), \lstinline{fakeCRS} (for \src{FAKECRS}), and \lstinline{simulator} (for \src{S}).
For the sake of space, we do not show the code of these processes here, we provide a list of all processes at the end of this section, together with a link to the code.

Below we present the code \lstinline{COM}, which we use to highlight how \DEEPSEC processes are essentially a complete transliteration (modulo syntactic difference) of their \rilc counterparts.
\begin{lstlisting}
let COM = 
    in(p2f,x22); let (=Commit,b) = x22 in
    out(f2a,Receipt);
    in(p2f,x23); let =Open = x23 in
    out(f2a,(Opened,b)).
\end{lstlisting}
The statement \lstinline{in(p2f,x22)} reads from a channel from the dummy party (\lstinline{p}) to the functionality (\lstinline{f}) and binds the result into variable \lstinline{x22}.
Then \lstinline{let (=Commit,b) = x22 in} checks that the message in \lstinline{x22} has the form \lstinline{(=Commit, b)}, for some value \lstinline{b}.
This is just like \lstinline[language=SRC]{rd Commit b from Read(PF)} in \Cref{lis:comm-prot-static-if}. 
Dually, \lstinline{out(f2a,Receipt)} writes to the channel from functionality (\lstinline{f}) to simulator (\lstinline{a}), just like \lstinline[language=SRC]{wr Receipt to Write(FS)} in \rilc (cfr \Cref{lis:comm-prot-static-if}). 

Unfortunately, the previous list of processes is not sufficient to successfully model the commitment protocol in \DEEPSEC because of the last limitation of the tool: its lack of confluence.
As they stand, there are interleavings of the execution that complicate the reasoning of trace equivalence.

Thus, we introduce another process, essentially an environment wrapper between the environment in \DEEPSEC and the processes we have defined.
This environment wrapper fixes the flow of messages from the environment, making the wrapped processes confluent.
This, in turn, eliminates the need to synchronise different entities (as done in \Cref{sec:commitment-static,sec:commitment-adaptive}), since the environment wrapper provides a synchronisation entity already.
The environment wrapper has the following code, which is an input-output alternation (we use the variable names to indicate what message is expected to be passed each time):
\begin{lstlisting}
let env = 
    in(z2e,x30corrupt);
    out(e2p,x30corrupt);
    in(p2e,x31ok);
    out(e2z,x31ok);
    in(z2e,x32getcrs);
    out(e2s,x32getcrs);
    in(s2e,x33pubstrings);
    out(e2z,x33pubstrings);
    in(z2e,x34commit);
    out(e2p,x34commit);
    in(q2e,x35receipt);
    out(e2z,x35receipt);
    in(z2e,x36open);
    out(e2p,x36open);
    in(q2e,x37opened);
    out(e2z,x37opened);
    0.
\end{lstlisting}
All wrapped processes talk to the environment via channel ends that end with \lstinline{e}, for example an input channel from the environment to the simulator is called \lstinline{e2s}.
The environment wrapper provides a binding for all those channels and forwards them through \lstinline{e2z} or \lstinline{z2e} channels, where \lstinline{z} is used to identify the \DEEPSEC environment.
The channels between the environment wrapper and all wrapped processes are set to private.

We compose the real and ideal world from these processes as follows:
\begin{lstlisting}
let realw = env | committer | receiver | CRS.
let idealw = env |  dummy_committer | dummy_receiver | COM | sim | fakecrs. 
\end{lstlisting}

In order to test that \lstinline{realw} and \lstinline{idealw} are trace equivalent, we issue the following \DEEPSEC command, which returns true, attesting to the equivalence of the two systems.
\begin{lstlisting}
query trace_equiv(idealw,realw).
\end{lstlisting}

The \DEEPSEC files can be found at the project page, \url{https://uc-is-sc.github.io/}, specifically at:
\begin{center}
    \url{https://uc-is-sc.github.io/deepsec/deepsec-commitment.zip}
\end{center}
and they include the following files.
Note that for simplicity, we split the proof of the corruption case in two files, one where the no-corruption case, and one for the case of the corruption of the committer:
\begin{description}
    \item[\texttt{com-nocor.dps}:] contains the proof for the static corruption case when no party is corrupted;
    \item[\texttt{com-corP.dps}:] contains the proof for the static corruption case when the committer is corrupted;
    \item[\texttt{com-adaptive.dps}:] contains the proof for the adaptive corruption case.
\end{description}

\subsection{Lessons Learned}\label{sec:lessons-learned}
One of the benefits of adding the \lstinline{env} process is that we can debug the trace equivalence proof step-by-step.
Essentially, the \lstinline{env} process is a sequence of input-output pairs, which identify what is the environment sending and what is it reading from whatever world (be it real or ideal) it is communicating with.
By considering such input-output pairs incrementally (i.e., first the first pair, then the first two pairs and so forth), we can incrementally check that trace equivalence holds.
And when it does not hold, we know precisely at which step did the equivalence fail.
We found ``attacks'' at almost every step, about half of them due simple typos, and the rest of them pointing out issues with the setup or the simulator related to the scheduling messages meant to achieve confluence.

The graphical user interface of \DEEPSEC allowed us to step through counter-examples in both real and ideal world, identifying the source of these issues. 
Analysis took a few seconds, so we could often fix them by trial-and-error.

In summary, the symbolic analysis of \DEEPSEC is an excellent starting point for mechanisation, and the automation obtained due to the Dolev-Yao model is very useful. 
Even though a cryptographic analysis would give stronger result, we observe that the majority of messages the environment receives is literally the same. 
Hence a hybrid approach that combines cryptographic reduction with symbolic execution could be beneficial. 

As previously mentioned, the confluence requirement of \rilc results in unnecessarily complicated and yields unnatural models of protocols.
This is a drawback that \rilc inherits from the classical \UC frameworks.
While confluence slightly simplifies the analysis (there are fewer traces to match between the real and ideal world, even though they are longer), \DEEPSEC was designed specifically to handle asynchronous communication, heavily exploiting parallel computing. 
Comparing our model with standard \DEEPSEC models, we see no reason why \DEEPSEC could not handle similar models with real asynchrony.
We are thus confident that future verification tools could easily draw from the algorithm of \DEEPSEC to handle asynchronous, non-deterministic communication.
Compared to \rilc, this would improve the generality and readability of the protocol models we analyse.

\section{Discussion}\label{sec:disc}

This section discusses the relation of our results to other notions in
\UC and \RC.
First, we briefly discuss other \UC notions and how they respect the axioms of \Cref{sec:connecting} (\Cref{sec:diff-uc}).
Then we discuss why our connection is the most precise, by showing that no other \RC notion connects with \UC, and that an existing connecting conjecture is false (\Cref{sec:wrong-conn}).
Finally, we argue the different choice of programming language style (reactive VS non-reactive, \Cref{sec:react-vs-nonreact-langs}).
\subsection{Different \UC Notions}\label{sec:diff-uc}

Several closely related notions of composable, simulation-based security exist in the literature, including GNUC~\cite{Hofheinz2011GNUC:-A-New-Uni}, IITM~\cite{kuestersIITMModelSimple2013}, EasyUC~\cite{canettiEasyUCUsingEasyCrypt2019}, \ILC~\cite{ilc}, iUC~\cite{iuc}, among others.
In this section we argue that the four axioms of \Cref{sec:connecting}, which that section discusses in the context of \UC as defined by \citet{canettiUniversallyComposableSecurity2001}, also hold in all of the previously mentioned models (and likely hold in other models from the literature).
As in the rest of this paper, we leave the many issues raised by \emph{computational} notions of \UC to future work.

To begin, we note that \Cref{ax:relation-uc} and \Cref{ax:canonical} hold by inspection for all of these systems: these two axioms merely require defining a trace model, $\ExecT$, and $\Exec$ commensurate with the computational model of the underlying notion of \UC.

\Cref{ax:nonprob}---the fact that non-probabilistic environments are complete---holds in all of these systems for the same reason it does in the \UC notion of \citet{canettiUniversallyComposableSecurity2001}: in all cases, the definition of simulation universally quantifies the environment \emph{after} binding the protocol and the adversary.
If there exists a probabilistic distinguishing environment, then there exists at least one set of random choices for which the environment succeeds, which we can hard-code into a deterministic environment.
For GNUC, see~\cite[Def.~7]{Hofheinz2011GNUC:-A-New-Uni}; for IITM, see~\cite[Def.~1]{kuestersIITMModelSimple2013}; for \ILC, see~\cite[\Ss6.3]{ilc}; for iUC, see~\cite[p.~8]{iuc}.
EasyUC mirrors Canetti's \UC definition in this regard, and Canetti explicitly states that non-probabilistic environments are complete~\cite[p.~47]{canettiUniversallyComposableSecurity2001}.

\Cref{ax:final-bit}---the fact that finite traces contain the final bit---also holds in all of these systems.
In GNUC, the environment's output is an arbitrary string, which is post-processed by a distinguisher that returns a bit~\cite[Defs.~6~and~7]{Hofheinz2011GNUC:-A-New-Uni}; this distinguisher is exactly the extraction function~$\beta$ of Axiom~4.
In IITM, either the environment writes a single bit to a \texttt{decision} tape or execution halts without writing such a bit, in which case the result is 0 by definition; the same holds in iUC~\cite[p.~7]{iuc}.
In ILC, \textsf{execUC} outputs a single bit~\cite[\Ss6.3]{ilc}.
In both EasyUC and Canetti's \UC, the environment terminates the trace with a final bit.

More generally, the axioms of \Cref{sec:connecting} appear to hold because of the \emph{structure} of composable security---the execution model, quantifier ordering, and distinguishing procedure.
Thus, we expect these axioms to hold for essentially all related notions of composable security.

\subsection{Wrong Connections}\label{sec:wrong-conn}
\rhpref is the only criterion we could have chosen to instantiate our connection.
Other candidates include other \RC criteria from the work of~\citet{rhc}, as well as a conjecture from \citet{conjecture}: fully-abstract compilation (\facref)~\cite{abadi}. 
This section first debunks the connection with other \RC criteria (\Cref{sec:wrong-other-rc}) before disproving the \facref conjecture (\Cref{sec:fac-disprove}).

\subsubsection{\rhpref and Other \RC Criteria}\label{sec:wrong-other-rc}
\citet{rhc} provide \RC criteria for preserving all known classes of hyperproperties, and arrange them in a partial order on their expressiveness.
In order to show that \rhpref, and only \rhpref, coincide with \UC, we prove that all neighbouring elements to \rhpref do not coincide with \UC.
These elements are:
\begin{itemize}
	\item \rghspref, the preservation of subset-closed hyperproperties (\schp);
	\item \rtpref, the preservation of arbitrary trace properties, which has been proven to be strictly weaker than \rhpref.
\end{itemize}

\paragraph{\rghspref Also Works}
\schp intuitively formalises the notion of refinement, and it has been proven to be what correct compilers preserve through compilation (in absence of attackers though)~\cite{journey-rel}.

Arbitrary hyperproperties are defined as sets of sets of traces, a hyperproperty \comsc{H} is \schp if for any set of set of traces \comsc{b} in \comsc{H}, then \comsc{H} also contains all subsets of \comsc{b}.
Formally: 
\[
	\schp \isdef \myset{ \comsc{H} }{ \forall\comsc{b'}\subseteq\comsc{b}\ldotp \pif\comsc{b}\in\comsc{H}\pthen\comsc{b'}\in\comsc{H} }
\]

There is a single difference between \rhpref and \rghspdef (below): while the former is a refinement notion, the latter is not. Instead, it is a coincidence, as the target and source behaviours are connected by a co-implication.
\begin{definition}[Robust Subset-Closed Hyperproperty Preservation]\label{def:rghsp}
	\begin{align*}
		\comp{\cdot}\vdash\rghspref \isdef 
			&\
			\forall\prgs{},\ctxt{}\ldotp\exists\ctxs{}\ldotp
			\behavt{\trg{\ctxt{}\hole{\comp{\prgs{}}}}}
			\subseteq
			\behavs{\ctxhs{}{\prgs{}}}
		\\
		\text{ or equivalently :}
			&\
			\forall\prgs{},\ctxt{}\ldotp\exists\ctxs{}\ldotp\forall\trace\ldotp
				\pif
				\trg{\ctxt{}\hole{\comp{\prgs{}}}}\semt\trace 
				\pthen
				\ctxhs{}{\prgs{}}\sems\trace
	\end{align*}
\end{definition}
Interestingly, this notion also coincides with \UC 
(proven in Isabelle/HOL in
\theoryTRef{RSCHPtoUC} 
and
\theoryTRef{UCtoRSCHP} ).
This suggests that with the \UC trace model, arbitrary hyperproperties and subset-closed ones collapse.
In fact, traces in \UC contain probabilities, and given a set of traces, the sum of their probabilities must be exactly 1.
\rghspref mandates that the traces of one system ($\trg{\ctxt{}\hole{\comp{\prgs{}}}}$) are contained in the other ($\ctxhs{}{\prgs{}}$), so the latter cannot contain any additional trace with non-zero probability.
Given that there cannot be any trace with zero probability (they cannot be emitted by any program, ever), the two sets must coincide, and thus the implication in \rghspref is equivalent to the co-implication of \rhpref.

Similar collapses of classes of hyperproperties are not novel, and there can be language operators that cause it (as discussed in \citet{rhc}).
Since they deter from the main point of the paper, we leave investigating additional collapses and the ramifications of this collapse for future work.

\paragraph{Other notions: Preserving Trace Properties, Hypersafety Properties and Relaxed Hypersafety Properties}
Trace properties, formally are defined as sets of traces~\cite{Schneider00,AbadiAAFKLS91,LamportS84},
capture capture many functional properties of programs.
Intuitively a compiler attains \rtpref if it preserves arbitrary trace properties through compilation~\cite{rhc}.

The difference between \rhpref and \rtpdef (below) is in the order of quantifiers: in the former the traces are quantified last, while in the latter, they are quantified \emph{before} the source context.
\begin{definition}[Robust Trace-Preserving Compilation]\label{def:rtp}
	\begin{align*}
		\comp{\cdot}\vdash\rtpref \isdef 
			\forall\prgs{},\ctxt{},\trace\ldotp 
			\exists\ctxs{} \ldotp
				\pif
				\trg{\ctxt{}\hole{\comp{\prgs{}}}}\semt\trace
				\pthen
				\ctxhs{}{\prgs{}}\sems\trace
	\end{align*}
\end{definition}

However, as reported by \citet{john1}, there exists a \UC-like notion whose quantifiers are in the same order as \rtpref: Reactive Simulatability (\comuc{RS})~\cite{rss}.
Indeed, at the formal level, the only thing that changes between \comuc{RS} and \UC is the quantifiers ordering:
\begin{definition}[Reactive Simulatability]\label{def:rs}
    \begin{align*}
        \prot{}\vdashrs \idfu{} \isdef
        \forall \patt{}, \env{}\ldotp  \exists \simu{} \ldotp
        \Exec[\env{},\patt{},\prot{}]
        \indistp
        \Exec[\env{},\simu{},\idfu{}]
    \end{align*}
\end{definition}
\citet{kuestersIITMModelSimple2013} state that \comuc{RS} and \UC coincide for \comuc{IITM} but not for the notion of \UC of \citet{uc}.
This would imply that \rtpref and \rhpref also coincide, given additional setting-specific assumptions.
We believe the connection between \comuc{RS} and \rtpref can be formally proven, though we leave investigating such specific assumptions for future work.

\smallskip

Additional collapses whose investigation we also leave for future work include arbitrary hypersafety properties and relaxed hypersafety properties.
Arbitrary \hs are hypersafety properties, which include meaningful security properties such as many flavours of non-interference~\cite{ClarksonS10}.
While safety properties are identified by a set of bad prefixes (that the safety property does not extend), hypersafety properties are identified by finite sets of sets of bad prefixes (which the hypersafety property does not extend).
Canonically, \hs are defined as finite sets of sets of prefixes, and in the finiteness of the first set lies the difference with \schp.
Assume we relax the notion of hypersafety to encompass \emph{infinite} sets of finite prefixes and call this set relaXed Hypersafety (\rhs)

\smallskip

We conjecture that in the specific trace model of \UC, \hs and \rhs coincide, since communication is bounded, value size is bounded and thus there cannot be an infinite set of traces.
We believe that \rhs and \schp collapse with the additional common assumptions on the trace model used in \UC.
This would just serve the purpose of illustrating what class does \UC morally correspond to, so we leave this for future work.

\subsubsection{Disproving The \facref Conjecture}\label{sec:fac-disprove}

Fully-abstract compilation (\facref) has been the de-facto standard for secure compilation until the recent proposal of \RC~\cite{scsurvey}.
Unlike \RC, \facref does not rely on a notion of traces, but on the notion of contextual equivalence~\cite{lcfConsidered} in order to specify security properties.
Two programs \prgc{1} and \prgc{2} are contextually equivalent (indicated as \comsc{\prgc{1}\ceq\prgc{2}}) if they co-terminate no matter what attackers (program contexts) they link against.
Formally, if we indicate single reduction steps as $\myred$ and $n$ such reductions as $\myred\redapp{n}$, termination of a (whole) program is defined as follows: 
\[
	\com{W\termc} \isdef \exists n\ldotp \com{W\myred\redapp{1}\cdots\myred\redapp{n} W'\nred}
\]
which leads to the following definition of contextual equivalence:
\[
	\comsc{\prgc{1}\ceq\prgc{2}} \isdef \com{\forall \ctxc{}\ldotp \ctxhc{}{\prgc{1}}\termc\piff\ctxhc{}{\prgc{2}}\termc }
\]

A compiler is fully abstract (indicated as $\vdash\comp{\cdot}:$\facdef) if it preserves (and reflects) contextual equivalence of source programs in their compiled counterparts.
Formally 
\[
	\vdash \comp{\cdot} : \facref \isdef \forall\prgs{1},\prgs{2}\ldotp \prgs{1}\ceqs\prgs{2}\piff\comp{\prgs{1}}\ceqt\comp{\prgs{2}}
\]
If we unfold the definitions of source and target contextual equivalence, we obtain the following statement:
\[
	\forall\ctxs{}\ldotp (\ctxhs{}{\prgs{1}}\termsl \piff \ctxhs{}{\prgs{2}}\termsl) \piff \forall \ctxt{}\ldotp (\trg{\ctxt{}\hole{\comp{\prgs{1}}}}\termt\piff\trg{\ctxt{}\hole{\comp{\prgs{2}}}}\termt)
\]
Thus, like \RC, \facref considers a robust notion of security, since it universally quantifies over all target program contexts.

\medskip

The simplest argument against the conjecture that \facref is \UC is counting arguments.
While \UC is \emph{propositional}, in that it talks about a single entity (i.e., the protocol or the ideal functionality) in each domain (concrete and abstract), \facref is \emph{relational}, in that it talks about pairs of source and target programs.
As such, when trying to derive an equivalence between the two notions, it is not possible to connect all of the elements: if \prgs{1} is \idfu{} and \comp{\prgs{1}} is \prot{}, what do \prgs{2} and \comp{\prgs{2}} correspond to?

More generally, \facref talks about \emph{arbitrary} program equivalences, since \prgs{1} and \prgs{2} can be chosen arbitrarily.
Instead, \UC talks about preserving the expected behaviour of an ideal functionality encoded in a protocol which interacts with real-world attackers.

\paragraph{What about Contextual Equivalence?}
If we model the `expected behaviour' as simply terminating or diverging, then we can see a similarity between the notion of contextual equivalence and \UC (despite its lack of an existentially-quantified simulator).
There, it is clearer what the two programs correspond to: \prgc{1} is the protocol while \prgc{2} is the ideal functionalty linked with the simulator.

Using contextual equivalence for proving \UC-like notions has been recently investigated by \citet{ipdl}.
We believe this approach is valid, and it can be seen as a specilisation of our result that uses termination as the environment final bit instead of traces.
In fact they rely on the dummy attacker theorem whose application we justify in \Cref{sec:dummy-for-rc}.
Additionally, since contextual equivalence only talks about a single language, it can only be applied to the setting where the language of the protocol and of the ideal functionality coincide.
While we have done the same here, our result is more general, and using \rhpref lets one use different languages for the protocol and for the ideal functionality.

\subsection{Reactive VS Non-Reactive Languages}\label{sec:react-vs-nonreact-langs}
We have seen how the connection can be set-up with a reactive language (\rilc), but as we mentioned before, we believe the connection holds also for non-reactive languages.

Recall that the main difference between reactive and non-reactive languages is that reactive ones have no notion of whole program, while non-reactive ones do.
A whole program has a well-defined entry point for the \mtt{main} and it cannot be effectively extended further.
Since all the dependencies are resolved, and the \mtt{main} is defined, any code that gets added to a whole program is essentially dead code.

In reactive languages (such as \rrilc), there is an explicit entity in the semantics that models the environment (a configuration with the write token: \comsc{\pair{\wrtk ; K ; \Xi ; C }}), and any communication with it yields a trace action.
With non-reactive programs, there is no such environment entity in the semantics, so one should ask what is a trace action in this setting.

To answer this question, notice that trace actions are what gets communicated on the environment interface, so we must identify the environment in the non-reactive setting.
We believe that in this setting the environment effectively gets merged with the attacker.
After all, as for the dummy attacker theorem, there is no need to have two universally-quantified entities (environment and attacker), and since there is no environment in this setting, both roles are covered by the attacker.
Thus, any communication on the interface between the attacker and the program needs to generate a trace action.

Notice, however, that the definition of traces is central to our development.
By changing the notion of traces and how they are defined, some of the presented axioms need changing.
We do believe, however, that none of the existing results break, and we now briefly discuss why.

\paragraph{\Thmref{thm:uc-is-sc}}
This result is untouched by the language setting.

\paragraph{\Thmref{thm:composition}}
In order to derive this result we rely on a number of composition operators.
First, $\linksymbol$ needs not return a whole program, lest its output be not composable with any other program.
Then $\linkwhole$ must also not return a whole program, just a complete one, the axioms that use it need not change (\Thmref{ax:inter-comp} and \Thmref{ax:const-elim}).
Finally, some technical machinery may be needed in order for \Thmref{ax:const-elim} to make sense.
Recall that its statement is: \( \prgs{}\relateAbs\prgt{} \pthen \prgo{} \linkwholeos \prgs{} \relateAbs \prgo{} \linkwholeot \prgt{} \).
Here, the programs of the premise would be whole programs, i.e., they define a \mtt{main} and they are complete.
It is therefore unclear what happens when something else links against them, so long as the new program defines the main, it can be called, otherwise it is dead code.
We expect the notion of the main entry point needs to be tweaked in order for this composition to make sense and leave studying this for future work.

\paragraph{\Thmref{thm:dummy-rc}}
Without a notion of environment, the dummy attacker theorem makes little sense: we cannot replace the only attacker entity with a dummy!
However, we can borrow existing results from work that used a trace model similar to those required here in order to obtain similar results to the dummy attacker theorem.

Recall that in this case we need traces to be those actions that are generated across the attacker-program interface.
Such trace semantics have been studied aplenty in the context of fully-abstract trace semantics~\cite{llfatr-j,javajr,Abadi:2012}, i.e., a trace semantics that is as precise as contextual equivalence.
A key element of those semantics is that they simplify reasoning by eliding the notion of attackers, which gets abstracted away.
Now suppose \T is equipped with a trace semantics that yields just those traces.
If the trace semantics of \T is fully-abstract wrt behavioural equivalence, we can use the trace semantics and eliminate the target attacker from the theorem statement.
This is essentially the same result we get in reactive languages when using the dummy attacker theorem.

With one such trace semantics for \T, we conjecture that we can follow a proof technique akin to the one used in reactive languages.
Essentially, from the target traces we can build the source-level attacker (or, simulator) \ctxs{}.
Then, we will have all source elements (program and simulator) and we can calculate whether they are trace-equivalent with the target compiled program.
This would be a simplification of an existing proof technique called trace-based backtranslation, that is often used in secure compilation work~\cite{catalinRSC,rsc,exorcising,scoo-j}.
We leave investigating this novel proof technique for future work.

\section{Related Work}\label{sec:rw}

\paragraph{Programming Languages Meets Cryptography}
Researchers have tried to connect programming languages and cryptography at length, as attested by the Dagstulh seminar of 2014~\cite{conjecture}.
Existing results span languages models as well as tools that bring the two worlds closer together.

From the language side, a number of languages provide more rigorous formalisation of those languages used by cryptographers.
\ILC is one such language, which the authors used as a way to encode the semantics of Interactive Turing Machines~\cite{ilc}.
As shown, \ILC is a good candidate to showcase our connection.
 
IPDL is another such language, which is equipped with an equational logic that can be used to reason about contextual equivalence of IPDL programs~\cite{ipdl}.
IPDL is also a good candidate for our connection, and its equational logic can likely be extended to reason about trace equivalence of IPDL programs (once they are lifted to the reactive setting).
In the case where the source and target languages differ, however, the IPDL logic would not be useful, since it is bound to a single language.

To show the high-level security of some low-level cryptographic primitives researchers have come up with high-level calculi~\cite{ABADI200237,appi,Abadi:2000:APC:325694.325734,abadiDecidingKnowledgeSecurity2006,fournetModularCodebasedCryptographic2011} or cryptographic-aware compilers~\cite{viaduct} -- a large body of work recently surveyed by \citet{sok-compile-smpc} .
Interestingly, \citet{viaduct} also discovered one half of our connection (\RC implies \UC), but were not interested in the other half\footnote{Personal communication.}, as we showcase in this work.

When it comes to tools, there are a number of verification tools for expressing cryptographic protocols and checking some of their properties.
These tools include
\DEEPSEC~\cite{chevalDEEPSECDecidingEquivalence2018}
EasyCrypt~\cite{pwhile}
CryptoVerif~\cite{blanchetComputationallySoundMechanized2008}
and Squirrel~~\cite{baelde:hal-03172119}.
All of them have a reactive formal language as the semantic foundation for their protocol specification language, but none is really built to prove \UC.
With the results of this work, we demonstrate that those tools that are built to prove (trace) equivalences can be used to provide \UC mechanised proofs.

\paragraph{\UC Works}

Universal composition was introduced in the seminal work of \citet{uc},
but the idea of describing security via an ideal functionality that
describes all non-attacking network traces via simulation goes back
much further~\cite{DBLP:journals/ipl/Goldreich90}.
We already discussed \UC and some of its descendants~\cite{iuc,ilc}.
Our work is a generalisation of this concept that removes the need to
specify the communication down to the machine model.
Our main result relates robust compilation to perfect emulation. For
a detailed discussion of the challenges concerning
\emph{computational} emulation, we refer to 
Hofheinz et al.~\cite{hofheinzPolynomialRuntimeComposability2013}.
Concepts from universal composability were transferred from \ITMs to other
languages, but also embeddings between different UC-like frameworks
have been shown (e.g., UC~\cite{uc} into IITM~\cite{rauschEmbeddingUCModel2022}).
One line of research considers the intricacies of \UC in the
applied-pi calculus~\cite{bohlSymbolicUniversalComposability2013,
delauneSimulationBasedSecurity2009}.
This line of research formulates emulation using process equivalences in the
applied-pi calculus and composition using a rewriting of the channels
used inside a process. 
This requires to track which channels are
used for network communication, which for the environment and which
are external. This requires several well-formedness conditions on the
processes and complicates the definitions, while essentially
describing the interfaces at the meta level rather than inside of the
language.
By contrast, the composition operator for \rilc (\cref{sec:rilc-module-sys}) 
benefits from a module system, combining defined interfaces in a straight-forward way.
In the end, both types of composition behave very similarly. The
former approach is directly compatible with existing tools such as
ProVerif and \deepsec (which do not support a module system),
albeit the well-formedness conditions need to be checked manually. Our
approach is conceptually simpler and separates more clearly between
language semantics,
module system
and meta theory. 
In terms of programming languages, \UC has been adopted to a fragment
of Java~\cite{kustersHybridApproachProving2015} and (concepts like
ideal functionality and emulation) to a fragment of F\#~\cite{fournetModularCodebasedCryptographic2011a}.
Each of these works considers \UC (or related concepts) within their
respective target language. Viewing \UC as a form of robust
compilation stands to simplify deriving similar results for other
languages and help relate these results across languages.
The results just mentioned come with entirely different
verification methods, the former using program dependency graphs to obtain
emulation from non-interference~\cite{kustersHybridApproachProving2015},
the latter using refinement types to obtain assert-rely-style
compositionality by type-checking~\cite{fournetModularCodebasedCryptographic2011a}.
At the protocol level, the \UC variants related to the
applied-pi calculus~\cite{bohlSymbolicUniversalComposability2013,
delauneSimulationBasedSecurity2009} rely on off-the-shelf checkers for
protocol equivalences.
\citet{delauneSurveySymbolicMethods2017} provided a survey about these in
2007; a more recent survey with
a more general scope also contains a section on equivalence
properties~\cite{barbosaSoKComputerAidedCryptography2021}.
By and large, protocol equivalence checkers can handle prototols with
an unbounded number of sessions if the `two worlds' are structurally
similar or if syntactic conditions can ensure that results for a small number
of sessions translate to the unbounded model --- sometimes called
\emph{small-model results}. For a bounded number of sessions, and in
the Dolev-Yao model there are multiple decision procedures, providing
convenience and automation, with \deepsec being the most recent and
most advanced.
This points to an interesting research question: can we structure the
composition operator so that small-model results apply and we can use
these decision procedures (in the Dolev-Yao model?).

In terms of direct mechanisations of \UC, we already discussed
a formalisation in the cryptography-centric theorem-prover EasyCrypt~\cite{canettiEasyUCUsingEasyCrypt2019}
in \cref{sec:howto-uc-proofs}.
Constructive Cryptography, a \UC-like
framework, was formalised in
Isabelle/HOL~\cite{basinAbstractModelingSystem2021a},
These two are the most advanced mechanisation efforts for \UC-like
models in the computational model.
Both require an enormous amount of effort and expertise for
modelling and proving, but provide a very high degree of assurance.
There is hope though: there are verification tools in the computational model that provide
more automation~\cite{blanchetComputationallySoundMechanized2008} by using program transformations
that are computationally justified. A related line of research
justifies deduction systems that perform symbolic reasoning in the computational
model~\cite{baelde:hal-03172119,bartheSymbolicMethodsComputational2019}, likewise promising to
increase the amount of automation.
Some of these~\cite{blanchetComputationallySoundMechanized2008,baelde:hal-03172119}
model protocols in a computational variant of the applied-pi calculus,
suggesting that the choice between the Dolev-Yao model and the
computational model is not such a fundamental questions.
The translation and the composition operators might not be all that
different and the most automated reasoning tools in the computational
model take inspiration from those in the Dolev-Yao model.

\citet{john1} present a number of \UC-like notions and compare their expressiveness.
This looks like a good starting point to further connect programming language notions with cryptographic ones, as hinted in \Cref{sec:wrong-other-rc}.

The idea of providing high-level abstractions for cryptographic
protocols in programming languages predates \UC, of course. 
\citet{ABADI200237}, for example, translate a high-level language with
an abstraction for channels to a low-level language that implements
this channel via encrypted and channel communication. 
On the first glance, this provides \emph{less} flexibility: the
abstraction result corresponds not to \UC as a concept, but to
a singular emulation result.
On the other hand, this approach allows, in principle, a tighter
integration into the language. Consider, for example, communication
protocols, which are better abstracted as a channel, versus
cryptographic primitives like encryption, which are better abstracted
to symbolic terms.
Exploring whether \UC results can be related to \RC this way – which
is competing to the approach presented in this work – remains an open
problem.

\paragraph{\RC Works}
The theory of robust compilation was recently devised by \citet{rhc} and later expanded to account for differences between the source and target trace models~\cite{journey-rel-j}.
We believe our connection does not need to account for source and target trace difference since \UC deals with the same language for protocols and functionalities.
 
A novel criterion for secure compilation, \RC has been compared to existing secure compilation notions such as fully-abstract compilation~\cite{foxhound}.
We reported that fully abstract compilation is not a good candidate for our connection in \Cref{sec:fac-disprove}.
 
\RC has been used to reason about the preservation of arbitrary safety properties through compilation~\cite{rsc,rsc-j} and through translation validation~\cite{busi-transval} even with mechanised proofs~\cite{akram}.
Additionally, it has been used to reason about the absence of leaks due to speculation attacks such as Spectre~\cite{exorcising}.

\section{Conclusion}\label{sec:conc}
This paper presented a striking connection between the cryptographic framework of Universal Composability (\UC) and the secure compilation criterion called Robust Hyperproperty Preservation (\rhpref).
This paper first formalised (and mechanised in Isabelle) this connection and then identified a the requirements for lifting this connection to an \rhpref-compiler between arbitrary languages.
Then, the paper presented a formal language that fulfills these requirements, encoded a single-commitment bit protocol in that language, and proved that the protocol is \UC by proving that the compiler generating that protocol attains \rhpref. 
Finally, the paper mechanised the \rhpref proof in Deepsec for both the statci and the dynamic corruption cases, providing a scalable, mechanised proof of \UC via our connection.

\begin{acks}
	The authors would like to thank
		Carmine Abate,
		Amal Ahmed,
		Dan Boneh,
		Deepak Garg,
		John Mitchell,
		Jeremy Thibault
	for useful feedback and discussions.
	This work was partially supported by 
		the Italian Ministry of Education through funding for the Rita Levi Montalcini grant (call of 2019).
\end{acks}

\newpage

\newpage
\appendix

\section*{Appendix Intro}
The appendices present auxiliary results for \rilc (\Cref{sec:rilc-aux}), for the commitments (\Cref{sec:comm-aux}), and the proofs for \Cref{sec:sec-comp-proofs} (\Cref{sec:proofs-aux}).
\section{Properties of \rrilc}\label{sec:rilc-aux}

\paragraph{Auxiliaries}

\begin{center}
	\typerule{Trace Alpha Equivalence}{}{
		\trace\aeq\trace'
	}{tr-aeq}
\end{center}
Traces are alpha-equivalent if they are the same up to the same consistent alpha-equivalent actions.
Actions are alpha-equivalent if they are the same but for the channel name.
We insist the actions are consistently alpha equivalence to ensure that some channel in one trace is renamed to the same channel in the other trace.

\begin{theorem}[Dummy Attacker does not Change Behaviour]\label{thm:dummy-beh}
	\begin{align*}
		\pif
			&
			\dummy{\ctxc{}}{\modc{}}
		\\
		\pthen
			&
			\behavc{\ctxc{}\linkc\modc{}} \aeq \behavc{\modc{}}
	\end{align*}
\end{theorem}
\begin{proof}
	The proof proceeds by contradiction.

	The sets being different means that there is a trace in one set but not in the other.

	Wlog we can talk about the different trace being different on a single action on a channel \chan{}, so we have 2 cases:
	\begin{itemize}
		\item $\trace\in\behavc{\ctxc{}\linkc\modc{}}$ and $\trace\notin\behavc{\modc{}}$.

		In this case, the trace is done by $\ctxc{}\linkc\modc{}$ but not by \modc{} alone.

		There are 2 cases: the action is done by \modc{} or by \ctxc{}:
		\begin{itemize}
			\item The traces done by \modc{} have actions with either the environment channels or the attacker channels, so we have two cases:
			\begin{itemize}
				\item The action is done on an environment channel.

				By definition, those are in \behavc{\mod{}}, but this is in contradiction with the premise.
				\contradiction

				\item The action is done on an attacker channel.
				
				By definition of dummy, \ctxc{} synchronises those actions over attacker channels and forwards them to the environment on a $\aeq$ channel.

				So, this generates an action that is $\aeq$ to the one done originally on the attacker channel.

				This contradicts the hypothesis that the traces is not done by \modc{} alone.
				\contradiction
			\end{itemize}
			\item The traces done by \ctxt{} are done on environment channels or on module ones:
			\begin{itemize}
				\item The action is done on a module channel.

				This does not generate an action, so there is no trace difference.
				\contradiction

				\item The action is done on an environment channel (proxying).

				This is analogous to the 2nd point above.
				\contradiction
				
				\item The action is done on an environment channel (not proxying).

				But this is not possible because \ctxc{} is the dummy.
				\contradiction
			\end{itemize}
		\end{itemize}

		\item $\trace\notin\behavc{\ctxc{}\linkc\modc{}}$ and $\trace\in\behavc{\modc{}}$

		In this case the trace is done by \modc{} alone, but it is not done by $\ctxc{}\linkc\modc{}$.

		Thus, the trace contains an action done over a channel \chan{} that \ctxc{} reads but does not forward on a channel $\aeq\chan{}$.

		This is not possible because \ctxc{} is the dummy.
		\contradiction

	\end{itemize}
\end{proof}

\BREAK

\begin{theorem}[Full Confluence]\label{thm:f-conf}
	\begin{align*}
		\pif
			&
			\confc{} \Xto{\trace} \confc{_1}
		\\
		\pand
			&
			\isterm{ \confc{_1} }
		\\
		\pand
			&
			\confc{} \Xto{\trace'} \confc{_2}
		\\
		\pand
			&
			\isterm{ \confc{_2} }
		\\
		\pand
			&
			\inps{\trace}=\inps{\trace'}
		\\
		\pthen
			&
			\confc{_1} = g(\confc{_2})
		\\
		\pand
			&
			\isterm{ g(\confc{_2}) }
		\\
		\pand 
			&
			\outs{\trace} = \outs{\trace'}
	\end{align*}
\end{theorem}
\begin{proof}
	By \Thmref{thm:f-g-conf} with $f = \lam{x}{x}$.
\end{proof}

\BREAK

\begin{theorem}[Full Generalisd Confluence]\label{thm:f-g-conf}
	\begin{align*}
		\pif
			&
			\confc{} \Xto{\trace} \confc{_1}
		\\
		\pand
			&
			\isterm{ \confc{_1} }
		\\
		\pand
			&
			f(\confc{}) \Xto{\trace'} \confc{_2}
		\\
		\pand
			&
			\isterm{ \confc{_2} }
		\\
		\pand
			&
			\inps{\trace}=\inps{\trace'}
		\\
		\pthen
			&
			\confc{_1} = g(\confc{_2})
		\\
		\pand
			&
			\isterm{ g(\confc{_2}) }
		\\
		\pand 
			&
			\outs{\trace} = \outs{\trace'}
	\end{align*}
\end{theorem}
\begin{proof}
	By \Thmref{thm:conf-bigs} with HPs 1 and 3 we get Thesis 1.

	By HP 4 with \Thmref{thm:term-ren} we get Thesis 2.
\end{proof}

\BREAK

\begin{lemma}[Termination]\label{thm:term-ren}
	\begin{align*}
		\pif
			&
			\isterm{\confc{}}
		\\
		\pthen
			&
			\isterm{g(\confc{})}
	\end{align*}
\end{lemma}
\begin{proof}
	Trivial.
\end{proof}

\BREAK

\begin{lemma}[Generalised Confluence]\label{thm:conf-bigs}
	\begin{align*}
		\pif
			&
			\confc{} \Xto{\trace} \confc{_1}
		\\
		\pand
			&
			f(\confc{}) \Xto{\trace'} \confc{_2}
		\\
		\pand
			&
			\inps{\trace}=\inps{\trace'}
		\\
		\pthen
			&
			\confc{_1} = g(\confc{_2})
		\\
		\pand 
			&
			\outs{\trace} = \outs{\trace'}
	\end{align*}
\end{lemma}
\begin{proof}
	The proof proceeds by induction on \Xto{\trace}.
	\begin{description}
		\item[Base]

		Trivial by \Cref{tr:rrilc-t-refl}.
		\item[Inductive]

		By \Thmref{thm:conf-sing} with the inductive hypothesis, this holds.
			\qedhere
	\end{description}
\end{proof}

\BREAK

\begin{lemma}[Generalised Confluence Single]\label{thm:conf-sing}
	\begin{align*}
		\pif
			&
			\confc{} \xto{\ac} \confc{_1}
		\\
		\pand
			&
			f(\confc{}) \xto{\ac'} \confc{_2}
		\\
		\pand
			&
			\inps{\ac}=\inps{\ac'}
		\\
		\pand
			&
			\confc{_1}.G = \confc{_2}.G
		\\
		\pthen
			&
					\confc{_1} = g(\confc{_2})
		\\
		\pand
			&
			\outs{\ac}=\outs{\ac'}
	\end{align*}
\end{lemma}
\begin{proof}
By induction on $\xto{\cdot}$ wh:
\begin{description}
	\item[Base] We have these cases:
	\begin{itemize}
		\item \Cref{tr:rrilc-e-read} 

		Trivial, since the third premise ensures that $\ac$ and $\ac'$ are the same and the cryptographic state is also changed in the same way between the two steps.

		\item \Cref{tr:rrilc-e-write}
		Trivial: the $v$ and $c$ are the same in \confc{} and in $f(\confc{})$, so the outputs in \ac and $\ac'$ are the same.
		Since the $v$ are the same, the change to the cryptographic state are also the same, so \confc{1} and \confc{2} are the same, for $g = f$.

		\item \Cref{tr:rrilc-e-ch}
		Analogous to the \Cref{tr:rrilc-e-read} case.
	\end{itemize}

	\item[Inductive]  

	By \Cref{tr:rrilc-e-rest}, this holds by \Thmref{thm:conf-sing-noreact}.
	\qedhere
\end{description}
\end{proof}

\BREAK

\begin{lemma}[Generalised Confluence Single Not Reactive]\label{thm:conf-sing-noreact}
	\begin{align*}
		\pif
			&
			C \xto{\noact} C_1
		\\
		\pand
			&
			f(C) \xto{\noact} C_2
		\\
		\pthen
			&
			\begin{aligned}[t]
				\peither
					&
					C_1 = g(C_2)
				\\
				\por
					& \exists C_3\ldotp 
					\begin{aligned}[t]
							&
							{C_1} \xto{\noact} {C_3}
						\\
						\pand
							&
							g({C_2}) \xto{\noact} {C_3}
					\end{aligned}
			\end{aligned}
	\end{align*}
\end{lemma}
\begin{proof}
	This is the same result proved by \citet{ilc}.
\end{proof}

\BREAK

\begin{lemma}[Linking Commutativity]\label{thm:link-comm}
	\begin{align*}
		\behavc{\ctxc{}\linkc\modc{}} = \behavc{\modc{}\linkc\ctxc{}}
	\end{align*}
\end{lemma}
\begin{proof}
	Trivial.
\end{proof}

\BREAK

\begin{theorem}[\rilc satisfies \Cref{ax:par-decomp}]\label{thm:rilc-sat-att-decomp}
	\begin{align*}
			&
			\behavc{\ctxc{0}\linksymbol\prgc{}} =
			\behavc{(\ctxc{1}\linksymbol\ctxc{2})\linksymbol\prgc{}}
	\end{align*}	
\end{theorem}
\begin{proof}
	Trivial, pick \ctxc{1} = \ctxc{0} and \ctxc{2} = \come.
\end{proof}

\BREAK

\begin{theorem}[\rilc satisfies \Cref{ax:inter-comp}]\label{thm:rilc-sat-intertwine}
	\begin{align*}
			&
			\behavc{(\ctxc{1}\linksymbol\ctxc{2})\linksymbol(\prgc{1}\linksymbol\prgc{2})}
			=
			\behavc{(\ctxc{1}\linksymbol\prgc{1})\linksymbol(\ctxc{2}\linksymbol\prgc2)}
	\end{align*}
\end{theorem}
\begin{proof}
	By \Thmref{thm:link-comm}.
\end{proof}

\BREAK

\begin{theorem}[\rilc satisfies \Cref{ax:const-elim}]\label{thm:rilc-sat-const-elim}
	\begin{align*}
		\text{ if }
			&
			\begin{aligned}[t]
				\text{ if }
					&
					\com{\wprgc{1} \semc (\trace,\rho)}
				\\
				\text{ then }
					&
					\com{\wprgc{2} \semc (\trace,\rho)}
			\end{aligned}
		\\
		\text{ then }
			&
			\begin{aligned}[t]
				\text{ if }
					&
					\com{ (\wprgc{} \linksymbol {\wprgc{1}}) \semc ({\trace'},\rho') }
				\\
				\text{ then }
					&
					\com{ (\wprgc{} \linksymbol {\wprgc{2}}) \semc ({\trace'},\rho') }
			\end{aligned}
	\end{align*}
\end{theorem}
\begin{proof}
	Assume by contradiction that 
	\[
		(\wprgc{} \linksymbol {\wprgc{1}}) \semc ({\trace_1},\rho_1) \qquad\mi{HPW1}
	\]
	and 
	\[
		(\wprgc{} \linksymbol {\wprgc{2}}) \semc ({\trace_2},\rho_2) \qquad\mi{HPW2}
	\]
	for $(\trace_1,\rho_1) \neq (\trace_2,\rho_2)$.

	We have two cases:
	\begin{itemize}
		\item $\trace_1 = \trace'\cdot\ac\neq \trace'\cdot\ac' = \trace_2$

		In this case, the program \( (\wprgc{} \linksymbol {\wprgc{1}}) \) does a common trace $\trace'$ and then a different action $\ac$, while \( (\wprgc{} \linksymbol {\wprgc{2}}) \) does the same common trace $\trace'$ and a different action $\ac'$.

		From the definition of $\cdots \sem \{\trace,\rho\}$ we know that $\cdots,\nrnd \sem \{\trace\}$

		We have these cases:
		\begin{itemize}
			\item $\ac = \snd{v}{\chan{}} \neq \snd{v'}{\chan{}} = \ac'$

				This is in contradiction with the semantics of environments, which can generate any values, so both \com{v} and \com{v'}.

			\item $\ac = \snd{v}{\chan{}} \neq \snd{v'}{\chan{}'} = \ac'$

				Analogous to the case above.
			\item $\ac = \snd{v}{\chan{}} \neq \snd{v}{\chan{}''} = \ac'$

				Analogous to the case above.

			\item $\ac = \sndb{v}{\chan{}} \neq \sndb{v'}{\chan{}} = \ac'$

				We have two cases:
				\begin{itemize}
					\item $\chan{}$ is defined in $\wprgc{}$

					This is a contradiction, since $\wprgc{}$ is the same in both runs and $\nrnd$ is the same.

					\item $\chan{}$ is defined in $\wprgc{1}$ and $\wprgc{2}$

					This is in contradiction with HP.
				\end{itemize}

			\item $\ac = \sndb{v}{\chan{}} \neq \sndb{v'}{\chan{}'} = \ac'$

				Analogous to the case above.
			\item $\ac = \sndb{v}{\chan{}} \neq \sndb{v}{\chan{}''} = \ac'$

				Analogous to the case above.

			\item $\ac = \snd{v}{\chan{}} \neq \sndb{v'}{\chan{}'} = \ac'$

				Cannot occurr by the inductive structure of traces.

			\item $\ac = \sndb{v}{\chan{}} \neq \snd{v'}{\chan{}'} = \ac'$

				Cannot occurr by the inductive structure of traces.
		\end{itemize}

		\item $\trace_1=\trace_2$ and $\rho_1\neq\rho_2$

			By the definition of \behavc{\cdot}, and of $\wprgc{},\secpam \sem \{(\trace',\rho')\}$ and of $\wprgc{},\secpam,\nrnd \sem \{\trace'\}$, the probability is calculated based on the size of \totrand and on the size of $\{\trace\}$.

			The size of \totrand does not change between programs, so the difference between $\rho_1$ and $\rho_2$ can only come from the size of $\{\trace'\}$.

			The traces in $\{\trace'\}$ come from the behaviour of $\wprgc{}$, from the behaviour of $\wprgc{1}$ and $\wprgc{2}$ and from the environment-generated actions (? actions).

			Since $\wprgc{}$ is the same in both programs, it cannot lead to different probabilities.

			From HP, we know that the behaviour of $\wprgc{1}$ and $\wprgc{2}$ is the same, so it cannot lead to different probabilities.

			Since the environment is non-probabilistic, we know that environment-generated actions cannot lead to different probabilities.

			Since nothing can lead to different probablities, we have a contradiction.
	\end{itemize}
\end{proof}

\newpage
\section{Proving the Axioms for the Commitments}\label{sec:comm-aux}

The single-bit static corruption commitment protocol uses authenticated channels.
As such there is no direct communication between the attacker and any party or ideal functionality and we get \Cref{ax:dummy-att} for free.
However, there is direct communication between the protocol and the environment, and so we need to prove \Cref{ax:dummy-prog} in order to derive \Cref{thm:dummy-rc}.
Since environment channel names are named with \trg{Z}, we use \trg{D} to indicate channel ends of the dummy program.
\begin{lstlisting}[language=TRG,label=lis:dummy-prog-comm-adaptive,caption=Dummy protocol for the static corruption protocol.,escapechar=|]
|\prgtdummy| = 
	fwd DZ to ZPre		// for the static corruption message 
	fwd PreZ to ZD		// for message `Ok'
	fwd DZ to ZCrs 		// for mesasge `GetCrs'
	fwd CrsZ to ZD 		// for message `PublicStrings ...'
	fwd DZ to ZP 			// for message `Commit b'
	fwd QZ to ZD			// for message `Receipt'
	fwd DZ to ZP 			// for message `Open'
	fwd QZ to ZD			// for message `Opened b'
\end{lstlisting}

\begin{proof}[Proof of \Cref{ax:dummy-prog} for the adaptive commitment]
Let $\prgt{}$ be the code of \Cref{lis:comm-prot-static-main}, let $\prgtdummy{}$ be the code of \Cref{lis:dummy-prog-comm-adaptive}, and let $\prgs{}$ be the code of \Cref{lis:comm-prot-static-if}.

We need to show
$\forall \ctxt{}\ldotp \exists \ctxt{'}, \ctxs{'}\ldotp$
    \[
        \ctxt{}\linktt\prgt{} \relateAbs \ctxt{'} \linktt (\prgtdummy{} \linktt \prgt{})
        \text{ and }
        \ctxt{} \linktt (\prgtdummy{} \linktt \prgs{}) \relateAbs \ctxs{'}\links\prgs{}
    \]
Since the source and target languages are the same, and the relation is symmetric, it is sufficient to prove: 
$\forall \ctxt{}\ldotp \exists \ctxt{'}\ldotp$
    \[
        \ctxt{}\linktt\prgt{} \relateAbs \ctxt{'} \linktt (\prgtdummy{} \linktt \prgt{})
    \]

We pick $\ctxt{'}$ to be all the code from $\ctxt{}$ that does not deal with the channels bound in $\prgtdummy$, we need to show that the two sides of the equation generate the same traces.

We reason on generated $\trg{\at?\at!}$ actions (the argument then builds up inductively for traces):
\begin{itemize}
	\item if the \trg{\at?} action appears on a channel in \prgtdummy, the message is forwarded to \trg{\prgt{}} (which is the same on both sides of the equatiob), so the same \trg{\at!} is generated;
	\item if the \trg{\at?} action does not appear on a chanel in \prgtdummy, the message stays within the attacker.
	
	By construction, the logic in \ctxt{'} is the same as in \ctxt{} for these messages, so the same \trg{\at!} is generated.
	\qedhere
\end{itemize}
\end{proof}

\BREAK

Differently, the single-bit adaptive corruption protocol of \Cref{sec:commitment-adaptive} talks with an attacker, so we need to devise a dummy attacker and prove \Cref{ax:dummy-att}.
Since there is no communication between the environment and the protocol directly, there is no dummy protocol that can be specified and we get \Cref{ax:dummy-prog} for free.

The dummy attacker is in \Cref{lis:dummy-att-comm-adaptive}.
Since attacker channel names are named with \trg{Z}, we use \trg{E} to indicate channel ends of the environment.
\begin{lstlisting}[language=TRG,label=lis:dummy-att-comm-adaptive,caption=Dummy attacker for the adaptive corruption protocol.,escapechar=|]
|\ctxtdummy| =
	fwd EZ to ZCrs 		// for message `GetCrs'
	fwd CrsZ to ZE 		// for message `PublicStrings ...'
	fwd EZ to ZP			// for message `Commit b'
	fwd PZ to ZE			// for message `Leak b'
	fwd EZ to ZP			// for message `DoCommit'
	fwd PZ to ZE			// for message `Commit''
	fwd EZ to ZQ			// for message `Commit''
	fwd QZ to ZE			// for message `Receipt'
	fwd EZ to ZP 			// for message `Open'
	fwd PZ to ZE 			// for message `Open''
	fwd EZ to ZQ			// for message `Open''
	fwd QZ to ZE			// for message `Opened'
\end{lstlisting}

\begin{proof}[Proof of \Cref{ax:dummy-att} for the adaptive commitment]
Let $\prgt{}$ be the code of \Cref{lis:comm-prot-adaptive-main}, let $\prgtdummy{}$ be the code of \Cref{lis:dummy-prog-comm-adaptive} and let $\ctxtdummy{}$ be the code of \Cref{lis:dummy-att-comm-adaptive}.

We need to show:
$\forall \ctxt{}\ldotp \exists \ctxt{'}\ldotp$
    \[
        \ctxt{}\linktt\prgt{} \relateAbs (\ctxt{'}\linktt\ctxtdummy)\linktt \prgt{}
    \]
In this case we can simply pick $\ctxt{'}$ to be all the code from $\ctxt{}$ that does not deal with the channels bound in $\ctxtdummy$, we need to show that the two sides of the equation generate the same traces.

We reason on generated $\trg{\at?\at!}$ actions (the argument then builds up inductively for traces):
\begin{itemize}
	\item if the \trg{\at?} action appears on a channel in \ctxtdummy, the message is forwarded to \trg{\prgt{}} (which is the same on both sides of the equatiob), so the same \trg{\at!} is generated;
	\item if the \trg{\at?} action does not appear on a chanel in \ctxtdummy, the message stays within the attacker.
	
	By construction, the logic in \ctxt{'} is the same as in \ctxt{} for these messages, so the same \trg{\at!} is generated.
	\qedhere
\end{itemize}
\end{proof}

\newpage

\section{Proofs}\label{sec:proofs-aux}

\begin{proof}[Proof \Thmref{thm:comm-comp-static} ]	

	When linked together, the simulator (\Thmref{lis:s-comm-prot-static-signature}) and the ideal functionality (\Thmref{lis:comm-prot-static-ideal-idx}), we have the same module signature of the protocol (\Thmref{lis:comm-prot-static-idx}).
	
	We have these traces, where \trg{A} is the interaction between \env{} and the \trg{Crs}.
	\begin{itemize}
		\item $(\trg{(CrptNo)? \cdot (Ok)! \cdot A \cdot (Commit~b)? \cdot (Receipt)! \cdot (Open)? \cdot (Opened~b)!}, 1/2)$

		\item $(\trg{(CrptP)? \cdot (Ok)! \cdot A \cdot (Commit'~y)? \cdot (Receipt)! \cdot (Open'~b~r)? \cdot (Opened~b)!}, 1/2\times1/\varphi)$

		\item $(\trg{(CrptP)? \cdot (Ok)! \cdot A \cdot (Commit'~y)? \cdot (Receipt)! \cdot (Open'~b~r)? \cdot (Error)!}, 1/2\times1/(1-\varphi))$

	\end{itemize}
	
	Execution in the protocol of \Thmref{lis:comm-prot-static-main} starts on line \Cref{line:comm-prot-static-start} and we have:
	\begin{itemize}
		\item \Cref{tr:rrilc-e-read} (\trg{M?}); \Cref{tr:ilc-sem-ex-bop} (==); 
	\end{itemize}

	Execution in the simulator of \Cref{lis:s-comm-prot-static-main} starts on line \Cref{line:s-comm-prot-static-start} and we have:
	\begin{itemize}
		\item \Cref{tr:rrilc-e-read} (\src{M?}); \Cref{tr:ilc-sem-ex-bop} (==); 
		
	\end{itemize}

	By construction, $\src{M} == \src{CrptNo}$ or $\src{M} == \src{CrptP}$.

	Given their equivalent probability, we can assume \src{M}=\trg{M} (HPM).

	From HPM we have that the traces are the same, so we have two cases:
	\begin{description}

		\item[No Corruption: (\trg{M} == \trg{CrptNO})] 

		The reductions happening in \Thmref{lis:comm-prot-static-main} from line \Cref{line:comm-prot-static-resume}:

		So far the probability is that of no-corruption: 1/2.
		\begin{itemize}
			\item \Cref{tr:ilc-sem-ex-iftet}; \Cref{tr:rrilc-e-write} (\trg{Ok!})
			\item \Cref{tr:rrilc-e-read} (\trg{GetCrs?})
			\item \Cref{tr:rrilc-e-take}; \Cref{tr:ilc-sem-ex-letin}; \Cref{tr:rrilc-e-take}; \Cref{tr:ilc-sem-ex-letin}; \Cref{tr:rrilc-e-take}; \Cref{tr:ilc-sem-ex-letin}; \Cref{tr:rrilc-sem-ex-key2}; \Cref{tr:ilc-sem-ex-letin}; \Cref{tr:rrilc-sem-ex-key2}; \Cref{tr:ilc-sem-ex-letin}; 
			\item \Cref{tr:ilc-sem-rw} (fcrs - p);  \Cref{tr:ilc-sem-rw} (p - fcrs) 
			\item \Cref{tr:rrilc-e-write} (\trg{PublicStrings\ s\ pk0\ pk1!})
			\item \Cref{tr:rrilc-e-read} (\trg{Commit~b?})
			\item \Cref{tr:ilc-sem-ch} (p - fcrs) %
			\item \Cref{tr:ilc-sem-rw} (fcrs - p) %
			\item \Cref{tr:rrilc-e-take}; \Cref{tr:ilc-sem-ex-bop} (==); 

			We have 2 options, each with probability 1/2:
			\begin{itemize}
				\item (\trg{b} == \trg{0}) \Cref{tr:ilc-sem-ex-iftet}; \Cref{tr:rrilc-sem-ex-prg}; 
				\item (\trg{b} == \trg{1}) \Cref{tr:ilc-sem-ex-iftef}; \Cref{tr:rrilc-sem-ex-prg}; \Cref{tr:rrilc-sem-xor-seal}
			\end{itemize}
			So here, if we reason about the traces symbolically, the probabilities of this fork are rejoined, and we are still at probability 1/2.
			\item \Cref{tr:ilc-sem-ex-letin}  
			\item \Cref{tr:ilc-sem-rw} (p - q); %
			\Cref{tr:ilc-sem-ch} (q - fcrs) %
			\item \Cref{tr:ilc-sem-rw} (fcrs - q) %
			\item \Cref{tr:rrilc-e-write} (\trg{Receipt!})
			\item \Cref{tr:rrilc-e-read} (\trg{Open?})
			\item \Cref{tr:ilc-sem-rw} (p - q); %
			\Cref{tr:ilc-sem-ex-bop} (==); \Cref{tr:rrilc-sem-ex-prg}; \Cref{tr:ilc-sem-ex-bop} (==); \Cref{tr:ilc-sem-ex-bop} (\&\&); \Cref{tr:ilc-sem-ex-bop} (==); \Cref{tr:rrilc-sem-ex-prg}; \Cref{tr:rrilc-sem-xor-seal}; \Cref{tr:ilc-sem-ex-bop} (==); \Cref{tr:ilc-sem-ex-bop} (\&\&); \Cref{tr:ilc-sem-ex-bop} ($\vee$); 
			\item \Cref{tr:ilc-sem-ex-iftet};
			\item \Cref{tr:rrilc-e-write} (\trg{Opened~b!})
		\end{itemize}

		Thus, symbolically, we have this trace:
		\begin{itemize}
			\item $\trg{(GetCRS?)\cdot(PublicStrings\ s\ pk0\ pk1!) \cdot (Commit~b)? \cdot (Receipt)! \cdot (Open)? \cdot (Opened~b)!}$
		\end{itemize}

		With probability 1/2 (no-corruption).

		\medskip

		The source reductions continue in \Cref{lis:s-comm-prot-static-main} from before on \Cref{line:s-comm-prot-static-resume} as:

		So far the probability is that of no-corruption: 1/2.
		\begin{itemize}
			\item \Cref{tr:ilc-sem-rw} (s - p); \Cref{tr:ilc-sem-ex-bop}; \Cref{tr:ilc-sem-ex-iftet}; \Cref{tr:ilc-sem-rw} (p - s)
			\item \Cref{tr:ilc-sem-rw} (s - fakecrs); \Cref{tr:rrilc-sem-ex-key2}; 

			\src{pk0} and \src{td0} are chosen nondeterministically, so we can assume \src{pk0} = \trg{pk0} (HPK0).

			\item \Cref{tr:ilc-sem-ex-letin}; \Cref{tr:rrilc-sem-ex-key2}; 

			\src{pk1} and \src{td1} are chosen nondeterministically, so we can assume \src{pk1} = \trg{pk1} (HPK1).

			\item \Cref{tr:ilc-sem-ex-letin}; \Cref{tr:rrilc-e-take}; \Cref{tr:ilc-sem-ex-letin}; \Cref{tr:rrilc-e-take}; \Cref{tr:ilc-sem-ex-letin}; 
			\item \Cref{tr:rrilc-sem-ex-prg}; \Cref{tr:rrilc-sem-ex-prg}; \Cref{tr:rrilc-sem-xor-seal}; 

			The $\sigma$ returned by \src{\prgen{}{}{}} are chosen nondeterministically, and so is the \src{s} returned by \src{xors}, so we can assume \src{s}=\trg{s} (HPS).

			\item \Cref{tr:ilc-sem-ex-letin}; \Cref{tr:ilc-sem-rw} (fakecrs - s)
			\item \Cref{tr:rrilc-e-write} \src{Ok!}
			\item \Cref{tr:rrilc-e-read} \src{GetCRS?}
			\item \Cref{tr:ilc-sem-rw} (fakecrs - p); \item \Cref{tr:ilc-sem-rw} (p - fakecrs);
			\item \Cref{tr:rrilc-e-write} \src{PublicStrings\ s\ pk0\ pk1!}	 

			From HPK0, HPK1, HPS we know this action is the same as its target counterpart.
			
			\item \Cref{tr:rrilc-e-read} (\src{Commit~b?} z - p); 

			This reduction is chosen nondeterministically, so we can assume \src{b}=\trg{b} (HPB).

			\item \Cref{tr:ilc-sem-rw} (p - f); \Cref{tr:ilc-sem-rw} (f - S); \Cref{tr:ilc-sem-rw} (S - Q)
			\item \Cref{tr:rrilc-e-write} (\src{Receipt!})
			\item \Cref{tr:rrilc-e-read} (\src{Open?}); \Cref{tr:ilc-sem-rw} (p - f); \Cref{tr:ilc-sem-rw} (f - S); \Cref{tr:ilc-sem-rw} (S - Q)
			\item \Cref{tr:rrilc-e-write} (\src{Opened~b!})

			From HPB, we know this action is the same as its target counterpart.
		\end{itemize}

		The source emits the same trace, so this case holds.

		\item[Corrupted \trg{p}] 
		Thus we have these reductions in \Thmref{lis:s-comm-prot-static-main} from \Cref{line:comm-prot-static-resume} and then \Cref{line:comm-prot-static-else}:

		So far the probability is that of corruption-p: 1/2.
		\begin{itemize}
			\item \Cref{tr:ilc-sem-ex-iftef}; \Cref{tr:rrilc-e-write} (\trg{Ok!}) 
			\item \Cref{tr:rrilc-e-read} (\trg{GetCRS?})
			\item \Cref{tr:rrilc-e-take}; \Cref{tr:ilc-sem-ex-letin}; \Cref{tr:rrilc-e-take}; \Cref{tr:ilc-sem-ex-letin}; \Cref{tr:rrilc-e-take}; \Cref{tr:ilc-sem-ex-letin}; \Cref{tr:rrilc-sem-ex-key2}; \Cref{tr:ilc-sem-ex-letin}; \Cref{tr:rrilc-sem-ex-key2}; \Cref{tr:ilc-sem-ex-letin}; 
			\item \Cref{tr:ilc-sem-rw} (fcrs - p);  \Cref{tr:ilc-sem-rw} (p - fcrs) 
			\item \Cref{tr:rrilc-e-write} (\trg{PublicStrings\ s\ pk1\ pk1!})
			\item \Cref{tr:rrilc-e-read} (\trg{Commit'~y?} z - p)
			\item \Cref{tr:ilc-sem-rw} (p - q)
			\item \Cref{tr:ilc-sem-ch} (q - fcrs) %
			\item \Cref{tr:ilc-sem-rw} (fcrs - q) %
			\item \Cref{tr:rrilc-e-write} (\trg{Receipt!})
			\item \Cref{tr:rrilc-e-read} (\trg{Open'~b~r?})
			\item \Cref{tr:ilc-sem-rw} (p - q); %
			\Cref{tr:ilc-sem-ex-bop} (==); \Cref{tr:rrilc-sem-ex-prg}; \Cref{tr:ilc-sem-ex-bop} (==); \Cref{tr:ilc-sem-ex-bop} (\&\&); \Cref{tr:ilc-sem-ex-bop} (==); \Cref{tr:rrilc-sem-ex-prg}; \Cref{tr:rrilc-sem-xor-seal}; \Cref{tr:ilc-sem-ex-bop} (==); \Cref{tr:ilc-sem-ex-bop} (\&\&); \Cref{tr:ilc-sem-ex-bop} ($\vee$); 
		\end{itemize}
		Now, on \Cref{line:comm-prot-static-if-q} we have two options:
		\begin{enumerate}
			\item The attacker is not making the party deviate from the protocol.

			This happens with probability $1/\varphi$, so the probability is $1/2 \times 1/\varphi$.

			We have two options, each with probability 1/2:
			\begin{enumerate}
				\item if $\trg{pk0}\mapsto\pair{td0, r, \sigma} \in \trg{G}$ and $\trg{y}=\sigma$ (HPW0) then $\trg{b} = \trg{0}$
				\item if $\trg{pk1}\mapsto\pair{td1, r, \sigma} \in \trg{G}$ and $\trg{\sigma}\mapsto\pair{s,vy}\in\trg{G}$ and $\trg{y}=\trg{vy}$ (HPW1) then $\trg{b} = \trg{1}$
			\end{enumerate}
			So here, if we reason about the traces symbolically, the probabilities of this fork are rejoined, and we are still at probability $1/2 \times 1/\varphi$.

			In both cases, the reductions continue as:
			\begin{itemize}
				\item \Cref{tr:ilc-sem-ex-iftet}; \Cref{tr:rrilc-e-write} (\trg{Opened~b!})
			\end{itemize}

			\item or the attacker tries to cheat.

			This happens with probability $1/(1-\varphi)$, so the probability is $1/2 \times 1/(1-\varphi)$.

			In this case we know $\lnot$(HPW0 $\wedge$ HPW1) (HPW).

			\begin{itemize}
				\item \Cref{tr:ilc-sem-ex-iftef}; \Cref{tr:rrilc-e-error}
			\end{itemize}

		\end{enumerate}

		We need to show that the simulator replicates either trace correctly.
		
		The code of \Thmref{lis:s-comm-prot-static-main} starts with these reductions from \Cref{line:s-comm-prot-static-resume} and then \Cref{line:s-comm-prot-static-else}:

		So far the probability is that of corruption-p: 1/2.
		\begin{itemize}
			\item \Cref{tr:ilc-sem-ex-iftet}
			\item \Cref{tr:ilc-sem-rw} (s - p); \Cref{tr:ilc-sem-ex-iftet}; \Cref{tr:ilc-sem-rw} (p - s); \Cref{tr:ilc-sem-rw} (s - fakecrs) %
			\Cref{tr:rrilc-sem-ex-key2}; \Cref{tr:ilc-sem-ex-letin}; \Cref{tr:rrilc-sem-ex-key2}; \Cref{tr:ilc-sem-ex-letin}; \Cref{tr:rrilc-e-take}; \Cref{tr:ilc-sem-ex-letin}; \Cref{tr:rrilc-e-take}; \Cref{tr:ilc-sem-ex-letin}; \Cref{tr:rrilc-sem-ex-prg}; \Cref{tr:rrilc-sem-ex-prg}; \Cref{tr:rrilc-sem-xor-seal}; \Cref{tr:ilc-sem-ex-letin}; \Cref{tr:ilc-sem-rw} (fakecrs - s)
			\item \Cref{tr:rrilc-e-write} (\src{Ok!}) 
			\item \Cref{tr:rrilc-e-read} \src{GetCRS?}
			\item \Cref{tr:ilc-sem-rw} (fakecrs - p); \item \Cref{tr:ilc-sem-rw} (p - fakecrs);
			\item \Cref{tr:rrilc-e-write} \src{PublicStrings\ pk0\ pk1!}	 

			The same reasoning as before applies.

			\item \Cref{tr:rrilc-e-read} (\src{Commit'~y?} z - p); 

			Since it is nondeterministically chosen, assume \src{y}=\trg{y} (HPY).

			The same holds for \src{G} and \trg{G} throughout execution (HPC).

			\item \Cref{tr:ilc-sem-rw} (p - s)
			\item \Cref{tr:rrilc-sem-ex-inv2-t} or \Cref{tr:rrilc-sem-ex-inv2-f}; 
			\item \Cref{tr:ilc-sem-ex-bop} (==); \Cref{tr:ilc-sem-ex-letin}; (\Cref{tr:rrilc-sem-ex-inv2-f} or \Cref{tr:rrilc-sem-ex-inv2-t}); \Cref{tr:rrilc-sem-xor-seal}; \Cref{tr:ilc-sem-ex-bop} (==); \Cref{tr:ilc-sem-ex-letin};
		\end{itemize}

		We are now on \Cref{line:s-comm-prot-static-g0true}, we have these cases:
		\begin{description}
			\item[ $\src{g0} == \trues$] 
				
				in this case, \src{y} = $\sigma$ and $\src{pk0}\mapsto\pair{td0,r,\sigma}\in\src{G}$ and we define a proof metavariable $\src{B}=0$ (HPY0)

				We also know that $\src{g1} == \falses$ by the semantics of prg and xors with $\sigma$.

				The probability here is $1/2 \times 1/\varphi \times 1/2$ (choice of 0 over1)

			\item[ $\src{g0} == \falses$ and $\src{g1} == \trues$ ]
				
				in this case, \src{y} = $vy$ and $\src{pk1}\mapsto\pair{td1,r,\sigma}\in\src{G}$ and $\sigma\mapsto\pair{s,vy}\in\src{G}$ and we define a proof metavariable $\src{B}=1$  (HPY1)

				The probability here is $1/2 \times 1/\varphi \times 1/2$ (choice of 0 over1)

			\item[ $\src{g0} == \falses$ and $\src{g1} == \falses$ ]

				In this case, \Cref{tr:env-cry-base,tr:env-cry-key,tr:env-cry-xor} tell how \src{G} is populated.

				The functionality is called with a fake parameter 0, but we know that both \src{g0} and \src{g1} are is \falses (HPG).

				The probability here is ($1/2 \times 1/(1-\varphi)$).
		\end{description}
		Instead of replicating the proof in each case, note that the three cases are mutually exclusive, so later on we will rely on having either HPY0 or HPY1 or HPG.

		The reductions continue as:
		\begin{itemize}
			\item \Cref{tr:ilc-sem-rw} (s - p); \Cref{tr:ilc-sem-rw} (p - f); \Cref{tr:ilc-sem-rw} (f - s); \Cref{tr:ilc-sem-rw} (s - q); 
			\item \Cref{tr:rrilc-e-write} \src{Receipt!}
			\item \Cref{tr:rrilc-e-read} \src{Open'~b~r?}

			Again, they're nondeterministically chosen, so assume \src{b} = \trg{b} (HPB) and \src{r} = \trg{r} (HPR).

			\item \Cref{tr:ilc-sem-rw} (p - s); \Cref{tr:ilc-sem-ex-bop} (==); \Cref{tr:ilc-sem-ex-bop} (\&\&); \Cref{tr:ilc-sem-ex-bop} (==); \Cref{tr:ilc-sem-ex-bop} (\&\&); \Cref{tr:ilc-sem-ex-bop} ($\vee$);
		\end{itemize}

		We now have these cases on \Cref{line:s-comm-prot-static-finalif}:
		\begin{description}
			\item[HPY0] From HPY0 and HPY and HPR and HPC, we fulfil the premise of Item 1.a, so from HPW0 we get \trg{b=0}

				From HPB we have \src{b}=0 and from HPY0 again we know that \src{g0}=\trues and that \src{B}=0

				The reductions continue as:
				\begin{itemize}
					\item \Cref{tr:ilc-sem-ex-iftet}
					\item \Cref{tr:ilc-sem-rw} (s - p); \Cref{tr:ilc-sem-rw} (p - f); \Cref{tr:ilc-sem-rw} (f - s); \Cref{tr:ilc-sem-rw} (s - q); 
					\item \Cref{tr:rrilc-e-write} \src{Opened~B!}				
				\end{itemize}
				Since \src{B}=\src{b}=\trg{b}, the same trace is emitted.

			\item[HPY1] From HPY1 and HPY and HPR and HPC we fulfil the premise of Item 1.b, so from HPW1 we get \trg{b}=1

				From HPB we have \src{b}=a and from HPY1 again we know that \src{g0}=\falses and \src{g1}=\trues and that \src{B}=1

				The reductions continue as:
				\begin{itemize}
					\item \Cref{tr:ilc-sem-ex-iftet}
					\item \Cref{tr:ilc-sem-rw} (s - p); \Cref{tr:ilc-sem-rw} (p - f); \Cref{tr:ilc-sem-rw} (f - s); \Cref{tr:ilc-sem-rw} (s - q); 
					\item \Cref{tr:rrilc-e-write} \src{Opened~B!}				
				\end{itemize}
				Since \src{B}=\src{b}=\trg{b}, the same trace is emitted.

				We can compact the probability of this trace (1) with the one above (0) and reason symbolically that this trace has probability $1/2 \times 1/\varphi$.

			\item[HPG] From HPG and HPW we know the target errors.

				The reductions continue as:
				\begin{itemize}
					\item \Cref{tr:ilc-sem-ex-iftef}; \Cref{tr:rrilc-e-error}
				\end{itemize}
				
				So the same trace is emitted, with probability $1/2 \times 1/(1-\varphi)$.
		\end{description}
	\end{description}
\end{proof}

\BREAK

\begin{proof}[Proof of \Thmref{thm:comm-comp-adaptive} ]
	We have the following reductions in \Cref{lis:comm-prot-adaptive-main}, for simplicity we treat the attacker channel as the environment, and elide a further forwarding on that interface:
	\begin{itemize}
		\item \Cref{tr:rrilc-e-read} (\trg{GetCrs?})
		\item \Cref{tr:rrilc-e-take}; \Cref{tr:ilc-sem-ex-letin}; \Cref{tr:rrilc-e-take}; \Cref{tr:ilc-sem-ex-letin}; \Cref{tr:rrilc-e-take}; \Cref{tr:ilc-sem-ex-letin}; \Cref{tr:rrilc-sem-ex-key2}; \Cref{tr:ilc-sem-ex-letin}; \Cref{tr:rrilc-sem-ex-key2}; \Cref{tr:ilc-sem-ex-letin}; 
		\item \Cref{tr:ilc-sem-rw} (fcrs - p; Waitcrs);  \Cref{tr:ilc-sem-rw} (p - fcrs; Waited) 
		\item \Cref{tr:rrilc-e-write} (\trg{PublicStrings\ s\ pk0\ pk1!})
		\item \Cref{tr:rrilc-e-read} (\trg{Commit\ b?})
		\item \Cref{tr:rrilc-e-write} (\trg{Receipt\ b!})
		\item \Cref{tr:rrilc-e-read} (\trg{DoCommit?})
		\item \Cref{tr:ilc-sem-rw} (p - fcrs; GetCrs);  \Cref{tr:ilc-sem-rw} (fcrs - q; WaitCrs);  \Cref{tr:ilc-sem-rw} (q - fcrs; Waited); \Cref{tr:ilc-sem-rw} (fcrs - p; Publicstrings); 
		\item \Cref{tr:rrilc-e-take}; \Cref{tr:ilc-sem-ex-letin}; \Cref{tr:ilc-sem-ex-bop} (==); 

		We have 2 options, each with probability 1/2:
		\begin{itemize}
			\item (\trg{b} == \trg{0}) \Cref{tr:ilc-sem-ex-iftet}; \Cref{tr:rrilc-sem-ex-prg}; 
			\item (\trg{b} == \trg{1}) \Cref{tr:ilc-sem-ex-iftef}; \Cref{tr:rrilc-sem-ex-prg}; \Cref{tr:rrilc-sem-xor-seal}
		\end{itemize}
		So here, if we reason about the traces symbolically, the probabilities of this fork are rejoined, and we are still at probability $1$.

		\item \Cref{tr:ilc-sem-ex-letin};
		\item \Cref{tr:rrilc-e-write} (\trg{Commit'\ y!})
		\item \Cref{tr:rrilc-e-read} (\trg{Commit'\ y'?})
		\item \Cref{tr:ilc-sem-rw} (q - fcrs; GetCrs); \Cref{tr:ilc-sem-rw} (fcrs - p; SyncOpen); \Cref{tr:ilc-sem-rw} (p - fcrs; Synched); \Cref{tr:ilc-sem-rw} (fcrs - q ; PublicStrings); 
		\item \Cref{tr:rrilc-e-write} (\trg{Receipt!})
		\item \Cref{tr:rrilc-e-read} (\trg{Open?})
		\item \Cref{tr:ilc-sem-rw} (p - fcrs; SyncOpen'); \Cref{tr:ilc-sem-rw} (fcrs - q; SynchOpen); \Cref{tr:ilc-sem-rw} (q - fcrs; Synched); \Cref{tr:ilc-sem-rw} (fcrs - p ; Synched); 
		\item \Cref{tr:rrilc-e-write} (\trg{Open'\ b\ r!})
		\item \Cref{tr:rrilc-e-read} (\trg{Open'\ b'\ r'?})

		\item \Cref{tr:ilc-sem-ex-bop} (==); \Cref{tr:rrilc-sem-ex-prg}; \Cref{tr:ilc-sem-ex-bop} (==); \Cref{tr:ilc-sem-ex-bop} (\&\&); \Cref{tr:ilc-sem-ex-bop} (==); \Cref{tr:rrilc-sem-ex-prg}; \Cref{tr:rrilc-sem-xor-seal}; \Cref{tr:ilc-sem-ex-bop} (==); \Cref{tr:ilc-sem-ex-bop} (\&\&); \Cref{tr:ilc-sem-ex-bop} ($\vee$); 
		
		Let $1/\varphi$ be $1/\varphi_y\times1/\varphi_b\times1/\varphi_r$ i.e., the probability that \trg{y'}=\trg{y}, \trg{b'}=\trg{b} and \trg{r'}=\trg{r}.

		We have 2 options, the first with probability $1/\varphi$, the latter with the remaining probability:
		\begin{itemize}
			\item \Cref{tr:ilc-sem-ex-iftet}
			\begin{itemize}
				\item \Cref{tr:rrilc-e-write} (\trg{Opened~b!})

				with probability $1/\varphi$.
			\end{itemize}
			\item \Cref{tr:ilc-sem-ex-iftet};
			\begin{itemize}
				\item \Cref{tr:rrilc-e-error}

				with probability $1-1/\varphi$.
			\end{itemize}
		\end{itemize}
	\end{itemize}
	\begin{align*}
		\trg{t_1}, 1/\varphi \isdef
		&\
		\trg{(GetCrs)? }\cdot \trg{(PublicStrings\ s\ pk0\ pk1)!}
		\\
		&\
		\cdot \trg{(Commit~b)?} \cdot \trg{(Commit'~y)!}\cdot \trg{(Commit'~y')?} \cdot \trg{(Receipt)!}
		\\
		&
		\cdot \trg{(Open)?} \cdot \trg{(Open'~b~r)!}\cdot \trg{(Open'~b',r')?} \cdot \trg{(Opened~b)!}
	\end{align*}
	\begin{align*}
		\trg{t_1^w}, 1-1/\varphi \isdef %
		&\
		\trg{(GetCrs)?} \cdot \trg{(PublicStrings\ s\ pk0\ pk1)!}
		\\
		&\
		\cdot \trg{(Commit~b)?} \cdot \trg{(Commit'~y)!}\cdot \trg{(Commit'~y')?} \cdot \trg{(Receipt)!}
		\\
		&
		\cdot \trg{(Open)?} \cdot \trg{(Open'~b~r)!}\cdot \trg{(Open'~b',r')?} \cdot \trg{Error!}
	\end{align*}

	The simulator reduces as follows:
	\begin{itemize}
		\item \Cref{tr:rrilc-e-read} (\src{GetCrs?})
		\item \Cref{tr:ilc-sem-rw} (fakecrs - s; StartS); \Cref{tr:ilc-sem-rw} (s - fakecrs; Fakesetup);
		\item \Cref{tr:ilc-sem-rw} (s - fakecrs); \Cref{tr:rrilc-sem-ex-key2}; 

		\src{pk0} and \src{td0} are chosen nondeterministically, so we can assume \src{pk0} = \trg{pk0} (HPK0).

		\item \Cref{tr:ilc-sem-ex-letin}; \Cref{tr:rrilc-sem-ex-key2}; 

		\src{pk1} and \src{td1} are chosen nondeterministically, so we can assume \src{pk1} = \trg{pk1} (HPK1).

		\item \Cref{tr:ilc-sem-ex-letin}; \Cref{tr:rrilc-e-take}; \Cref{tr:ilc-sem-ex-letin}; \Cref{tr:rrilc-e-take}; \Cref{tr:ilc-sem-ex-letin}; 
		\item \Cref{tr:rrilc-sem-ex-prg}; \Cref{tr:rrilc-sem-ex-prg}; \Cref{tr:rrilc-sem-xor-seal}; 

		The $\sigma$ returned by \src{\prgen{}{}{}} are chosen nondeterministically, and so is the \src{s} returned by \src{xors}, so we can assume \src{s}=\trg{s} (HPS).

		\item \Cref{tr:ilc-sem-ex-letin}; \Cref{tr:ilc-sem-rw} (fakecrs - s ; Fake); \Cref{tr:ilc-sem-rw} (s - fakecrs; Ok); \Cref{tr:ilc-sem-rw} (fakecrs - p; WaitCrs); \Cref{tr:ilc-sem-rw} (p - fakecrs; Waited);
		\item \Cref{tr:rrilc-e-write} (\src{PublicStrings\ s\ pk0\ pk1!})

		From HPK0, HPK1, HPS we know this action is the same as its target counterpart.

		\item \Cref{tr:rrilc-e-read} (\src{Commit\ b?})

		This reduction is chosen nondeterministically, so we can assume \src{b}=\trg{b} (HPB).

		\item \Cref{tr:rrilc-e-write} (\src{Receipt\ b?})

		From HPB we have this is the same as its target counterpart.

		\item \Cref{tr:rrilc-e-read} (\src{DoCommit?})
		\item \Cref{tr:ilc-sem-rw} (p - s); \Cref{tr:ilc-sem-rw}; 
		\item \Cref{tr:rrilc-e-take}; \Cref{tr:ilc-sem-ex-letin}

		We have 2 options, each with probability 1/2:
		\begin{itemize}
			\item (\src{b} == \src{0}) \Cref{tr:ilc-sem-ex-iftet}; \Cref{tr:rrilc-sem-ex-prg}; 
			\item (\src{b} == \src{1}) \Cref{tr:ilc-sem-ex-iftef}; \Cref{tr:rrilc-sem-ex-prg}; \Cref{tr:rrilc-sem-xor-seal}
		\end{itemize}
		So here, if we reason about the traces symbolically, the probabilities of this fork are rejoined, and we are still at probability $1$.

		\item \Cref{tr:ilc-sem-ex-letin}; \Cref{tr:ilc-sem-rw} (s - q; Commit') ;
		\item \Cref{tr:rrilc-e-write} \src{Commit'\ y!}

		Since $\sigma$ are nondeterministically chosen, we have \src{y} == \trg{y}.
		
		\item \Cref{tr:rrilc-e-read} \src{Commit'\ y'?}
		\item \Cref{tr:ilc-sem-rw} (p - s; Commit'); \Cref{tr:ilc-sem-rw} (s - q; Receipt) 
		\item \Cref{tr:rrilc-e-write} \src{Receipt!}
		\item \Cref{tr:rrilc-e-read} \src{Open?}
		\item \Cref{tr:ilc-sem-rw} (p - s; Open); \Cref{tr:ilc-sem-rw} (s - p; Open); \Cref{tr:ilc-sem-rw} (p - f ; Open); \Cref{tr:ilc-sem-rw} (f - s; Opened); \Cref{tr:ilc-sem-rw} (s - q; Open' b r);
		\item \Cref{tr:rrilc-e-write} \src{Open'\ b\ r!}

		From HPB we know \src{b}=\trg{b}.

		Since we have the same random distributions fed in input to the runs, we get \src{r}=\trg{r} (HPR)
		
		\item \Cref{tr:rrilc-e-read} \src{Open'\ b'\ r'?}
		\item \Cref{tr:ilc-sem-rw} (p - s; Open'); 
		\item \Cref{tr:ilc-sem-ex-bop} (==); \Cref{tr:ilc-sem-ex-bop} (==); \Cref{tr:ilc-sem-ex-bop} (==); \Cref{tr:ilc-sem-ex-bop} (\&\&); \Cref{tr:ilc-sem-ex-bop} (\&\&); 

		Let $1/\varsigma$ be $1/\varsigma_y\times1/\varsigma_b\times1/\varsigma_r$ i.e., the probability that \src{y'}=\src{y}, \src{b'}=\src{b} and \src{r'}=\src{r}.

		Since \src{y'} and \src{y} are taken from the same space as \trg{y'} and \trg{y}, $\varsigma_y$ = $\varphi_y$.

		Since \src{b'} and \src{b} are taken from the same space as \trg{b'} and \trg{b}, $\varsigma_b$ = $\varphi_b$.

		Since \src{r'} and \src{r} are taken from the same space as \trg{r'} and \trg{r}, $\varsigma_r$ = $\varphi_r$.

		Thus, $\varsigma = \varphi$ and the probabilities of the source traces coincide with the probabilities of the target ones.

		We have 2 options, the first with probability $1/\varphi$, the latter with the remaining probability:
		\begin{itemize}
			\item \Cref{tr:ilc-sem-ex-iftet}
			\begin{itemize}
				\item \Cref{tr:rrilc-e-write} (\src{Opened~b!})

				with probability $1/\varphi$.

				From HPB we know \src{b}=\trg{b}.
			\end{itemize}
			\item \Cref{tr:ilc-sem-ex-iftet};
			\begin{itemize}
				\item \Cref{tr:rrilc-e-error}

				with probability $1-1/\varphi$.
			\end{itemize}
		\end{itemize}
	\end{itemize}

	So we get exactly the same traces, with the same probabilities, in both cases.
\end{proof}

\newpage 
\bibliography{../refs.bib,../RCisUC.bib}
\end{document}